\documentclass{aa}  
\usepackage{hyperref}
\usepackage{graphicx}

\usepackage{txfonts}
\usepackage{siunitx}
\usepackage{multirow}
\usepackage{placeins}
\usepackage{subfigure}
\usepackage{bm}

\newcommand{\Msun}{M$_{\odot}$}
\newcommand{\Rsun}{R$_{\odot}$}
\newcommand{\Lsun}{L$_{\odot}$}
\newcommand{\um}{\si{\micro\meter}}
\def\Hii{H\,{\sc ii} }

\begin{document}

   \title{The onset of stellar multiplicity in massive star formation: A search for low-mass companions of massive young stellar objects with $L'$-band adaptive optics imaging\thanks{Based on observations collected at the European Organisation for Astronomical Research in the Southern Hemisphere under ESO programme 090.C-0207(A).}}

   \author{E. Bordier\inst{1,2}, W.-J. de Wit\inst{1}, A. J. Frost\inst{1,2}, H. Sana\inst{2}, T. Pauwels\inst{2}, E. Koumpia\inst{1}}

   \institute{\inst{1}ESO, Alonso de Cordova 3107, Vitacura, Santiago 19, Chile \\
              \email{emma.bordier@protonmail.com} \\
              \inst{2} Institute of Astronomy, KU Leuven, Celestijnenlaan 200D, B-3001 Leuven, Belgium \\}

   \date{Received Month xx, xxxx; accepted Month xx, xxxx}

    \titlerunning{MYSO low-mass companions with $L'$-band adaptive optics imaging}
   \authorrunning{E.~Bordier et al.}

 
  \abstract
   {Given the high incidence of binaries among mature field massive stars, it is clear that multiplicity is an inevitable outcome of high-mass star formation. Understanding how massive multiples form requires the study of the birth environments of massive stars, covering the innermost to outermost regions. }
   { We aim to detect and characterise low-mass companions around massive young stellar objects (MYSOs) during and shortly after their formation phase. By the same means, we also probed the 3.8-\um\, emission that surrounds these massive protostars, in order to link the multiplicity to their star-forming environment.}
   {To investigate large spatial scales, we carried out an $L'$-band high-contrast direct imaging survey seeking low-mass companions (down to $L_{\text{bol}}\approx 10$\Lsun\, or late A-type) around thirteen previously identified MYSOs using the VLT/NACO instrument. From those images, we looked for the presence of companions on a wide orbit, covering scales from 300 to 56,000~au. Detection limits were determined for all targets and we tested the gravitational binding to the central object based on chance projection probabilities.}
   {We have discovered a total of thirty-nine potential companions around eight MYSOs, the large majority of which have never been reported to date. We derived a multiplicity frequency (MF) of $62\pm13$\% and a companion fraction (CF) of $3.0\pm0.5$. The derived stellar multiplicity and companion occurrence are compared to other studies for similar separation ranges. The comparisons are effective for a fixed evolutionary stage spanning a wide range of masses and vice versa. We find an increased MF and CF compared to the previous studies targeting MYSOs, and our results match the multiplicity rates derived among more evolved populations of massive stars. For similar separation ranges, we however confirm a higher multiplicity than that of T-Tauri stars ($\sim$30\%), showing that the statement in which multiplicity scales with primary mass also extends to younger evolutionary stages. The separations at which the companions are found and their location with relation to the primary star allow us to discuss the implications for the massive star formation theories.}
   {Our findings do not straightforwardly lift the uncertainty as to the formation process of massive stars as a whole but we rather examine the likely pathways for individual objects. However, the wide distance at which companions are detected rather supports core fragmentation or capture as the main mechanisms to produce wide multiples. We find hints of triggered star formation for one object and discuss the massive star against stellar cluster formation for other crowded fields. }
   


   \maketitle
%

\section{Introduction}

Ever since it has been unanimously established that a significant fraction of stars, regardless of their mass, evolve in pairs or multiples, the study of star formation has taken a major turning point. Yet, there is not a consensus on the physical processes and mechanisms that settle multiplicity, and it is of prime importance to include multiple formation into star formation studies, in order to comprehensively understand star formation as a whole. The studies of young populations of low-mass stars aim to quantify the intrinsic physical properties of multiple star systems and the distribution of binary parameters accurately, and to unravel the complex processes and history of binary formation. The presence and abundance of these low-mass (proto-)stars in our close environment (within 1~kpc), in clusters or as field stars and visible in the optical light, has made it possible to converge on reliable statistics for this mass regime \citep[see for instance:][]{Duchene+2007,Connelley+2008,Raghavan+2010,Kraus+2011,Duchene+2013,WardDuong+2015,Elliott+2015,Tobin+2016,ElliottBayo+2016,Moe+2017,Kounkel+2019,Zuniga+2021,Tobin+2022}. These studies converge towards a fraction of 50\% of solar-type stars in the field with a companion, with an increased incidence in star-forming regions, indicating that many low-mass stars are born in binaries. These statistics have been derived from a wide range of observing techniques, ranging from spectroscopy, interferometry, sparse aperture masking, and adaptive optics (AO) assisted imaging, and thus they provide a comprehensive view of binarity from the innermost to the outermost environments (from 0.1~au to a couple dozen $10^{4}$~au). 

The question of the formation of massive binaries that populate the upper part of the Hertzprung-Russell (HR) diagram (M$_{\text{i,primary}}>8$~\Msun) is all the more crucial as massive stars are almost entirely observed while on or already off the zero-age main sequence (MS) phase, as a result of their rapid evolution. As such, most surveys about multiplicity refer to fully formed MS O and B stars, for which nearly 100\% are known to belong to a multiple system with a high tendency towards close ($\lesssim 1$~au) orbits \citep{Sana+2011,Sana+2012,Sana+2013,Sana+2014,Dunstall+2015,Moe+2017}. Despite the observational obstacles to image young massive stars, teams have combined efforts to address binarity in the younger stages of massive stars (\citealt{Apai+2007,Pomohaci+2019,Koumpia+2019,Koumpia+2021,Shenton+2023}). Binarity in the high-mass regime is known to significantly impact the evolution and fate of the systems, and as such constitutes an important axis of study for clarifying the origin of gravitational waves and understanding the dynamical and chemical structure of our Galaxy. Nonetheless, even with multiplicity being a ubiquitous feature of massive stars, binarity rates and orbits derived from MS populations do not necessarily reflect the primordial binary properties as the latter is impacted by secular evolution and dynamical processes \citep[see e.g. the review of][]{Offner+2022}. An important point to address concerns the physical processes and mechanisms that settle multiplicity. 

Numerous theories try to explain how binaries and multiples are formed. They include disk fragmentation, core fragmentation, and capture in the protostellar phase \citep{Tohline2002,Kratter+2008,Krumholz+2009,Meyer+2018}. These proposed pathways do not solely apply to massive binaries, but rather explain the possible pairing mechanisms regardless the mass of the primary. 
Dynamical capture occurs when, in dense stellar environments where close interactions are frequent, an object initially born single and unbound to any other system may capture passing protostars to form a multiple system \citep{Tohline2002}. This process is expected to take place later on, during the dissolution of the star-forming region \citep{Fujii+2011,Parker+2014}. 

The two most prominent theories involve fragmentation within the gravitationally collapsing parent cloud. However, the timescales and scales at which fragmentation occurs is still very uncertain. Fragmentation of a turbulent pre-stellar core occurs when density inhomogeneities caused by turbulence create clumps above the local Jeans mass that collapse faster than the background main core, ultimately forming a multiple system with separations 100~au\,$\lesssim a \lesssim 10,000$~au \citep{Goodwin+2004,Offner+2010,Myers+2013}. In disk fragmentation theories, multiplicity is presumably generated by an accretion disk that is subject to strong gravitational instabilities causing subsequent fragmentation \citep{Bonnell+1994}. Numerical simulations involving this latter process have been efficient in forming companions around massive protostars \citep{Kratter+2006,Krumholz+2007,Oliva+2020,Mignon-Risse+2023}. The observations of fragmenting accretion disks suggest that this mechanism tends to form rather close systems with separations $a\lesssim 100$~au, and provides an ideal framework for inward migration that may result in spectroscopic binaries \citep{Sana+2017,Meyer+2018}. 

In order to gain a better understanding of the formation of massive multiples as well as to predict their future evolution, we must have a clear description of the multiplicity properties, the mass ratio, and the initial separation distribution of forming massive stars \citep{Sana+2012}. The massive young stellar objects
(MYSO) phase is a short-lived evolutionary stage towards the end of the hot core phase in which the protostar experiences accretion events necessary to grow in mass. Evidence of circumstellar accretion disks is, for instance, observed in \citet{Kraus+2010} and \citet{Frost+2019} as well as bipolar outflows \citep{Cooper+2013} and a surrounding dusty envelope, which is opaque at visible wavelengths \citep{Wheelwright+2012}. Keplerian-like disks that could be associated with accretion in larger scales are also reported in various studies \citep{Johnston+2015,Zapata+2019,Maud+2019}. Unlike low-mass protostars, the large distances at which they are located, their high dust extinction, and rarity has made systematic studies challenging. It is only recently with the advent of high angular resolution and high-contrast observations assisted with AO that the study of multiplicity among small populations of MYSOs has been made possible, slowly lifting the veil on the confusion that reigns as to their formation. We can, for example, quote the studies of \citet{Pomohaci+2019} whose work probed for wide ($>$ 1000 au) multiple companions of 32 MYSOs, assessing a field binary fraction of 31$\pm$8\%, or the recent study of \citet{Koumpia+2021} that reported a fraction of close binaries as low as 17\% 17$\pm$15\% (2~mas\,$\lesssim a \lesssim 300$~mas) among a sample of six MYSOs. Our present study aims to complement the pilot study led by \citet{Pomohaci+2019}, using a smaller sample of MYSOs with a slightly longer wavelength filter (3.8~\um), ensuring the detection of even younger companions that might not have been visible in the $K-$band images. Expanding these studies is crucial to find a consensus on the multiplicity and the initial orbital parameters in order to determine which of the scenarios presented above is the most likely to form multiple massive systems. These tabulated values over the full stellar mass and separation range can serve to feed numerical and analytical models or as a comparison for massive star formation. 

Here, we study the environment around thirteen young massive stars using mid-infrared (MIR) wavelengths, with the ultimate goal to address their multiplicity. We investigate the presence of potential companions in wide orbits with AO-supported observations at 3.8~\um. In Section \ref{section:obs}, we present the observations and the reduction routine. Section \ref{section:datanalysis} deals with the data analysis, ranging from source detection, point-spread function (PSF) fitting, detection limits, and chance projection probability. The results per object are described in Section \ref{section:results}. The multiplicity and companion fractions and the implications of our findings in the context of star formation are discussed in Section \ref{section:Discussion}. We summarise the results and conclude in Section \ref{section:Conclusions}. 





\section{Observations and data reduction}
\label{section:obs}

Our target list of 13 MYSOs is taken from the Leeds Red MSX Source (RMS) survey, where MSX is the Midcourse Space Experiment telescope \citep{Lumsden+2013}. This survey aimed to search for MYSOs across the Milky Way. Colour cuts were used to find the initial 2000 candidate MYSOs from the survey data, following the criteria of \citet{lums2002}. Follow-up observations at different wavelengths were performed to exclude other objects that satisfied the colour requirements such as planetary nebulae (PNe), UCHII regions and evolved stars to allow the robust identification of the MYSOs. The survey ultimately found $\sim$800 (candidate) massive young stellar objects and \Hii regions within 18~kpc and $L>$ $2\times10^{4}$~\Lsun \citep{Lumsden+2013}. The target selection from this database for our sample was based on bolometric luminosity, a 4\-kpc distance cut, with declinations ($\delta<10^{\degr}$) accessible to ESO's Very Large Telescope facility.

Our observations aimed to detect putative companions to a sample of MYSOs. In order to do so, we chose to probe multiplicity in the mid-IR wavelength region at the highest possible angular resolution. This wavelength choice was motivated by minimising the effects of dust extinction and the envelope's dust emission. Depending on the source, dust emission from the envelope dominates the Spectral Energy Distribution (SED) from about 5 microns onwards. Hence, $L'$ was chosen as the optimal wavelength to be able to detect companions to the central MYSO that are up to a factor of 1000 fainter.  

\begin{table*}[!t]
\caption{Log of the NACO observations including observing conditions. }
    \label{tab:NACO_journal}
    \centering
    \begin{tabular}{llllllccccc}
    \hline
    \hline  \\ [-1.5ex]
    RMS Name & $\alpha$ (J2000) & $\delta$ (J2000) & MJD & $K_\mathrm{S}$ & $L^\mathrm{'}$ & Distance & Lbol & Seeing & Contrast & Mult. \\
     & (h m s) & (\degr\ \arcmin\ \arcsec) &  &  &  & (kpc) & (\Lsun) & (mas) & ($L'$mag)  & \\
    \hline \\ [-1.5ex]
    
    G194.9349-01.2224 & 06 13 16.12 & +15 22 43.8 & 56271.2 & 9.8 & 6.8 & 2.0\tablefootmark{a} & 2620 &  0.82 & 2.0 & N \\

    G203.3166+02.0564 & 06 41 10.12 & +09 29 33.7 & 56307.2 & 4.9 & 1.7 & 0.74\tablefootmark{b} & 1080 & 1.25 & 7.2 & Y \\

    G232.6207+00.9959 & 07 32 09.86 & $-$16 58 13.0 & 56295.3 & 8.3 & 5.3 & 1.68\tablefootmark{c} & 11270 & 0.76 & 3.3 & Y \\

    G254.0548-00.096108 & 08 17 52.51 & $-$35 52 50.2 & 56317.1 & 9.5 & 7.2 & 2.75 & 4910 & 1.27 & 2.2 & N \\

    G263.7434+00.1161 & 08 48 48.74 & $-$43 32 29.0 & 56257.3 & 9.0 & 3.8 & 0.7 & 520 & 0.68 & 6.9 & N \\

    G263.7759-00.4281 & 08 46 34.90 & $-$43 54 31.6 & 56295.3 & 9.6 & 5.7 & 0.7\tablefootmark{d} & 880 & 0.68 & 3.9 & Y \\

    G265.1438+01.4548 & 08 59 27.40 & $-$43 45 03.7 & 56307.2 & 9.6 & 3.8 & 0.7\tablefootmark{d} & 670 & 0.85 & 3.6 & Y \\

    G268.3957-00.4842 & 09 03 25.15 & $-$47 28 27.1 & 56308.3 & 8.3 & 3.9 & 0.7\tablefootmark{d} & 720 & 0.95 & 7.2 & Y \\

    G269.1586-01.1383A & 09 03 32.1 & $-$48 28 43.3 & 56314.3 & 10.7 & 7.4 & 0.7\tablefootmark{d} & 220 & 1.20 & 1.4 & Y \\

    G305.2017+00.2072A & 13 11 10.45 & $-$62 34 38.6 & 56314.2 & 9.4 & 5.5 & 4.0\tablefootmark{e} & 30320 & 0.95 & 2.5 & Y \\

    G310.0135+00.3892 & 13 51 37.92 & $-$61 39 07.5 & 56309.3 & 4.9 & 2.3 & 3.25 & 54680 & 1.42 & 7.1 & Y \\

    G314.3197+00.1125 & 14 26 26.3 & $-$60 38 31.5 & 56313.3 & 10.6 & 4.7 & 3.6\tablefootmark{f} & 12870 & 1.06 & 3.2 & N \\

    G318.0489+00.0854B & 14 53 42.7 & $-$59 08 58.9 & 56319.4 & 11.7 & 7.7 & 3.36\tablefootmark{f} & 6810 & 1.03 & 2.6 & N \\
    
    \hline
\end{tabular} 
\tablefoot{The first three columns list the names, Right Ascension ($\alpha$) and Declination ($\delta$), taken from the RMS survey \citep{Lumsden+2013}. The $K_\mathrm{S}$ magnitude is taken from 2MASS, considering that the nearest 2MASS source to the MSX position is assumed to be associated. The estimated $L^\mathrm{'}$ magnitudes are extrapolated from the WISE $W1$-band and $W2$-band magnitudes. Distances are taken from the RMS catalogue \citep{Urquhart+2011} and we add further references in the notes of this table. Bolometric luminosities were derived in \citet{Mottram+2011}. The uncertainties on distances are rather large: of the order of $\sim$1~kpc and of about 20 to 35 per cent for the bolometric fluxes. The $L'-$contrast magnitudes are derived following a method that consists in injecting artificial sources and that is described in Section \ref{subsection:contrasts}. The last column named 'Mult.'(for multiple) indicates the detection of any companion in the NACO images: \emph{N} stands for NO while \emph{Y} stands for YES. A more detailed analysis is given in Sect. \ref{section:results}.\\
\tablefoottext{a}{\citet{Kawamura+1998};}
\tablefoottext{b}{H\textsubscript{2}O maser: \citet{Kamezaki+2014};}
\tablefoottext{c}{Maser parallax distance: \citet{Reid+2009};}
\tablefoottext{d}{\citet{Liseau+1992,Netterfield+2009};}
\tablefoottext{e}{Parallax of 6.7 GHz methanol maser: \citet{Krishnan+2017};}
\tablefoottext{f}{Kinematic distance using the source velocity: \citet{Green+2011,Green+2012}.}
}
\end{table*}

We selected MYSOs with $L'$ magnitudes comprised between 7 and 10. All 13 objects are classified as MYSOs in the RMS catalogue and a summary is provided in Table \ref{tab:NACO_journal}. 
Direct adaptive-optics (AO) assisted imaging is to date the most relevant method to explore the wide-orbit environments around massive protostars. To study the multiplicity and the environment of the objects listed in Table\,\ref{tab:NACO_journal} therefore, we made use of the capabilities the NAOS \citep[Nasmyth Adaptive Optics System]{Rousset+2003} adaptive optics system integrated with the CONICA \citep[COudé Near IR Camera]{Lenzen+2003} NIR imaging camera (or NACO), an instrument decommissioned in 2019 from ESO's facilities.

The $L'$-band observations were acquired in service mode over 9 different nights spread between November 26, 2012 and January 27, 2013. The seeing was variable during these nights, ranging from 0.68\arcsec to 1.42\arcsec in the visible, meeting the requirements for good quality products. For the purpose of our study we used the camera mode L27 with the narrow band filter Lp (3.8~\si{\micro\meter}), providing a medium resolution of $\sim$27~mas/pixel and resulting in a field of view (FoV) of $28\arcsec \times 28\arcsec$. The pixel size of $0.027"$ corresponds to 54~au given the average distance of 2~kpc of our sources. However, the survey probes the separation range from the minimum achieved full width at half-maximum (FWHM, calculated from the detected new sources) of $\sim$160~mas up to the full FoV of $28 \times 28$ arcsec. For a typical distance of $\sim$2~kpc, this translates in separations ranging from $\sim$320~au to 56,000~au. The detection limit of low-mass companions ranged up to $L'=16$, given the high extinction of the targeted MYSOs. The average $L'$ magnitude of the targets is 5.1 (see Table \ref{tab:NACO_journal}).
With $K<10.5$~mag, all the targets could be used as their own reference sources for wavefront sensing (the magnitude limit for wavefront sensing being $K<13$~mag). We used the infrared wave-front sensor (IR WFS) with the $JHK$ dichroic that passes 100\% of the NIR light to the NAOS IR WFS and forwards all the $L'$ light to CONICA onto the science detector. The observations were obtained in short-exposure cube mode that stores each individual co-add image instead of a median-combined one. It allows for frame selection with integration times of less than 0.2s.

In order to get rid of the background contamination and obtain a precise bad pixel map, each target was observed using a jitter sequence (a dozen images) with a fixed sky offset. This moves the telescope alternatively between object and sky positions within a predefined box of $20$~\arcsec$\times 20$~\arcsec, allowing the sky from all the observations to be estimated. For each source, a total of two cycles were observed, each cycle consisting of a sequence of object and sky observations following the jitter pattern. 

We reduced the data initially using the standard ESO NACO pipeline recipes (public version 4.4.11). However, the recipes were found to be unsuitable for proper estimation of the background using dedicated \emph{sky} frames. We opted instead to custom-made our own reduction routine for the pre-processing of the raw images. We downloaded the processed calibration files from the ESO archive. All science and sky cubes were bias subtracted, flat-fields corrected and bad pixels masked. Given the large number of frames per observation, we performed a quick test throughout the cube (both for object and sky) to identify and remove bad frames (based on the cross-correlation of each frame to the median of all frames in a given cube). This led to the removal of $\sim$3\% of all images on average, with mainly the first 3 frames' quality being below average, likely due to the start of the adaptive optics loop or instrumental effects. For each of the 2 cycles, we median-combined together sets of 4 consecutive object (resp. sky) images, leading to 2 separate \emph{object} (resp. sky) stacked images. Each object image was sky subtracted by its corresponding combined sky image. We aligned the two reduced images using the offset information stored in the header and the master image was obtained by stacking the two images. The reduced images are presented in Figs \ref{fig:mYSO_comp_1},\ref{fig:mYSO_comp_2},\ref{fig:mYSO_comp_3} and \ref{fig:mYSO_single}.

\section{Data analysis}
\label{section:datanalysis}

We have imaged a total of 13 young massive stars for which we aim to identify possible companions between 300$-\sim$60,000~au and characterise the thermal infrared environment. To achieve this purpose, we follow the same data analysis procedure for each system. Our analysis consists of; 1) measuring the background, 2) identifying all possible point sources and 3) determining the astrometry and photometry of these detected objects. We provide all the details about the analysis in this section and discuss the sensitivity of our observations. 

\subsection{Background and noise estimation}
It is essential to derive an accurate estimate of the background emission level to measure the photometry of our targets. Similarly, the significance of source detection and photometric errors are affected by the background noise. We use the sigma clipping technique that removes the pixels that peak above a specified sigma level from the median and recalculates the statistics until convergence is reached. We fixed $\sigma$ at 3 and the median filter is calculated over each image following the procedure written by \citet{Bradley+2022}. However, the background level may vary across the image. To probe background level variation across the detector, we generated for each science observation a 2D image of the background. We computed sigma-clipped statistics on a grid to create a low-resolution background image. The meshes have a size of $50\times50$ pixels ($1.3\arcsec\times1.3\arcsec$), which is small enough to catch any background variations and larger than the size of the sources. We obtained the final background map by interpolating the low-resolution image. The background map was subsequently subtracted from the co-added images to generate the final reduced one.

\subsection{Companion detection}
\label{subsection:detection}
The identification of astronomical sources in the images is a precondition to performing photometry and astrometry measurements. In order to identify physical companions in the frames, we first need to pinpoint all possible point sources. A first visual approach and inspection of the reduced images confirm the presence of nearby and rather bright sources cohabiting with fainter point sources. We developed our own routine to numerically identify star-like objects in the FoV around each central MYSO. The code mainly uses the \textsc{python} affiliated package \textsc{photutils} \citep{Bradley+2022} of \textsc{astropy} that provides an implementation of the DAOFIND algorithm. The latter is designed to detect astronomical objects in crowded and non-crowded background-subtracted images. A similar approach has been used in \citet{Bodensteiner+2020}, for example, and has proven its efficiency. We initialised the star finder to search for objects with FWHMs of around three pixels and peak at least 5$\sigma$ above the background signal (background map). We first set standards on the roundness such that only star-like objects are returned and blob-like-shaped objects are discarded. We visually examined the images and looked for any missed detection. After inspection, it seems that no object was left behind, but given the possible extended nature of MYSOs and keeping in mind that most of the companions might be young forming stars, this parameter was later on left free in order to allow the detection of any exotic shape. In practice, no additional objects were found when relaxing the roundness parameter. The returned parameters contain the results of the search, including the source locations and peak amplitude. These parameters feed the PSF fitting code that performs photometry measurements on the detected sources, which we explain in the following subsection.

\subsection{PSF fitting}
\begin{figure}
    \centering
    \includegraphics[scale=0.18]{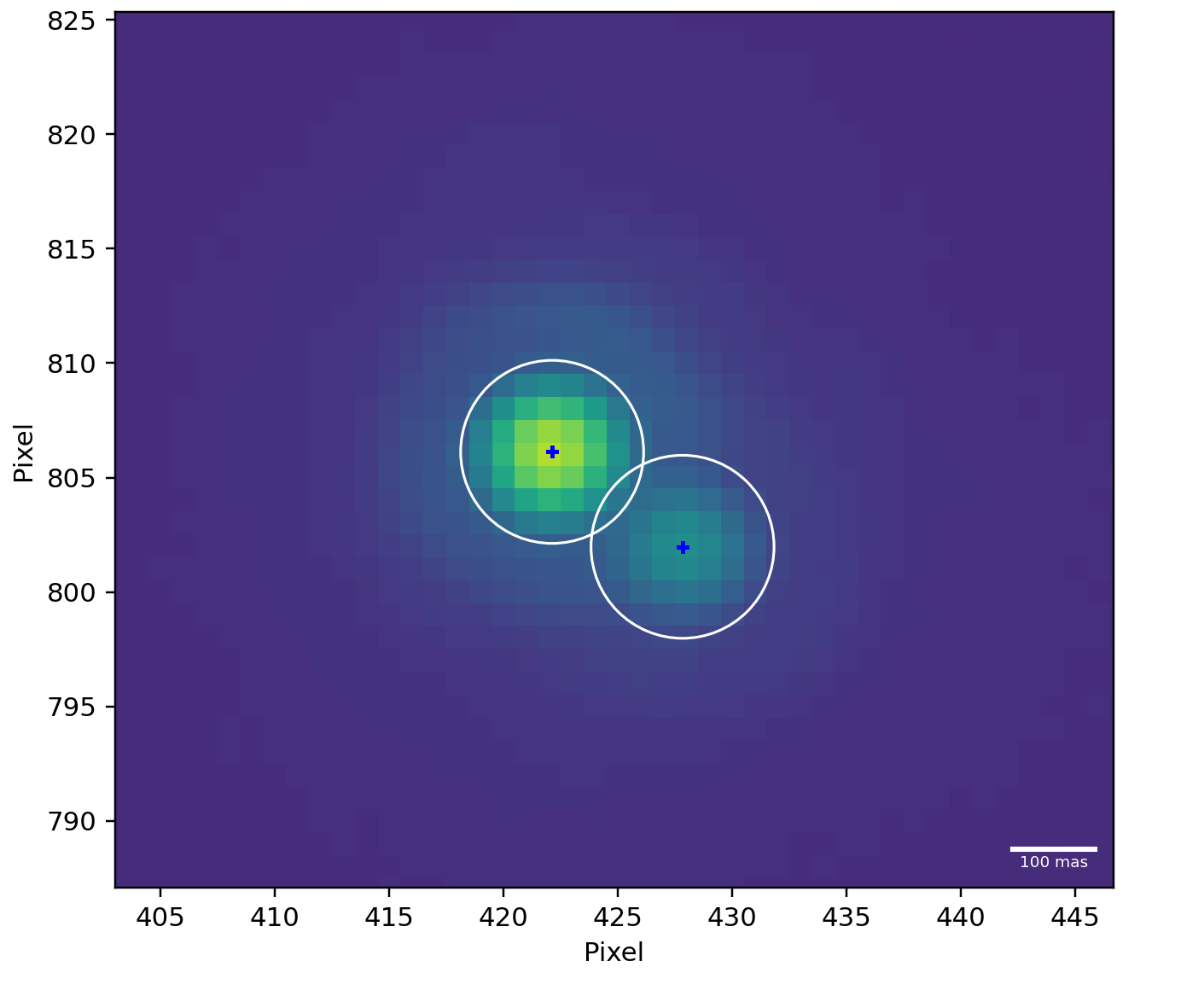}
    \caption{Zoom over the northern companion system for target G268.3957, showing two distinct components. The white solid lines is the 4-pixel aperture applied in the aperture photometry method and the markers show the position of the centroids as found with DAOFIND. For reference, north is up and east is left.}
    \label{fig:zoom_binary}
\end{figure}

We follow the PSF-fitting methods as detailed in \citet{Bodensteiner+2020} and \citet{Rainot+2020}. From the companion detection routine, we obtain accurate centroid positions for each peak detected in the image. The low source density in the images allows simple aperture photometry as a first estimation. We build a set of aperture objects (typically a circular aperture with a radius of three pixels) from the peak locations and perform the photometry taking into account the overlap of apertures on the pixel grid. The fractional overlap of the aperture is calculated and the aperture sum for each detected peak is derived accordingly. However, this method is not exact and some flux contribution outside the extraction area might be missed when summing the photons. We therefore also used a second method consisting of PSF-fitting, that is, modelling the radial shape of each star in the image. Simultaneous PSF-fitting photometry better handles stars with close neighbours, which is the case in some of our images (see Sect. \ref{section:results}) for the achieved spatial resolution. Indeed, some companions appear to be in such pairs. Figure \ref{fig:zoom_binary} shows an example of an overlapping aperture originally detected as a single source without the DAOFIND method. For such systems, the PSF-fitting method derives more accurate fluxes. To generate the PSF, we use the high signal-to-noise primary star and use the derived aperture photometry as a first instance. None of the central stars appears to be resolved into a close binary. This PSF model was fit to each source individually whose positions are the fixed centroids derived from the source detection procedure (see Sect. \ref{subsection:detection}). In this way, we obtain estimates of the total flux and flux errors. 


\subsection{Photometry and astrometry}
The fluxes and positions derived from both methods, aperture and PSF fitting, are consistent within error bars for isolated sources, i.e. non-blended sources. In the case of close binary systems, derived centroids are unchanged and consistent with the PSF method. However, the photometry is equally balanced between the two components when modelling the brightness with a PSF. For the remainder of the paper, we adopt all fluxes and source locations derived via the PSF-fitting method. 

The fitted fluxes of the newly detected sources are converted into $L'$ magnitudes relative to the brightness of the primary star, for which we have an estimation of the $L'$ mag (see Table \ref{tab:NACO_journal}). No photometric calibrators were contained in our observations thus we derive relative photometry with respect to the central star. For every single point source that is detected, we calculate the angular separation from the primary, first in detector coordinates that we then convert into angular units, applying the pixel scale of 0.027~\arcsec$/$pixel, as stated in the NACO user manual\footnote{VLT-MAN-ESO-14200-2761\_v90.pdf} and confirmed in the header file of the targets. The physical distances and position angles are straightforwardly computed using the distances reported in the RMS Catalogue \citep{Lumsden+2013}. We additionally checked the GAIA eDR3 \citep{Bailer-Jones+2021} distances for every single source. Massive young stars are expected to become visible once they have reached or are close to the birthline. Given their strong IR emission rising from their circumstellar material, the young high-mass objects are often opaque to GAIA bands, reducing our chances to find reliable parallaxes. Among our sample of 13 MYSOs, we find reliable GAIA distances for three objects (G265.1438, G268.3957 and G318.0489) whose coordinates matched the RMS catalogue within 2 arcsec. For these three targets, the GAIA distances are consistent with the RMS distances. We use the RMS distances for the remainder of the paper.


\subsection{Sensitivity limits}
\label{subsection:contrasts}

\begin{figure*}[!t]
\centering
\includegraphics[scale=0.85]{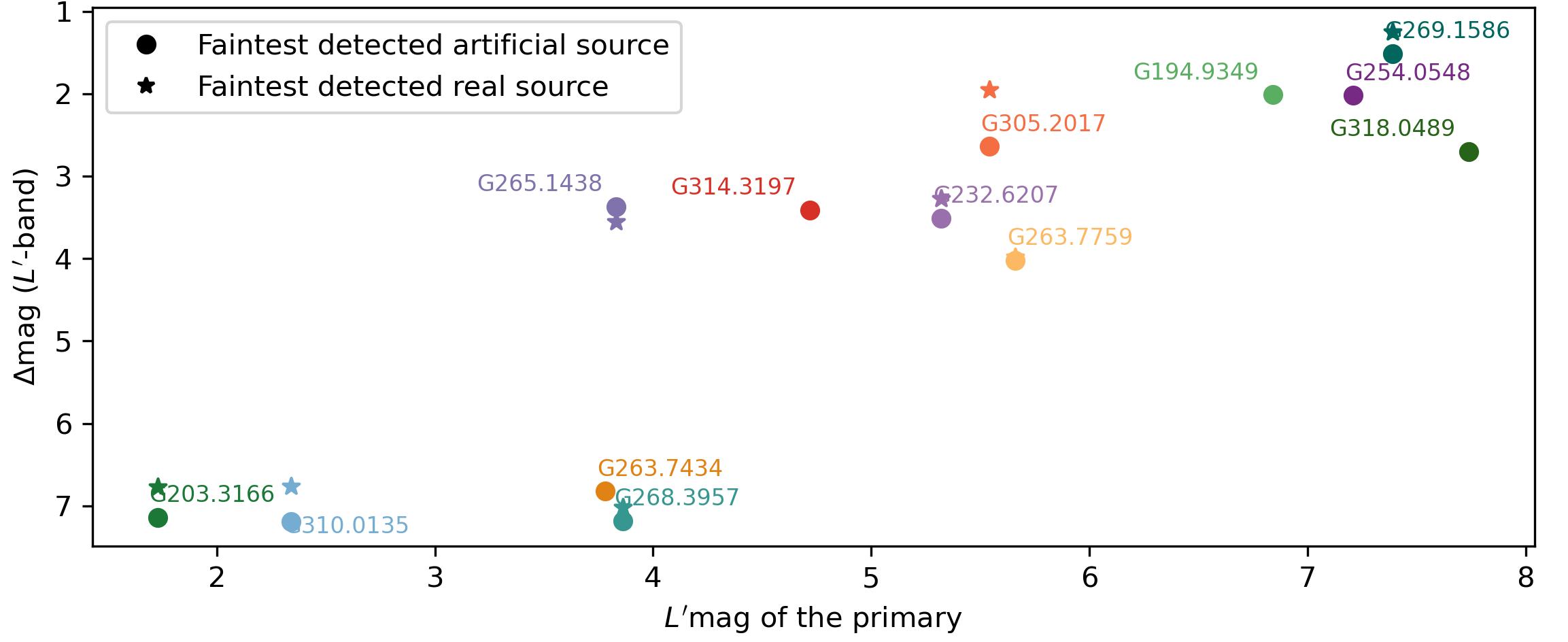}
\caption{Detection limits for each observed field. The star symbol denotes the $\Delta$mag of the faintest source detected in the NACO image (if any) while the dot symbol refers to the $\Delta$mag of the faintest detected artificial source, injected into the image. The minimum angular separation at which companions are found is about 2.7\arcsec which translates to $\sim$5000~au at a typical distance of 2~kpc. However, the faintest companion of each field is found at a typical separation of 8.7\arcsec ($\sim$17,000~au).}
\label{fig:deltamag_vs_sep}
\end{figure*}

We determine our contrast limits following the methods of \citet{Pomohaci+2019} and \citet{Rainot+2020}. The sensitivity of the observations is determined for each target in terms of flux ratio with respect to the central object. We computed the contrast limits by injecting 300 artificial stars into each image. The brightness varies from a thousandth to a hundredth of the central object and the simulated objects are randomly distributed across the image. Each artificial star is a 2D Gaussian source that is built following with the same FWHM as the PSF. We then run the same procedure as for the companion detection, and set the minimum flux at which the simulated source was detected at the 5\-$\sigma$ level above background noise as the limiting flux. The results of a single iteration are shown in Appendix \ref{appendix:detection_limits}. Since the artificial sources have random brightness and are placed at random distances across the image, we repeat the process over 20 iterations and the final detection limits recorded in Table \ref{tab:NACO_journal} are issued following the bootstrapping procedure. Figure \ref{fig:deltamag_vs_sep} shows the detection limits as a function of the $L'$mag of the primary. For each system, we display both the contrast limit measured in the real images and the one derived from the injection of artificial sources. The contrast varies from 7 to 1 depending on the $L'$mag of the central star but is roughly constant in terms of the apparent companion magnitude $L'_{\text{comp}}\sim$8, showing that the images are rather background limited. The noise does become indistinguishable from background noise at an angular separation of about 4\arcsec from the central star.

\section{Results}
\label{section:results}
We have imaged 13 MYSOs in the mid-IR at a spatial resolution of $\sim$160~mas in order to probe the presence of companions.
We identified a total of eight multiple systems from 13 targets, with a total of 41 newly detected mid-IR companions candidates, summed over all fields. We test the probability of physical association of the new sources to the central object in Section \ref{subsection:spurious_association}. We also describe below each of the 13 sources together with the results of our analysis. 

For four of them, we compare our findings to those of previous AO-assisted observations, published in \citep{Pomohaci+2019}. For a thorough comparison with their study, we quickly present the similarities and differences with regard to the sample size, specific observations and techniques used in their study. Their work aimed at probing the widest companions (with a physical separation range of 400$-$46 000~au) around 32 MYSOs classified in the RMS Catalogue using $K$-band observations with NACO. Their observations were carried out roughly three years later than the $L'$-band ones, and provide $K-$band detections and their relative binary parameters. Despite a shorter wavelength filter and narrower FoV that might have lead to a higher spatial resolution probe (120~mas to 7~arcsec), they only achieve a full coverage for distances up to 3~arcsec away from the main source, as a result of technical issues during the observations that affected the bottom left quadrant of the detector. They reach an average limiting magnitude of 14 and are able to probe sources 5.5 mag fainter than the primary source, in the $K$ band. They infer the chance projection following a similar method but using the 2MASS point-source catalogue in the $K$ band to derive the background density of sources. 

Further archival and important notes on individual sources are provided in Appendix \ref{appendix:Notes}. The list of detected companion candidates together with their properties are given in Table \ref{tab:companion_parameters}. We point out that the sources are arbitrarily classified and are not sorted according to their proximity to the primary star or their brightness. We simply follow the order returned to us by the PSF-fitting procedure.

\subsection{Eliminating spurious associations}
\label{subsection:spurious_association}
 An important step in confirming the nature of the systems is to assess whether the detected companions share a physical connection with the central object. The analysis of common proper motion and orbital motion would unambiguously discard that a visual pair is just a chance alignment. However, with only one observation in hand, we lack sufficient temporal coverage to detect whether the companion candidates are co-moving. To overcome the lack of multi-epoch observations, we employ statistical methods to determine the likelihood of a given source being bound to the central object. We follow the procedure described in \citet{Rainot+2020,Pauwels+2023}. 
 
 At first glance, the NACO images are not densely populated and candidate companions are not uniformly distributed across the observed field. The denser the field, the greater the chance of spurious associations. Quantitatively, we estimate the potential contamination from fore- and background sources given the target's galactic coordinates and the limiting magnitudes of the observations. As surveys and catalogues in the $L'-$band do not exist, we simulate a collection of background stars by running the Besançon model of stellar population synthesis of the Galaxy \citep{Robin+2003} in the $L-$band, from 0-150~kpc and spanning a surface of 1~deg$^{2}$ in the direction of the object. The model predicts between 0.2 and 4 stars in a $28\arcsec\times28\arcsec$ area, depending on the magnitude properties of each source. We define the probability of spurious association ($P_{\text{spur}}$) of a given companion candidate $i$ as the probability that, given the local number density of stars ($\Sigma(L_{i})$) as bright as companion $i$, at least one source is  found by chance within a circle of radius $\rho_{i}$ (i.e. separation of companion $i$). For consistency, we use a Monte Carlo simulation to generate 1000 populations of stars featuring these properties ($\Sigma(L\leq L_{i})$) and we sum the iterations where at least one source is spotted at $\rho\leq\rho_{i}$. The final $P_{\text{spur}}$ is given by the result of the sum, divided by the total number of simulations. 
 
 The results are provided in the last column of Table \ref{tab:companion_parameters}. We consider $P_{\text{spur}}=20\%$ as a limit to disprove any physical association \citep{Pomohaci+2019}. All of the 41 detected sources but two have $P_{\text{spur}}\leq 8\%$ suggesting that they most likely are real companions. Companion candidate B of G310.0135 and A of G268.3957 have respectively over 30\% and 69\% chance of being galactic contaminants. 

 Among the stars around which no source is detected, we perform a statistical test to find out if the non-detections are in agreement with the $P_{\text{spur}}$ estimates of the other sources with detected companions. We proceed in the same way by simulating a background stellar density and using the $L'-$band detection limits to calculate the probability that a source is detected in a given field. We find probabilities ranging from 5-23\% to find a source in the five fields where our tools failed in detecting sources, showing the real lack of $L'-$bright companions around these MYSOs or that NACO is not sensitive enough to detect fainter companions.

\subsection{MYSOs within a distance of 2~kpc}

\subsubsection{G203.3166+02.0564}
G203.3166+02.0564 (hereafter G203.3166) also known as AFGL989, NGC2264 IRS1 or Allen's object \citep{Allen+1972} is one of the massive members of NGC2264, a young \citep[$\sim$3~Myr]{Venuti+2019} star forming region located at 760~pc \citep{Dahm+2008}. G203.3166 has been the subject of several observation campaigns, ranging from NIR to sub-mm, we gather the most prominent results in Appendix \ref{appendix:Notes} \citep{Thompson+1998,Schreyer+2003,deWit+2009,Grellmann+2011}. Six companions were discovered in 1998 thanks to $K-$band images \citep{Thompson+1998} and two additional ones were reported in \citet{Schreyer+2003}. 

Our NACO $L'-$band images allow separations between 120 and 15,000~au to be explored for this source. We confirm the presence of these companions, tagged from A to G, and we clearly detect the binary (A and B objects, in Table \ref{tab:companion_parameters} and Figure \ref{fig:mYSO_comp_1}) in the south-west (SW) direction, that was described as the extended `object 8' and proposed to be a binary in \citealt{Schreyer+2003}. We also detect 2 additional objects, H and I, that were not previously referenced in the literature. The geometric arrangement of companions in line, on one side of the central object is intriguing. Whether they could be knots in a jet could be further investigated upon the acquisition of new data. \citet{Thompson+1998} had noticed this distribution of objects and suggested that it shows evidence of a triggered star formation. We discuss this scenario in the next section. 

\subsubsection{G232.6207+00.9959}
G232.6207+00.9959 (hereafter G232.6207) lies at a distance of about 1.68~kpc as estimated by \citet{Reid+2009} using the trigonometric parallax measurements of the 12 GHz methanol masers. The associated IRAS point source is IRAS 07299-1651 for which we find multi-wavelength studies in the literature (see Appendix \ref{appendix:Notes}). 

The image of G232.6207 shows a bright nebulosity in the $L'-$band surrounding the targeted source, which could be associated with the outflows detected at longer IR wavelength, for example with SOFIA \citep{DeBuizer+2017}. With NACO we probe separations ranging from $\sim270-33,000$~au. We detect seven sources around this young star at separation ranges between $\sim$8000 to $\sim$25000~au, with flux ratios ranging from $\sim$5\% to almost 70\%. All of these candidate companions are statistically likely to be bound to the central object, or at the slightest fault belong to the same cluster. \citet{Pomohaci+2019} report the detection of two candidate companions in the $K-$band with a magnitude contrast of 5.0 and 2.6 but these are likely to be merely visual binaries given their high probability of spurious associations. Being located at $\sim$6.4\arcsec and 6.7\arcsec\, from the central source, these detections are most likely different stars than the ones we find. Objects G and E are aligned with the direction of the jet axis. Thus, one cannot ignore the possibility that they could be knots of material that have formed due to outflow processes (e.g. \citealt{bally1995}). Spectroscopic study or time-series imaging could assist in determining the definite nature of these sources, but that is outside the scope of this work.

\subsubsection{G263.7434+00.1161}
G263.7434+00.1161 (hereafter G263.7434) has been classified as a candidate MYSO and belongs to the Vela Molecular Ridge (VMR) \citep{Mottram+2007}. \citet{Urquhart+2007} report that the protostar 
has a spherical and unresolved morphology. The 10.4~\um~TIMMI2 image displayed in \citep{Mottram+2007} shows a single object while the 2MASS $K-$band images are slightly more crowded.

With a distance of 0.7~kpc, NACO can detect companions from $\sim$110 to 14,000~au. No source apart from the central one is detected in the $L'-$band images. We found a $5\sigma$ detection limit of $\rm \Delta=6.8^{m}$.

\subsubsection{G263.7759-00.4281}
G263.7759-00.4281 (hereafter G263.7759) is better known under the name IRAS 08448-4343 and is, like G263.7434, located in the VMR. The protostar has a luminosity equivalent to that of a main sequence spectral type B2 \citep{Navarete+2015} and is located at a distance of $\sim$0.7~kpc, which means that any companion in the sensitivity of NACO can be detected between $\sim$110 and 10,000~au. The resolved 20~$\mu$m images of \citet{Wheelwright+2012}, as well as the $\rm H_{2}$ continuum subtracted images \citep[]{Giannini+2005}, agree on the extended emission of the central source, showing clear signs of a structure likely to be a bipolar jet. 

The extended emission is also noticeable in our $L'-$band images when performing the source detection code. We relaxed the roundness parameter giving more flexibility to the algorithm to detect the central source. NACO's FoV allows the detection of companions between $110-14,000$~au. We detect three additional objects, a binary system at $\sim$10\arcsec\, in the SW direction and a single object at $\sim$3\arcsec\, pointing to the north-west (NW) direction. After visual inspection of the L,M,N images of \citet{Giannini+2005}, the position of object 40 seems to coincide with that of the binary. Object 40 is a sub-complex of six resolved peaks in the $K-$band, characterised as embedded protostars, and proposed to be the driving sources of the outflow. Such a sub-cluster is not resolved in our images, apart from the binary-like system. Nonetheless, none of the two sources cited above report the detection of the NW object. Given its direction that seems aligned with the $H_{2}$ jet detected by \citet{Giannini+2005}, we cannot rule out the hypothesis that this object is a knot of ejected material from the central object.  

We point out the extended shape of our detected object C, closer to the primary than the double detection AB. The peak emission is broader than the PSF and can be of a different origin. Most MYSOs are expected to be extended and all the sources in the sample are classified as MYSOs. Being even more extended than the others would imply that this companion is particularly embedded, hence of a younger age. Given the relatively low spurious association probability, it may also be a young and massive forming star whose central object has brightened to the point where its surrounding material (likely envelope and disk) is better illuminated than the other sources. In the latter case, such radiation could be associated with accretion. Finally, object C could also be fortuitously inclined, producing a broader emission, which could also be a distance effect since it is particularly close. However, it seems that other equally close sources (e.g. G265.1438) do not show any extended geometry beyond the PSF.




\subsubsection{G265.1438+01.4548}
G265.1438+01.4548 (hereafter G265.1438) is a MYSO and member of the massive star-forming region RCW 36 located at 0.7~kpc, in Cloud C of the VMR \citep{Ellerbroek+2013}. Unlike the rather void 20~$\mu$m images of the central source \citep{Wheelwright+2012}, we find no fewer than eleven objects in the FoV ($110-14,000$~au), making G265.1438 the source for which the most sources are found. The sources are located along the NW to SE direction, usually above $\sim$10~mas around the central source, with the exception of sources G and F which are located about 7 mas in the orthogonal direction (see Appendix \ref{appendix:images}). We find source J, which is slightly brighter than the MYSO, at the edge of the FoV. Chance alignment is low ($<2.3$\%) so one could rightfully assume the physical association with the central source. However, looking back at \citet{Ellerbroek+2013}'s images, the region seems rather crowded, with a lot of Class 0/I/II objects. The latter authors and \citet{Wheelwright+2012} report the presence of an outflow. No more details about the direction of the outflow can be found in the literature but the arrangement of the detected objects in the $L'-$band images can be indicative of a NW to SE direction. Constraining the origin of the outflow, the true nature of the detected objects (real stars or knots of ejected material) and their physical association would require more observations.

\subsubsection{G268.3957-00.4842}

G268.3957-00.4842 (hereafter G268.3957) is an MYSO candidate located at approximately $\sim$0.7~kpc, a rather short distance allowing for source detection from $110-14,000$~au with NACO. The NACO $L-$band image shows the presence of five companion candidates, including a relatively NIR bright binary system (resolved as Objects D and E) in the north vicinity of the central object. All but one (object A in Table \ref{tab:companion_parameters}, which we discard in the remainder of the discussion) have spurious associations lower than 1.1\%, pointing to the fact that they are most likely physical companions. The companions have projected separations ranging from $\sim$1350~au to $\sim$5800~au for the binary system. The projected separation of the two components forming the binary amounts to 0.15\arcsec which translates to $\sim107$~au. All companions have also been detected in the NACO $K-$band study of \citet{Pomohaci+2019} with angular separations matching our findings with a difference of less than 1\% for objects B and C and of about ~7\% for the components of the binary (objects D and E). However, they find a chance projection of the binary greater than 48\% and conclude that the system is rather a fore- or back-ground multiple which is inconsistent with our derived probabilities of less than 1\%.

\subsubsection{G269.1586-01.1383A}
G269.1586-01.1383A (hereafter G269.1586) is classified as an MYSO candidate and associated with an \Hii region, approximately 7\arcsec to the north-east (NE) \citep{Urquhart+2007}. Located at a distance of 0.7~kpc, NACO is sensitive to source detection ranging from $110-14,000$~au. We detect two additional relatively bright objects, both at around $\sim$8\arcsec\ (which translates to nearly 6000~au) from the protostar and with low spurious association probabilities. Both objects are respectively located towards the NE and NW directions from the central protostar, as shown on the Figure \ref{fig:mYSO_comp_1}. The newly detected sources are approximately $\sim$32\% and $\sim$77\% as bright as the primary.

\subsection{MYSOs above a distance of 2~kpc}

\subsubsection{G194.9349-01.2224}
G194.9349-01.2224 (hereafter G194.9349, also known as RAFGL 5186) is an MYSO located 2.0~kpc away \citep{Lumsden+2013}. G194.9349 has been part of NIR and FIR spectroscopic campaigns \citep[see Appendix \ref{appendix:Notes} for further notes]{Cooper+2013,Cunningham+2018} but we could not find NIR nor MIR imaging archives in the literature. We did not find indications for a companion star in our $L'-$band images and the probed separations of $320-40,000$~au (see Appendix \ref{appendix:images}). We derive a shallow detection limit of $\rm \Delta=2.0^{m}$. 

\subsubsection{G254.0548-00.096108}
G254.0548-00.096108 (hereafter G254.0548) has also been classified as an MYSO located at a distance of 2.75~kpc. This protostar is among the faintest of our sample, with a $L'$=7.2mag. Similarly to G263.7434, a 10.4~\um\, the image shows a single central object while the 2MASS $K-$band image indicated a more crowded region \citep{Mottram+2007}. 

No signature of a companion star was found in the NACO $L'-$band images (see Appendix \ref{appendix:images}), for the probed separation ranging from $\sim$440 to 54,500~au. We derive a $5\sigma$ detection limit of $\rm \Delta=2.2^{m}$.

\subsubsection{G305.2017+00.2072A}
 G305.2017+00.2072A (hereafter G305.2017) is located at a distance of $\sim$4~kpc and is thus the most distant MYSO of our sample. With such a large physical distance, NACO probes for separations between 640 and 56,000~au. We focus here on the multiplicity of G305.2017 but additional notes are provided in Appendix \ref{appendix:Notes} as the protostar is a broadly studied object and its geometry is well constrained\citep{Frost+2019,Krishnan+2017}

\citet{Liu+2019} review all the detections surrounding G305.2017, as of 2019, for a large wavelength coverage, ranging from sub-mm to NIR. Their high-spatial-resolution observations in the MIR (8.8-24.5~\um) show an emission component in the NE of the massive protostar lying at about 1\arcsec ($\sim$4000~au), that they name G305B2. However, their study shows that this emission is no longer visible in shorter wavelength filters, at 3.8~\um, for example, which is our band of interest. Their MIR observations probing a wider environment also show strong emissions towards the east of the central object, that peak around $\sim$~30~\um\,  and that they call G305A. Our high-angular resolution NACO observations look over similar separation ranges and have the capacity to resolve both objects. Nonetheless, images do not show the presence of a NE-4000~au-orbiting companion and rather confirm the absence of such NIR emissions corresponding to G305B2 or G305A. \citet{Pomohaci+2019} do not mention the presence of any additional sources near G305.2017 in their NACO $K-$band images. We report the detection of two new companions ($L'-$band), with projected separations of respectively $\sim$1.5\arcsec ($\sim$6000~au) and $\sim$8.8\arcsec ($\sim$35000~au) and flux ratios of around $17\%\pm1\%$. Both objects have relatively low spurious associations resulting from a low stellar surface density in this direction. None of these two objects is consistent with the detections presented by \citet{Liu+2019}.
The fact that two additional objects are detected at longer wavelengths might indicate that the emitting source is in a much younger and more embedded stage (likely a hot core phase) than the central G305.2017. 


\subsubsection{G310.0135+00.3892}
G310.0135+00.3892 (hereafter G310.0135) has been extensively studied under the name of IRAS 13481-6124. The object is classified as a high-mass YSO of an equivalent ZAMS spectral type O9. The central object has an estimated mass of 20~\Msun \, \citep{Grave+2009}. It is one of the brightest sources in our sample, with $L'=2.3^{m}$. We provide some extra important facts about this system in Appendix \ref{appendix:Notes}, as they do not directly impact our results but give a comprehensive and consistent picture of the system. However, it is worth comparing our multiplicity results to those of \citet{Pomohaci+2019} in the near-IR $K$-band. 

Located at 3.25~kpc, $L'-$bright sources with separations ranging from 520 to 60,000~au should be detectable. \citet{Pomohaci+2019} detected a companion located at a distance of 2.56\arcsec~which coincides, within errorbars, with our detected $L'$-band source at $\sim$2.6\arcsec. Using the 2MASS catalogue in the $K$-band, they derived a chance projection of 16\% yielding a likely binding connection between the central star and the companion. We derive a chance projection of only 0.11\% in the $L'$-band given the low background surface number density of L-bright stars along the line of sight. \citet{Pomohaci+2019} infer a mass of 7.6~\Msun when using foreground extinction. We detect a second object lying at $\sim16$\arcsec~in the NW direction. The presumed companion is most likely a background object ($P_{\text{spur}}=69\%$ and is discarded for the rest of the analysis. \citet{Wheelwright+2012} present 20~$\mu$m  images of G310.0135, showing a resolved and slightly extended object in the NE/SW direction that matches the direction of the outflow modelled in \citet{Kraus+2010}. The contours of the MIR emission also seem to show emission in the NE direction of the MYSO, approximately 2-3\arcsec~away. However, the authors do not comment on this localised emission. Several studies probed the astronomical-unit scales by means of high-angular resolution interferometry: none of which reported the presence of an inner companion. It seems that the companion discovered by \citet{Pomohaci+2019} whose detection is confirmed here, is the closest object ($\sim$8400~au) orbiting the MYSO.

\subsubsection{G314.3197+00.1125}
G314.3197+00.1125 (hereafter G314.3197) is an MYSO located at a distance of $\sim$3.6~kpc. Both the 20~$\mu$m  and 10.4~$\mu$m  images show a single object, without any MIR bright companion in the $\sim$10\arcsec surroundings \citep{Mottram+2007,Wheelwright+2012}. Three NIR-bright sources are seen in the 2MASS $K-$band images. 

We do not detect any extra source in the $28\arcsec\times28\arcsec$ $L'-$band NACO images ($\sim$580$-$50,000~au). We determine companion detection limits on a $5\sigma$ level of $\rm \Delta=3.4^{m}$.

\subsubsection{G318.0489+00.0854B}
G318.0489+00.0854B (hereafter G318.0489) is an MYSO associated with two \Hii regions whose radio counterparts have been established by \citet{Urquhart+2007, Navarete+2015}. G318.0489 is the faintest object in our sample with $L'=7.7^{m}$. The 20~$\mu$m image from \citet{Wheelwright+2012} shows a single resolved object with a slight cometary-like morphology towards the east direction. 

Within the probed separation range (540 to 65,000~au), no companion source was detected around G318.0489 in the $L'-$band NACO images. We place a $5\sigma$ detection limit of $\rm \Delta=2.7^{m}$.

\begin{table*}[!t]
\caption{Details of individual detections. }
    \centering
    \label{tab:companion_parameters}
    \begin{tabular}{c c c c c c c c c}
    \hline
    \hline \\ [-1.5ex]
        Source ID & Detect. sources & Sep. & Proj. Sep. & PA & $f_{r}$ & $L'$ & $\Delta mag$ & $P_{\text{spur}}$ \\ 
         & & (arcsec) & (au) & ($\deg$) & (\%) & (mag) & (mag) & (\%) \\ 
        \hline \\ [-1.5ex]
        \multirow{9}{*}{G203.3166+02.0564} & A & 12.41 & 9185 & 45.8 & $1.65\pm0.19$ & 6.2 & 4.5 & 0.37 \\ 
        ~ & B & 12.3 & 9099 & 46.7 & $4.58\pm0.28$ & 5.0 & 3.3 & 0.18 \\ 
        ~ & C & 5.113 & 3784 & 11.7 & $1.06\pm0.07$ & 6.6 & 4.9 & 0.11 \\ 
        ~ & D & 3.848 & 2848 & 317 & $0.62\pm0.04$ & 7.2 & 5.5 & 0.075 \\ 
        ~ & E & 4.605 & 3408 & 288 & $3.05\pm0.20$ & 5.5 & 3.8 & 0.036 \\ 
        ~ & F & 6.36 & 4706 & 276 & $0.50\pm0.03$ & 7.4 & 5.7 & 0.24 \\ 
        ~ & G & 2.738 & 2026 & 276 & $4.25\pm0.27$ & 5.1 & 3.4 & 0.011 \\ 
        ~ & H & 2.445 & 1809 & 165 & $0.69\pm0.04$ & 7.1 & 5.4 & 0.029 \\ 
        ~ & I & 9.226 & 6827 & 110 & $0.20\pm0.02$ & 8.5 & 6.8 & 1.4 \\ 
        
        \hline \\ [-1.5ex]
        \multirow{7}{*}{G232.6207+00.9959} & A & 14.22 & 23880 & 32.2 & $38.06\pm1.30$ & 6.3 & 1.0 & 0.83 \\ 
        ~ & B & 15.77 & 26500 & 57.2 & $10.53\pm0.38$ & 7.7 & 2.4 & 3.4 \\ 
        ~ & C & 8.575 & 14410 & 21.2 & $4.94\pm0.40$ & 8.6 & 3.3 & 2.2 \\ 
        ~ & D & 9.392 & 15780 & 48.7 & $18.47\pm0.66$ & 7.1 & 1.8 & 0.75 \\ 
        ~ & E & 8.403 & 14120 & 309 & $69.04\pm2.60$ & 5.7 & 0.4 & 0.17 \\ 
        ~ & F & 5.114 & 8591 & 43.9 & $9.90\pm0.39$ & 7.8 & 2.5 & 0.37 \\ 
        ~ & G & 9.182 & 15430 & 133 & $21.94\pm0.80$ & 6.9 & 1.6 & 0.57 \\ 

        \hline \\ [-1.5ex]
        \multirow{3}{*}{G263.7759-00.4281} & A & 10.28 & 7199 & 45.5 & $3.09\pm0.22$ & 9.5 & 3.8 & 1.2 \\ 
        ~ & B & 10.33 & 7229 & 46.3 & $2.57\pm0.20$ & 9.7 & 4.0 & 1.4 \\ 
        ~ & C & 2.812 & 1968 & 123 & $13.35\pm1.10$ & 7.9 & 2.2 & 0.03 \\ 

        \hline \\ [-1.5ex]
        \multirow{11}{*}{G265.1438+01.4548} & A & 13.51 & 9456 & 354 & $9.87\pm0.70$ & 6.3 & 2.5 & 1.5 \\ 
        ~ & B & 14.27 & 9986 & 331 & $6.37\pm0.51$ & 6.8 & 3.0 & 2.3 \\ 
        ~ & C & 12.65 & 8853 & 313 & $11.58\pm0.82$ & 6.1 & 2.3 & 1.2 \\ 
        ~ & D & 10.41 & 7289 & 314 & $5.27\pm0.43$ & 7.0 & 3.2 & 1.4 \\ 
        ~ & E & 10.32 & 7222 & 313 & $3.79\pm0.37$ & 7.4 & 3.6 & 1.8 \\ 
        ~ & F & 7.199 & 5039 & 50.4 & $4.21\pm0.33$ & 7.2 & 3.4 & 0.83 \\ 
        ~ & G & 7.153 & 5007 & 198 & $5.40\pm0.46$ & 7.0 & 3.2 & 0.67 \\ 
        ~ & H & 14.03 & 9822 & 119 & $13.05\pm0.96$ & 6.0 & 2.2 & 1.3 \\ 
        ~ & I & 13.49 & 9443 & 125 & $25.14\pm1.80$ & 5.3 & 1.5 & 0.75 \\ 
        ~ & J & 16.32 & 11430 & 134 & $107.10\pm7.60$ & 3.7 & -0.1 & 0.32 \\ 
        ~ & K & 9.092 & 6364 & 328 & $14.08\pm1.10$ & 5.9 & 2.1 & 0.52 \\ 

        \hline \\ [-1.5ex]
        \multirow{5}{*}{G268.3957-00.4842} & A & 9.047 & 6333 & 275 & $0.16\pm0.01$ & 10.9 & 7.0 & \textbf{32} \\ 
        ~ & B & 1.929 & 1350 & 168 & $8.58\pm0.45$ & 6.6 & 2.7 & 0.05 \\ 
        ~ & C & 4.769 & 3338 & 183 & $1.97\pm0.10$ & 8.2 & 4.3 & 1.1 \\ 
        ~ & D & 8.202 & 5741 & 197 & $8.88\pm1.40$ & 6.5 & 2.6 & 0.81 \\ 
        ~ & E & 8.355 & 5848 & 198 & $41.94\pm1.70$ & 4.8 & 0.9 & 0.29 \\

        \hline \\ [-1.5ex]
        \multirow{2}{*}{G269.1586-01.1383A} & A & 8.474 & 5932 & 208 & $77.45\pm2.90$ & 7.7 & 0.3 & 2.4 \\ 
        ~ & B & 8.544 & 5981 & 156 & $31.55\pm1.30$ & 8.7 & 1.3 & 5.5 \\

        \hline \\ [-1.5ex]
        \multirow{2}{*}{G305.2017+00.2072A} & A & 1.485 & 5940 & 78.5 & $16.64\pm1.20$ & 7.4 & 1.9 & 0.24 \\ 
        ~ & B & 8.77 & 35080 & 165 & $17.18\pm1.10$ & 7.4 & 1.9 & 7.8 \\ 

        \hline \\ [-1.5ex]
        \multirow{2}{*}{G310.0135+00.3892} & A & 2.604 & 8464 & 225 & $9.91\pm0.68$ & 4.8 & 2.5 & 0.11 \\ 
        ~ & B & 16.32 & 53040 & 133 & $0.20\pm0.01$ & 9.1 & 6.8 & \textbf{69} \\

        \hline
        
    \end{tabular}
\tablefoot{We provide their separations both in arcsec and au and their apparent $L'$-band magnitudes determined relative to the primaries. Errors on the physical separations are based on the distance uncertainties provided in the RMS catalogue and following \citet{Reid+2009} models, i.e. 10\%. $\Delta mag$ is the magnitude difference between the secondary and the primary, directly retrieved from the flux ratio. The last column provides the probabilities of chance projection of all the detected sources.}
\end{table*}

\section{Discussion}
\label{section:Discussion}

Our sample composed of 13 MYSOs offers an exclusive opportunity to study multiplicity among massive young stars for separations ranging from $\sim$300-50,000~au. In this section, we aim to compare our multiplicity results to those found for a wide range of masses ($\sim$0.05$-$25~\Msun), evolutionary stages (from (M)YSO to PMS/ZAMS and MS) and probed separations (0.1$-10^{4}$~au). We further discuss the implications for massive multiple formation.

\subsection{Multiplicity of MYSOs}

We have identified 8 potential multiples (1 binary, 2 triples, 1 quadruple and 4 higher-order multiple systems) from 13 targets. We rejected 2 detected sources that have a high probability of chance projection (P\textsubscript{spur} > 20 per cent: G310.0135B and G268.3957A). This results in a master sample of 39 physical companions, detected around 13 MYSOs. Our newly discovered companions comprise a significant minority of all known systems: only a few of the detected companions have already been reported in the literature. A compilation of the measured physical properties of the multiple systems is provided in Table \ref{tab:companion_parameters}. 

Multiplicity and companion fractions (MF, CF) can be computed in different ways but we adopt here the quantities recently presented in the review by \citet{Offner+2022}. For consistency, we have defined two ranges of physical separations in the following analysis. We derived the physical separations of the multiple systems using the distances (kinematic, parallax and H\textsubscript{2}O masers) shown in Table \ref{tab:NACO_journal}. We provide a global MF and CF for the whole separation range probed in this study, 600-35,000 au. For this parameter space, we consider 39 companion detections. We define another range as 600-10,000 au, resulting in a lower sample of 31 companion detections. However, the number of multiples remains unchanged, consequently, the frequency of multiples is identical for both subgroups. 

\begin{figure*}[!t]
\centering
\includegraphics[scale=0.55]{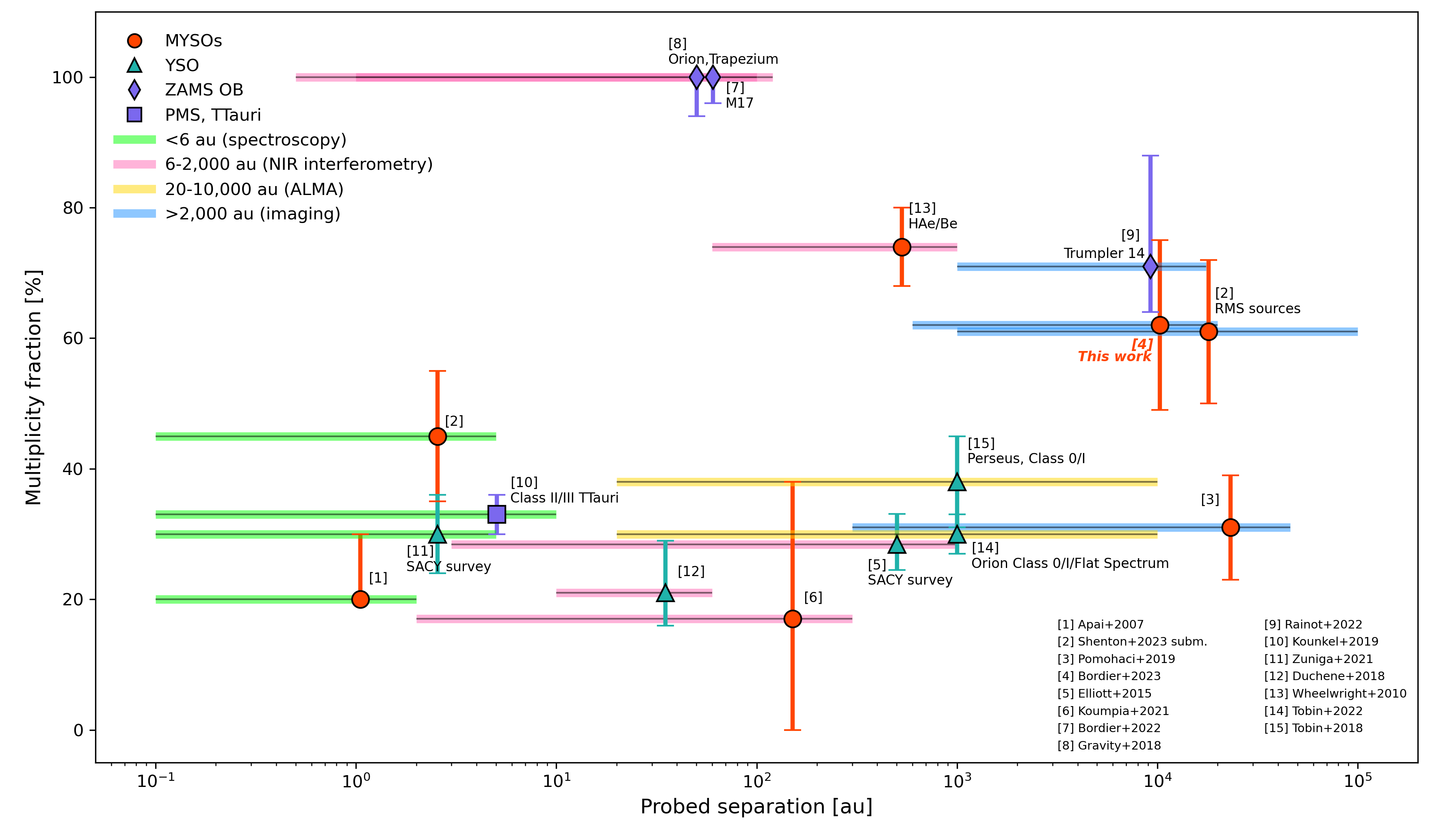}
\caption{Multiplicity frequency for young stars (low- to high-mass stars) for different spatial scales: from 0.1 to several hundreds of thousands of au. The horizontal bars of different colours indicate the separation intervals specific to each study. The different symbols refer to a category of young stars (see legends). }
\label{fig:multiplicity_all}
\end{figure*}

We estimate a raw multiplicity frequency of MF\textsubscript{600-35,000}=MF\textsubscript{600-10,000}=$62\pm13\%$ and a companion fraction of CF\textsubscript{600-35,000}=$3.0\pm0.5$ for the complete dataset and CF\textsubscript{600-10,000}=$2.4\pm0.4$ for the restricted group, agreeing with each other within the errors. The uncertainties on the MF follow binomial statistics while the uncertainties on the CF follow Poissons statistics, as described in \citet{Sana+2014} and \citet{Moe+2017}. Our survey probes the multiplicity of massive stars at their youngest stage and complements well-existing surveys which focus on closer orbits, from spectroscopic to interferometric regimes. Comparing our findings derived from a group of MYSOs with those of different classes of objects and evolutionary stages is not a trivial exercise and has some limitations. Hence, it is important to consider statistics tracing similar ranges of separations, but also the specific observations and techniques used in each study. One should keep in mind that the statistics are highly affected by the homogeneity of the samples, the sample size and the sensitivity limits. As a result, the statistics associated with low-mass stars campaigns are much more robust than for their massive counterparts. In fact, the samples contain several hundreds or even thousands of low-mass objects while the studies on young massive stars contain at most a few tens of objects. It is for these various reasons that we decided to split our results into two different separation ranges, for more consistency and a fairer comparison with the existing studies, especially the one of \citet{Pomohaci+2019}, for which the comparison is among the most robust ones we can refer to. Figure \ref{fig:multiplicity_all} gathers a handful of multiplicity results for different populations of young stars, in clusters and field, targeting different mass ranges and separations.

For similar separation ranges (600-10,000+~au), the previous survey of \citet{Pomohaci+2019} reports MF\textsubscript{600-10,000}=$31\pm8\%$, lower than our results despite a larger sample size and average distance (3.1~kpc). Unlike their conclusions, we notice a significant difference in the MF over distance. Among the seven sources lying at distances shorter than 2~kpc, six are multiples yielding MF\textsubscript{$<$2~kpc}=$85\pm13\%$ while we measure MF\textsubscript{$\geq$2~kpc}=$33\pm19\%$ for sources further away than 2~kpc. The distance cut we used here seems to show that at further distances, the fainter companions are not detected. A similar comparison can be drawn for the MF as a function of the bolometric luminosity. We count nine MYSOs whose bolometric luminosity is under 10,000~\Lsun and four brighter than 10,000~\Lsun, yielding MF\textsubscript{$<$10000\Lsun}=$55\pm17\%$ and MF\textsubscript{$\geq$10000\Lsun}=$75\pm22\%$ respectively. For both luminosity ranges, the MFs are consistent with each other within the errors. Keeping the comparison strictly to the MF regardless of the distance towards the source, a recent study lead by R. Shenton shows notable consistency with our derived MF, for similar separation probes (see Fig. \ref{fig:multiplicity_all}). Using $K-$band data from VVV, UKIRT and UKIDSS, they report a MF varying from 54 to 70 per cent, for a population of 402 MYSOs and spanning separations of $10^{3}-10^{5}$~au with a mean physical separation of companions of $\sim$18,000~au \citep{Shenton+2023}. The similarity with our work is all the more encouraging given their much higher sample size. The main asset of their analysis is the deeper limiting magnitude at which they can detect companions, however, their spatial resolution stays below the NACO's spatial resolution presented in this work or in \citet{Pomohaci+2019}: the inner $\sim$1 to 1.5\arcsec\, of each MYSO is essentially a blank spot and only the brightest companions between 1.5 to 2\arcsec are detected, with a peak of overall detected companions between 3 to 6\arcsec. 

Zooming in at the interferometric scales, the results of \citet{Koumpia+2021} whose sample was composed of six MYSOs with masses ranging from 8.6~\Msun to 15.4~\Msun, are somewhat smaller than ours but in agreement with those of \citet{Pomohaci+2019}. They report a rather low MF of $17{\substack{+21 \\ -17}}$\%. These statistics are in stark contrast with those of \citet{GravityCollab+2018} and \citet{Bordier+2022} who report an MF of 100\% for relatively small samples of pre-main-sequence massive stars (4 and 6 stars respectively), for similar traced separations. The latter results are consistent with the more evolved populations of massive stars \citep[see also Fig. \ref{fig:multiplicity_all}]{Sana+2014} but raise a major question about the evolution of multiplicity between the youngest and more evolved stages. Dynamic rearrangements within dense clusters must be at play to shape the final structures of the systems and could point towards an agreement with the competitive accretion models which require the stars to interact via mass or momentum exchanges in order to shape the final structure of the cluster. This point is further discussed in Section \ref{subsection:star_formation_models}.

Moving on to spectroscopic separations, \citet{Kounkel+2019} and \citet{Zuniga+2021} studied the low-mass regime of young stars, the former focusing on a sample of Class II/III pre-main-sequence T-Tauri stars and the latter on a sample of 410 nearby YSOs ($< 200$~pc). They respectively derive a multiplicity fraction of $30\pm6$~\% (averaged across their three groups) and $33\pm3$~\%. We also mention the results of \citet{Shenton+2023} who investigate close binarity of 29 YSOs by the means of high-resolution spectroscopy with X-SHOOTER/VLT and find a multiplicity occurrence varying from 35 to 55\%. All these values are in alignment with the findings of \cite{Tobin+2018,Tobin+2022} that derived similar protostellar multiplicity statistics ($30\pm3$~\% and $38\pm7$~\%), considering separations as small as 20~au and up to $10^{4}$~au (ALMA observations), within their sample of young low-mass stars in Perseus and Orion star-forming regions, respectively. Meanwhile, the close binary fraction within a sample of Herbig Ae/Be protostars, whose masses lie between low-mass YSOs and the MYSOs we study here, inferred by \citet{Apai+2007} reaches $20{\substack{+10 \\ -0}}$\%. This is within the errors of the above-mentioned results, but considerably lower than the wide binary fraction derived by \cite{Wheelwright+2010} for the same class of protostars ($74\pm6$~\%). 

Figure \ref{fig:multiplicity_all} does not show a clear trend in the multiplicity statistics of young stars, independently of their mass and environment. It seems that two trends are emerging: a significant number of studies \citep{Apai+2007,Elliott+2015,Duchene+2018,Tobin+2018,Kounkel+2019,Koumpia+2021,Zuniga+2021,Tobin+2022} have recently reviewed the multiplicities of several samples and whatever the mass and separation explored, the multiplicity seems to vary between 20 and 50\% (bottom part of the figure). The other trend appears for multiplicities above 50\%. We note that no spectroscopic study unambiguously derives a multiplicity fraction above 50\%. Assuming that this class of stars is observable with the current technical means and that they meet the requirements of instruments' sensitivity, the low spectroscopic multiplicity found among MYSOs is questionable. One could naturally conclude that 1~Myr old massive stars do not have any close companions. The present study shows that companions can be much fainter than the primary, which might also be the case in the spectroscopic regime. The most massive star could be two or three orders of magnitude brighter than the faint companion in which case the primary would overrule the spectroscopic emission and hide the spectroscopic features from the companion, preventing one to proceed with a double spectroscopic binary analysis. Similarly, a low-mass object would not significantly affect the orbit of a higher-mass primary, falling off the radial velocities detection limits. In the scenario in which there is a dearth of spectroscopic companions as it was shown in M17 \citep{Sana+2017,Ramirez+2021}, the companions lying at intermediate separations \citep[as observed by]{GravityCollab+2018,Bordier+2022} can eventually experience an inward migration on Myr timescales, as theorised by \citet{Sana+2017}. On the other hand, most interferometric studies, whether conducted with ALMA or VLTI, converge towards rather high multiplicities in the 1-200~au scales, \citep{Wheelwright+2010,GravityCollab+2018,Bordier+2022}, with the exception of the study conducted by \citet{Koumpia+2021}, for which the MF is lower but the evolutionary stage differs slightly as their objects are all MYSOs.
 These latter results have to be taken with more caution as the probed stars tend to have already reached the birthline. Here again, the physical distances to the object are an important source of error and we recall that unlike young low-mass stars, the majority of massive stars are found at large distances. Thus, the average distance of the \citet{Pomohaci+2019} study for instance, is 3.3~kpc while that of our study is slightly more than 2~kpc. In contrast, \cite{Tobin+2018}'s studies are settled for regions at 200~pc from the Sun, a factor of 10 closer.


\begin{figure*}[!t]
\centering
\includegraphics[scale=0.50]{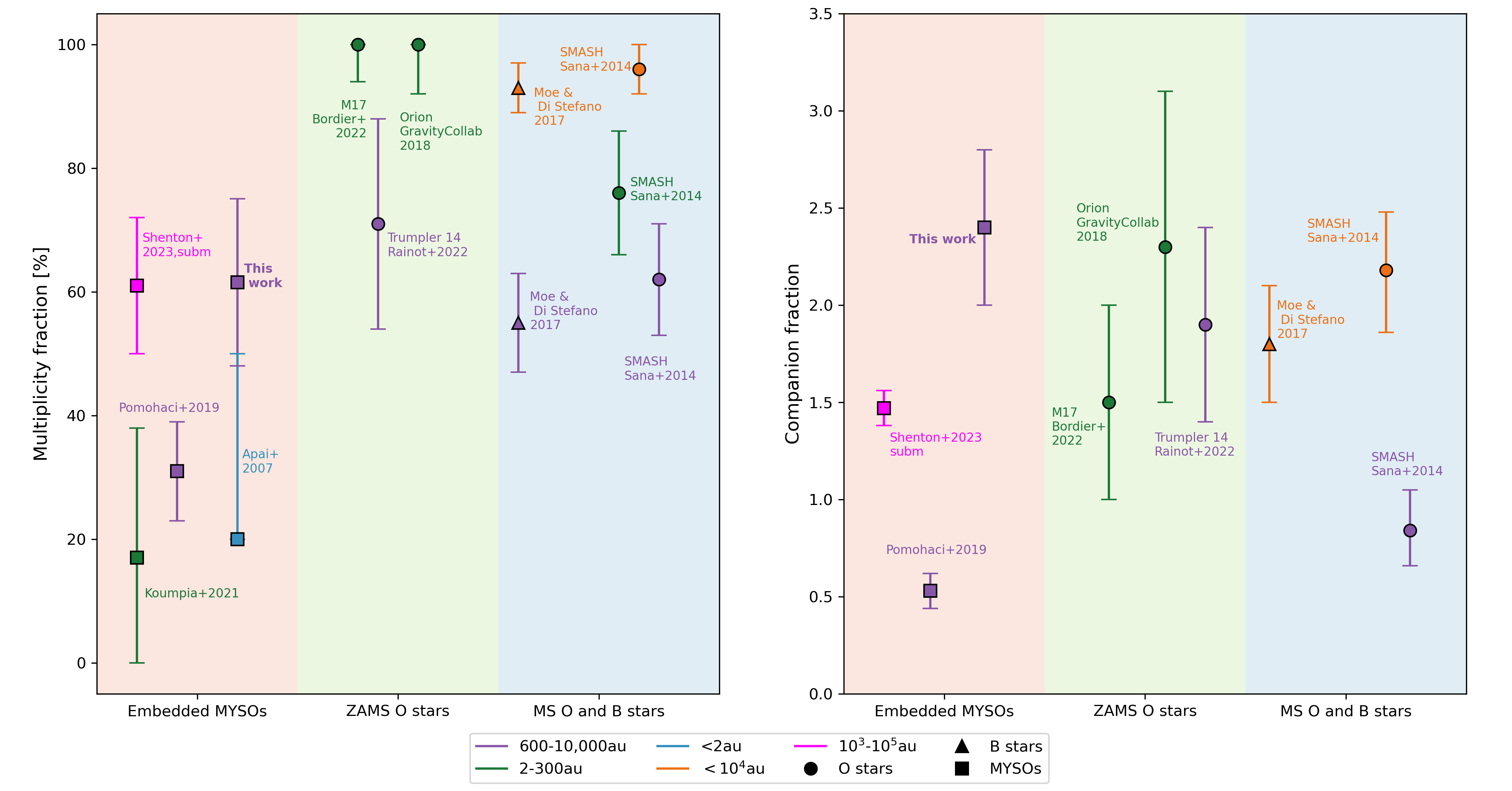}
\caption{Multiplicity statistics, including the MF (right) and CF (left) for massive stars and different evolutionary stages. The different colours and symbols stand for the different probe separation ranges, as stated in the legend box.}
\label{fig:MF_CF_Massive}
\end{figure*}

We further investigate our findings for a specific class of objects: massive stars. To do so, we relax the evolutionary stage parameter and compare our results to those of different studies \citep{Apai+2007,Sana+2014,GravityCollab+2018,Pomohaci+2019,Koumpia+2021,Rainot+2022,Bordier+2022}. The results are shown in Fig. \ref{fig:MF_CF_Massive}. We keep the comparison consistent by providing statistics tracing similar separation ranges. From 600 to 10,000~au we note that the multiplicity frequency is consistent across all the evolutionary stages and within error bars, with the exception of the study conducted by \citet{Pomohaci+2019} who found a lower fraction, almost two times lower, for the same probed separations. However, looking at the companion rate panel, the frequency of companions at wide separations seems to drop across the evolution, while it is higher when studies go down to the spectroscopic regime. Indeed, the SMASH survey that assesses the companion fraction among main-sequence O stars shows that the overall companion fraction (resolved+unresolved) is more than a factor of two higher than the statistics for the sole group of resolved companions \citep{Sana+2014}. It may indicate that the presence of close companions is preponderant when deriving the companion rates for main-sequence massive stars, while it does not seem to be the case for young massive stars. For instance, the spectroscopic statistics derived by \citep{Apai+2007} show a rather low limit for the binary fraction. This trend is consistent with the idea that massive star formation does not result in a tight, close binary system, but rather forms at large distances and experiences dynamical effects over time \citep{Ramirez+2021}. Overall, we note that our derived statistics are in better agreement with those derived for older ages than the statistics derived for similarly aged populations \citep{Pomohaci+2019,Koumpia+2021}. It is clear from Fig. \ref{fig:multiplicity_all} and Fig. \ref{fig:MF_CF_Massive} that the multiplicity statistics are strongly dependent on the mass of the targeted sources.

\begin{figure}
\centering
\includegraphics[scale=0.40]{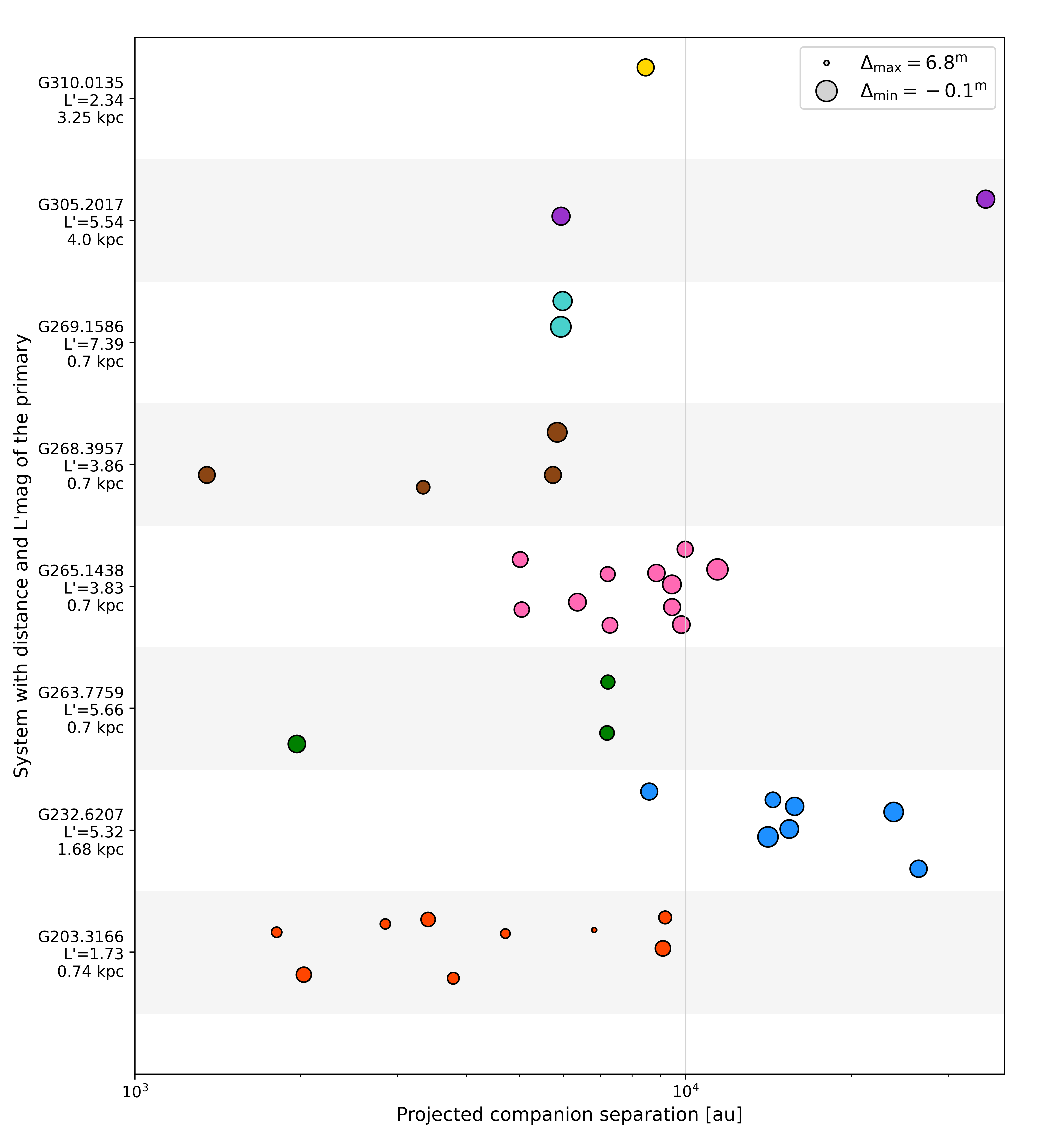}
\caption{Companion separation per system with given their L' magnitude and distance in kpc. The colours indicate different systems and the marker size scales with the magnitude difference in relation to the primary. The vertical grey line indicates the 10,000~au separation that we used as a cut earlier for a fairer comparison with other studies. Small variations in the y-direction are to avoid overlapping markers. }
\label{fig:Mag_sep}
\end{figure}

We calculated the physical separations of the multiple systems using the RMS kinematic distances, displayed in Table \ref{tab:NACO_journal}. Figure \ref{fig:Mag_sep} shows the projected physical separation of the detected companions per system. The size of the markers refers to the brightness ratio in relation to the primary. The projected separation is highly dependent on the physical distance towards the primary star. The closest companion detected is G268.3957B, at around 1350~au of its primary. The plot suggests a preferred separation range for companion stars, which likely indicates the separation at which they were formed given the young age of the central object. We clearly observe that the vast majority of the companions (80\%) are detected within 10,000~au around their central MYSO and at least one companion per system is observed in this parameter space. For a given system, the nature of the detected companion is rather heterogeneous, faint companions are found at a separation of $\sim$5000~au as well as brighter companions. Figure \ref{fig:Mag_sep} shows the powerful nature and effectiveness of AO-imaging for detecting new and faint companions, down to a factor of about 500 times fainter than the massive central YSO in the present study ($\rm \Delta_{max}=6.8^{m}$ for G203.3166-I at a distance of $\sim$6800~au). 

In view of the results, there is no clear correlation between the separation and the type of companion that is formed in the vicinity of the massive primary. It is important to remind that we observe the $L'$ emission in the images, which includes for instance the dust emission, whose radiation comes from the natal envelope and the circumstellar disk. Thus, $L-$band magnitudes cannot be trivially used as a proxy for stellar mass companions as it is commonly the case for $K-$band magnitudes (\citealt{Pomohaci+2019,Rainot+2020,Reggiani+2022,Pauwels+2023}). In fact, we lack crucial information such as a mass-luminosity relation for the $L'$ band, an accurate distance, and both the extinction (interstellar and circumstellar) and embeddedness of the systems. Masses are commonly estimated using comparison with evolutionary tracks \citep{Bordier+2022} or directly from the observed spectral type of the primary, which strongly depends on the extinction correction \citep{Beltran+2016}. All these unknowns could be hypothesised, such as assuming that the extinction towards the primary spreads for the whole system, but this is less likely to be valid for distant companions (as it is the case here) since these companions may not be obscured by the same amount of dust as the primary and such an assumption would result in an overcorrection. Combined with the uncertainty about the age of the systems, the derived masses would include large error bars which would not allow us to have a clear view as to the real mass of the formed companions or their spectral type. In light of the present results, the only conclusive statement we can make about the mass of the companions is that they appear less massive than the central massive protostar, assuming that the companions are shrouded by the same or lesser amount of dust than the primary. 

That being said, companion separation alone allows us to discuss the most likely scenarios for the formation of such systems, which we highlight in the next section.

\subsection{Implications for star formation models}
\label{subsection:star_formation_models}

As predicted by the initial mass function (IMF), massive stars (M$_{\text{i,primary}}>8$~\Msun) make up $<$ 1\% of the total stellar population, meaning that many MYSOs are most likely the most massive objects locally. Observationally, regions of massive star formation in our Galaxy are located in more remote regions than those of low-mass star formation, which has the effect of considerably reducing the level of detail that we can probe and rendering their study arduous. Unlike low-mass star formation, obtaining a clear view of the formation process of massive stars with a high level of detail is difficult. As such, defining the frontier of similarities and differences between low-mass and high-mass star formation is not a trivial task, with an added complexity when multiplicity is at play. We mainly rely on (3D) hydrodynamic and magneto-hydrodynamic (MHD) simulations of collapsing molecular clouds to understand all the physical processes involved \citep{Krumholz+2007,Myers+2013,Meyer+2018,Oliva+2020,Mignon-Risse+2021a,Mignon-Risse+2023}. Because massive binary formation involves many complex processes, various observational campaigns aiming at probing different regions around MYSOs as well as studying the substructures reinforced the numerical works and yield a better understanding of individual effects at one time. 

Several theories may explain the origin of massive binaries. A complete review of the current models and modes of fragmentation is provided in \citet{Offner+2022}. Here we provide the three main and independent scenarios for multiples formation and discuss the most likely channels for our observations. Owing to the probed spatial scales in this work that set a limit on multiple system separations, our results can only place constraints on formation mechanisms at core scales ($>$ 1,000~au). 

The gravitational fragmentation of massive cloud cores or filaments into two Jean's unstable over-densities, providing a sufficient reservoir of gas, may explain the formation of wide binaries ($>$1,000~au). In that specific scenario, the main triggers of fragmentation are rotation and turbulence which create density and velocity asymmetries. The Jeans-radius being of the order of 10,000~au, two wide and bound components can be generated. Some characteristics may be imprinted during formation that are still detectable when catching young multiple systems right after the birth parent cloud has dispersed. The initial separation is one of those predicted signatures: a lower limit is set at $\sim$100~au, strictly fixed by the effects of magnetic support, angular momentum and tidal forces \citep{LeeHennebelle2018}, and 0.1~pc is proposed as an upper limit as two fragments have nearly no chance to be initially bound at such separations and will inevitably break off. Turbulent fragmentation also tends to produce randomly distributed alignments of stellar spins, accretion disks, and protostellar outflows \citep{Offner+2016,Bate2018,Lee+2019}. \citet{Lee+2019}'s numerical simulations confirmed the existence of such wide embedded and accreting multiples for a short amount of time, before they migrate towards closer orbits, at a couple hundreds of au. \citet{Pomohaci+2019} and the present study also shows observational evidence for such wide companions. At this stage, we cannot confirm that core fragmentation alone explains the observed structures,  as we would need to probe the innermost regions of MYSOs to seek for the presence of closer companions and conclude as to the other mechanisms that could simultaneously take place. We point out the case of G282.2988 around which \citet{Koumpia+2019} showed the presence of an interferometrically resolved companions at $\sim$50~au while \citet{Pomohaci+2019}'s study had already reported two wide companions at $\sim$5800~au and $\sim$10,000~au respectively. The complementary of the probed spatial scales here, shows the existence of a wide range of companions, however, given the very young age of that source, core fragmentation would fail at explaining the origin of the 50-au close companion. This brings us to the next theory that may explain the presence of closer companions. 

Independently from the previous mechanism and during the monolithic collapse of a massive core, the required massive circumstellar disks which need to feed the central object to high masses are prone to gravitational instabilities and lead to fragmentation and subsequent formation of companions \citep{Krumholz+2007,Kuiper+2010,Meyer+2018,Oliva+2020,Mignon-Risse+2023}. The main physical mechanism responsible for the creation of spiral arms where fragments are created resides in the so-called Toomre instabilities \citep[a gravitationally unstable disk will have Toomre Q parameter $<1$,][]{Toomre1964}. In fact, it has been demonstrated that early in evolution, disks reach a sufficient mass to become self-gravitating, form spirals and fragment. An increasing amount of observational evidence and characterisation of disks around MYSOs and its substructures, including the detection of a fragment in the outskirts of the accretion disk, have been reported since the advent of high-angular resolution facilities such as ALMA and the VLTI \citep[see for e.g.][]{Kraus+2010,Cesaroni+2017,Beuther+2017,Ginsburg+2018,Ilee+2018,Frost+2019,Frost+2021a,Maud+2019,Johnston+2020}. These freshly created fragments can grow in mass through secondary smaller disks that form locally. If they successfully overcome the fragmentation epoch imprinted with many mechanisms of destruction such as shearing, collisions, accretion bursts or being accreted by the central protostar, they can become actual companions. Ultimately, simulations show that the central massive protostar gains sufficient mass through different episodes of accretion and the orbits of the surviving fragments shrink, which provides an ideal framework for the formation of spectroscopic binaries \citep{Kratter+2006,Oliva+2020,Tokovinin+2020}. Low-mass cluster member stars would thus be formed all within the size of the accretion disk, giving rise to an unstable small-N system, prone to (partial) dynamical disruption. In order to explain binaries with separations smaller than 10~au, the scenario \emph{fragmentation then migration} has gained popularity since massive disks inefficiently fragment at distances smaller than 50$-$100~au from the central object \citep{Krumholz+2009,Kratter2010,Zhu+2012}. In this way, binaries can shrink their orbit by one or two orders of magnitude after they have formed at wider separations \citep{Moe+2018}. 

The observational predictions supporting disk fragmentation is undoubtedly the presence of companions within the expected disk size around MYSOs (up to 2,000~au, refer to the previous references for disk sizes around massive stars). Another reliable test would be to measure the alignment of the accretion disk with that of the binary orbit. In fact, disk fragmentation would lead to a final orbit within the same plane as the accretion disk. We detect four companions within 2,000~au, in three different systems (see Fig. \ref{fig:Mag_sep}). NACO offers sufficient angular resolution to probe closer regions but these companions are not found. Several possibilities may explain the observed lack of closer companions: the systems may be very young such that these companions have just formed in the outskirt of the disk and have not experienced, if any, inward migration nor any other dynamical interaction. Conversely, disk fragmentation could have been successful at forming companions within the accretion disk, but substantial torque generated by small individual secondary disk in addition to the ongoing accretion could push the companion outwards on a wider orbit, as demonstrated by \citet{Munoz+2019}, in the case of binaries embedded in disks.  Another explanation is that companions do not form in the 100-2,000~au but rather at larger scales or in the innermost regions that are not probed in the present study, but that were detected in \citep{Kraus+2017,Koumpia+2019} for example. Moreover, any companion that would have already cleared up its dusty cocoon might be opaque at the probed wavelength. Though this assumption is less likely as the central source might evolve faster than the surrounding lower-mass companions. Lastly, we cannot rule out the possibility of highly eccentric orbits \citep{MaizApellaniz+2017,Tokovinin_ecc+2020}. As a deduction, the companion spends a consequent amount of time at a large separation before reaching the periastron location, which can be much closer to the MYSO. Statistically and with only one snapshot in hand, it is most likely that we are overestimating the separation to the primary. Only high-precision astrometric measurements can provide accurate eccentricity measurements and eliminate this hypothesis. Even though such measurements may also provide key constraints on the binary formation mechanisms, they are however challenging to obtain as binaries with separations of $\sim$1000~au have orbital periods of several thousand years, which would translate into very tiny variations of the overall orbit by performing a typical monitoring with a  baseline of a few years \citep{Hwang+2022}. 
    
Dynamical interactions and capture may occur during the dissolution of the forming regions, that is, post fragmentation, where an initially single and unbound star is paired with or attracts another nearby star or system, due to energy and angular momentum exchange.  The high degree of clustering in most star-forming regions facilitates the interactions between stars, readjusting the hierarchy, multiplicity and orbital parameters of the systems formed via the two above-mentioned channels. The timescales and upper limit scales at which capture occurs are uncertain. This mechanism would explain the formation of both the tightest and widest systems. \citet{Murillo+2016} demonstrate that a significant fraction of high order multiple systems are non co-eval, showing evidence for different episodes of core fragmentation and possible reconfigurations via dynamical interactions. Most of our targets do not seem to belong to particularly crowded regions, as per the derived local stellar densities in the $L-$band (see Sect. \ref{section:datanalysis}, at the exception of G265.1438). It also questions the real binding connection between the central object and the detected sources and their association to the cluster rather than to the system. At this point, it is not an easy task to assess whether all these associations shared a common origin, due to the absence of proper motion measurements. 

Orbital hardening through disk-driven migration, causes the shrinkage of the inner orbits and is likely associated with the ejection of outer low-mass companions. Dynamical-friction forces are also at play and could efficiently reduce orbital separations from thousands of au to hundreds or even tens of au, in less than a few Myr \citep{Ramirez+2021}. Such inward migration is consistent with disk fragmentation simulations and produces spectroscopic binaries \citep{Oliva+2020}. Kozai-Lidov cycles coupled with tidal friction are yet the most efficient way to produce the tightest binaries, caused by the presence of an inclined outer tertiary \citep{Kozai1962,Lidov1962,Michaely+2014,Toonen+2020}. All these dynamical interactions are the ultimate opportunity to shape the final multiplicity distribution observed among pre-MS and more evolved field star populations, however, they highly depend on the birth local environment, such as radiation and protostellar outflows. 

On that latter point, we want to point out the case of G203.3166. \citet{Parker+2022} recently published a study aiming at tracing back the formation history of NGC2264, star-forming region that hosts G203.3166. They isolate two main sub-clusters of about 2~pc radii, based on the derived age dichotomy: S Mon ($\sim$5~Myr) and IRS 1/2 ($\sim$2~Myr) sub-clusters. G203.3166 (IRS1 in their study) is thus the main driving force of one of the sub-clusters. They report that NGC 2264 might have experienced multiple episodes of star formation, given the different age of the sub-cluster without further iterating on the preferred pathways for their formation. \citet{Flaccomio+2023} add that NGC 2264 is not dynamically relaxed and its present configuration is the result of multiple dynamical processes that are most likely still ongoing. If we focus on G203.3166 itself and our results, providing that the observed objects are physically associated with each other, if one looks closely at the disposition of companions around G203.3166, one should notice the distribution of the objects in an arc on the SW side of the MYSO. This specific arrangement would result in a high-pressure front or shock wave formed by a trigger that is most likely G203.3166 being the brightest and most luminous object in the field \citep{Allen+1972,Thompson+1998}. The reported presence of an outflow further supports that hypothesis, in which G203.3166 would form first and the remaining stars are formed within the next 2,000~yr, of the onset of the outflow assuming a typical outflow velocity of 10 km$/$s and a distance of 4,000~au. After that estimated time, most of the gas and dust reservoir needed to form new stars would have been swept away. This timescale argues strongly for the concept of triggered star formation.




\section{Conclusions}
This study was designed to witness one of the earliest stages in the lives of massive stars, after the formation of the star itself in order to tackle the uncertainty as to the formation process by characterising the epoch of the onset of stellar multiplicity. In this work we present the multiplicity statistics in the $L'-$band for a sample composed of thirteen MYSOs, from AO-assisted imaging VLT/NACO observations. The parameters from individual observations including night conditions are gathered in Table \ref{tab:NACO_journal} and a summary of the detected companions per system is provided in Table \ref{tab:companion_parameters}. We have derived multiplicity properties of our sample within the limited angular range of 0.16\arcsec$-$28\arcsec, corresponding to projected separations of $\sim$120~au for the closest targets ($\sim$0.7~kpc) to $\sim$60,000~au for the widest ones. These results are compared to other populations by conscientiously selecting the comparison criteria (evolutionary stage, mass, and separation range) and seeking statistical similarities or differences. Our main results can be summarised as follows:
\begin{enumerate}
    \item We report the identification of eight multiple systems (one binary, two triples, one quadruple, and four higher-order systems), with a total of 39 companions detected from the observation of thirteen MYSOs. 
    \item The detected sources cover a wide contrast range of between $\rm \Delta=-0.1^{m}$ and  $\rm \Delta=6.9^{m}$ but the majority of them are found in the $\rm \Delta=2.5^{m}-5^{m}$ range. The poor local source density in the $L$ band in most of our fields leads to a negligible chance of spurious association apart from two candidates that we discarded in the statistics.
    \item No clear-cut trend could be found in the radial location of the sources, or in their brightness ratios. However, we note that nearly 70\% of the companions are to be found between $3,000-10,000$~au and that at least one companion per system is to be found within 10,000~au. 
    \item When no companion is found within the field, we provide a detection limit computed from the artificial injection of fake companions, demonstrating that NACO reaches the necessary capabilities to detect faint ($L'\sim9$) companions over the entire FoV of our observations. 
    \item The detection of companions at different scales informs us on possible pathways for massive star formation and evolution, regardless of their separation. The presence of wide companions ($\gtrsim$1000~au) indicates that they rather formed via fragmentation during the initial free-fall collapse, in the earliest stages of massive star formation. For closer companions and given the large scales that the accretion disk can reach around MYSOs, fragmentation can occur later on in the outskirts of the disk and produce companions. This specific category of companion is most likely to be affected by dynamical interactions such as inward migration and meet the orbital properties measured in older populations of massive stars.
\end{enumerate}


Given the data currently available, including our results and the broader literature, it is still too early to properly conclude whether the observed structures preferentially follow one formation scenario over the other. Further attempts to observe these objects at higher angular resolutions with more sensitive instruments such as the VLTI/GRAVITY will allow us to cover the missing innermost separation range and as such detect the substructures involved in the formation of the central star (accretion disk and outflows) or closer companions. In addition, observing these sources in a band with a well-known mass-luminosity relation would help in assessing the mass and spectral type of the detected companions, thus discarding some dynamical interactions that are less likely to occur for a given mass ratio. Probing the full range of separations and having proper mass estimates would assuredly help us better understand the interactions between the disk, the host star, and its companions.

\label{section:Conclusions}

\begin{acknowledgements}
E.B. gratefully acknowledges support from the studentship programme founded by the European Southern Observatory. This project has further received funding from the European Research Council under European Union's Horizon 2020 research programme (MULTIPLES, No 772225). This publication is based on observations collected at the European Organisation for Astronomical Research in the Southern Hemisphere under ESO programme 090.C-0207(A).
This paper made use of information provided on the Red MSX Source archive at \url{http://rms.leeds.ac.uk/}. This research has made use of the SIMBAD database and VizieR catalogue access tool, operated at CDS, Strasbourg, France. Part of our code made use of Astropy (\url{http://www.astropy.org}) a community-developed core Python package for Astronomy \citep{astropy:2013, astropy:2018} and of the python code \textsc{photutils}. We used the internet-based NASA Astrophysics Data System for bibliographic purposes.
\end{acknowledgements}

\bibliographystyle{aa}
\bibliography{main}

\begin{thebibliography}{130}
\expandafter\ifx\csname natexlab\endcsname\relax\def\natexlab#1{#1}\fi

\bibitem[{{Allen}(1972)}]{Allen+1972}
{Allen}, D.~A. 1972, \apjl, 172, L55

\bibitem[{{Apai} {et~al.}(2007){Apai}, {Bik}, {Kaper}, {Henning}, \& {Zinnecker}}]{Apai+2007}
{Apai}, D., {Bik}, A., {Kaper}, L., {Henning}, T., \& {Zinnecker}, H. 2007, \apj, 655, 484

\bibitem[{{Astropy Collaboration} {et~al.}(2018){Astropy Collaboration}, {Price-Whelan}, {Sip{\H{o}}cz}, {G{\"u}nther}, {Lim}, {Crawford}, {Conseil}, {Shupe}, {Craig}, {Dencheva}, {Ginsburg}, {Vand erPlas}, {Bradley}, {P{\'e}rez-Su{\'a}rez}, {de Val-Borro}, {Aldcroft}, {Cruz}, {Robitaille}, {Tollerud}, {Ardelean}, {Babej}, {Bach}, {Bachetti}, {Bakanov}, {Bamford}, {Barentsen}, {Barmby}, {Baumbach}, {Berry}, {Biscani}, {Boquien}, {Bostroem}, {Bouma}, {Brammer}, {Bray}, {Breytenbach}, {Buddelmeijer}, {Burke}, {Calderone}, {Cano Rodr{\'\i}guez}, {Cara}, {Cardoso}, {Cheedella}, {Copin}, {Corrales}, {Crichton}, {D'Avella}, {Deil}, {Depagne}, {Dietrich}, {Donath}, {Droettboom}, {Earl}, {Erben}, {Fabbro}, {Ferreira}, {Finethy}, {Fox}, {Garrison}, {Gibbons}, {Goldstein}, {Gommers}, {Greco}, {Greenfield}, {Groener}, {Grollier}, {Hagen}, {Hirst}, {Homeier}, {Horton}, {Hosseinzadeh}, {Hu}, {Hunkeler}, {Ivezi{\'c}}, {Jain}, {Jenness}, {Kanarek}, {Kendrew}, {Kern}, {Kerzendorf}, {Khvalko}, {King}, {Kirkby}, {Kulkarni},
  {Kumar}, {Lee}, {Lenz}, {Littlefair}, {Ma}, {Macleod}, {Mastropietro}, {McCully}, {Montagnac}, {Morris}, {Mueller}, {Mumford}, {Muna}, {Murphy}, {Nelson}, {Nguyen}, {Ninan}, {N{\"o}the}, {Ogaz}, {Oh}, {Parejko}, {Parley}, {Pascual}, {Patil}, {Patil}, {Plunkett}, {Prochaska}, {Rastogi}, {Reddy Janga}, {Sabater}, {Sakurikar}, {Seifert}, {Sherbert}, {Sherwood-Taylor}, {Shih}, {Sick}, {Silbiger}, {Singanamalla}, {Singer}, {Sladen}, {Sooley}, {Sornarajah}, {Streicher}, {Teuben}, {Thomas}, {Tremblay}, {Turner}, {Terr{\'o}n}, {van Kerkwijk}, {de la Vega}, {Watkins}, {Weaver}, {Whitmore}, {Woillez}, {Zabalza}, \& {Astropy Contributors}}]{astropy:2018}
{Astropy Collaboration}, {Price-Whelan}, A.~M., {Sip{\H{o}}cz}, B.~M., {et~al.} 2018, \aj, 156, 123

\bibitem[{{Astropy Collaboration} {et~al.}(2013){Astropy Collaboration}, {Robitaille}, {Tollerud}, {Greenfield}, {Droettboom}, {Bray}, {Aldcroft}, {Davis}, {Ginsburg}, {Price-Whelan}, {Kerzendorf}, {Conley}, {Crighton}, {Barbary}, {Muna}, {Ferguson}, {Grollier}, {Parikh}, {Nair}, {Unther}, {Deil}, {Woillez}, {Conseil}, {Kramer}, {Turner}, {Singer}, {Fox}, {Weaver}, {Zabalza}, {Edwards}, {Azalee Bostroem}, {Burke}, {Casey}, {Crawford}, {Dencheva}, {Ely}, {Jenness}, {Labrie}, {Lim}, {Pierfederici}, {Pontzen}, {Ptak}, {Refsdal}, {Servillat}, \& {Streicher}}]{astropy:2013}
{Astropy Collaboration}, {Robitaille}, T.~P., {Tollerud}, E.~J., {et~al.} 2013, \aap, 558, A33

\bibitem[{{Bailer-Jones} {et~al.}(2021){Bailer-Jones}, {Rybizki}, {Fouesneau}, {Demleitner}, \& {Andrae}}]{Bailer-Jones+2021}
{Bailer-Jones}, C.~A.~L., {Rybizki}, J., {Fouesneau}, M., {Demleitner}, M., \& {Andrae}, R. 2021, \aj, 161, 147

\bibitem[{{Bally} {et~al.}(1995){Bally}, {Devine}, {Fesen}, \& {Lane}}]{bally1995}
{Bally}, J., {Devine}, D., {Fesen}, R.~A., \& {Lane}, A.~P. 1995, \apj, 454, 345

\bibitem[{{Bate}(2018)}]{Bate2018}
{Bate}, M.~R. 2018, \mnras, 475, 5618

\bibitem[{{Beltr{\'a}n} \& {de Wit}(2016)}]{Beltran+2016}
{Beltr{\'a}n}, M.~T. \& {de Wit}, W.~J. 2016, \aapr, 24, 6

\bibitem[{{Beuther} {et~al.}(2017){Beuther}, {Walsh}, {Johnston}, {Henning}, {Kuiper}, {Longmore}, \& {Walmsley}}]{Beuther+2017}
{Beuther}, H., {Walsh}, A.~J., {Johnston}, K.~G., {et~al.} 2017, \aap, 603, A10

\bibitem[{{Bodensteiner} {et~al.}(2020){Bodensteiner}, {Sana}, {Mahy}, {Patrick}, {de Koter}, {de Mink}, {Evans}, {G{\"o}tberg}, {Langer}, {Lennon}, {Schneider}, \& {Tramper}}]{Bodensteiner+2020}
{Bodensteiner}, J., {Sana}, H., {Mahy}, L., {et~al.} 2020, \aap, 634, A51

\bibitem[{{Boley} {et~al.}(2016){Boley}, {Kraus}, {de Wit}, {Linz}, {van Boekel}, {Henning}, {Lacour}, {Monnier}, {Stecklum}, \& {Tuthill}}]{Boley+2016}
{Boley}, P.~A., {Kraus}, S., {de Wit}, W.-J., {et~al.} 2016, \aap, 586, A78

\bibitem[{{Bonnell} \& {Bate}(1994)}]{Bonnell+1994}
{Bonnell}, I.~A. \& {Bate}, M.~R. 1994, \mnras, 271, 999

\bibitem[{{Bordier} {et~al.}(2022){Bordier}, {Frost}, {Sana}, {Reggiani}, {M{\'e}rand}, {Rainot}, {Ram{\'\i}rez-Tannus}, \& {de Wit}}]{Bordier+2022}
{Bordier}, E., {Frost}, A.~J., {Sana}, H., {et~al.} 2022, \aap, 663, A26

\bibitem[{Bradley {et~al.}(2022)Bradley, Sipőcz, Robitaille, Tollerud, Vinícius, Deil, Barbary, Wilson, Busko, Donath, Günther, Cara, Lim, Meßlinger, Conseil, Bostroem, Droettboom, Bray, Bratholm, Barentsen, Craig, Ginsburg, Rathi, Pascual, Perren, Georgiev, de~Val-Borro, Kerzendorf, Bach, \& Quint}]{Bradley+2022}
Bradley, L., Sipőcz, B., Robitaille, T., {et~al.} 2022, astropy/photutils: 1.6.0

\bibitem[{{Caratti o Garatti} {et~al.}(2015){Caratti o Garatti}, {Stecklum}, {Linz}, {Garcia Lopez}, \& {Sanna}}]{Caratti+2015a}
{Caratti o Garatti}, A., {Stecklum}, B., {Linz}, H., {Garcia Lopez}, R., \& {Sanna}, A. 2015, \aap, 573, A82

\bibitem[{{Caratti o Garatti} {et~al.}(2016){Caratti o Garatti}, {Stecklum}, {Weigelt}, {Schertl}, {Hofmann}, {Kraus}, {Oudmaijer}, {de Wit}, {Sanna}, {Garcia Lopez}, {Kreplin}, \& {Ray}}]{Caratti+2016}
{Caratti o Garatti}, A., {Stecklum}, B., {Weigelt}, G., {et~al.} 2016, \aap, 589, L4

\bibitem[{{Cesaroni} {et~al.}(2017){Cesaroni}, {S{\'a}nchez-Monge}, {Beltr{\'a}n}, {Johnston}, {Maud}, {Moscadelli}, {Mottram}, {Ahmadi}, {Allen}, {Beuther}, {Csengeri}, {Etoka}, {Fuller}, {Galli}, {Galv{\'a}n-Madrid}, {Goddi}, {Henning}, {Hoare}, {Klaassen}, {Kuiper}, {Kumar}, {Lumsden}, {Peters}, {Rivilla}, {Schilke}, {Testi}, {van der Tak}, {Vig}, {Walmsley}, \& {Zinnecker}}]{Cesaroni+2017}
{Cesaroni}, R., {S{\'a}nchez-Monge}, {\'A}., {Beltr{\'a}n}, M.~T., {et~al.} 2017, \aap, 602, A59

\bibitem[{{Connelley} {et~al.}(2008){Connelley}, {Reipurth}, \& {Tokunaga}}]{Connelley+2008}
{Connelley}, M.~S., {Reipurth}, B., \& {Tokunaga}, A.~T. 2008, \aj, 135, 2496

\bibitem[{{Cooper} {et~al.}(2013){Cooper}, {Lumsden}, {Oudmaijer}, {Hoare}, {Clarke}, {Urquhart}, {Mottram}, {Moore}, \& {Davies}}]{Cooper+2013}
{Cooper}, H.~D.~B., {Lumsden}, S.~L., {Oudmaijer}, R.~D., {et~al.} 2013, \mnras, 430, 1125

\bibitem[{{Cunningham} {et~al.}(2018){Cunningham}, {Lumsden}, {Moore}, {Maud}, \& {Mendigut{\'\i}a}}]{Cunningham+2018}
{Cunningham}, N., {Lumsden}, S.~L., {Moore}, T.~J.~T., {Maud}, L.~T., \& {Mendigut{\'\i}a}, I. 2018, \mnras, 477, 2455

\bibitem[{{Dahm}(2008)}]{Dahm+2008}
{Dahm}, S.~E. 2008, in Handbook of Star Forming Regions, Volume I, ed. B.~{Reipurth}, Vol.~4, 966

\bibitem[{{De Buizer} {et~al.}(2017){De Buizer}, {Liu}, {Tan}, {Zhang}, {Beltr{\'a}n}, {Shuping}, {Staff}, {Tanaka}, \& {Whitney}}]{DeBuizer+2017}
{De Buizer}, J.~M., {Liu}, M., {Tan}, J.~C., {et~al.} 2017, \apj, 843, 33

\bibitem[{{de Wit} {et~al.}(2009){de Wit}, {Hoare}, {Fujiyoshi}, {Oudmaijer}, {Honda}, {Kataza}, {Miyata}, {Okamoto}, {Onaka}, {Sako}, \& {Yamashita}}]{deWit+2009}
{de Wit}, W.~J., {Hoare}, M.~G., {Fujiyoshi}, T., {et~al.} 2009, \aap, 494, 157

\bibitem[{{Duch{\^e}ne} {et~al.}(2007){Duch{\^e}ne}, {Bontemps}, {Bouvier}, {Andr{\'e}}, {Djupvik}, \& {Ghez}}]{Duchene+2007}
{Duch{\^e}ne}, G., {Bontemps}, S., {Bouvier}, J., {et~al.} 2007, \aap, 476, 229

\bibitem[{{Duch{\^e}ne} \& {Kraus}(2013)}]{Duchene+2013}
{Duch{\^e}ne}, G. \& {Kraus}, A. 2013, \araa, 51, 269

\bibitem[{{Duch{\^e}ne} {et~al.}(2018){Duch{\^e}ne}, {Lacour}, {Moraux}, {Goodwin}, \& {Bouvier}}]{Duchene+2018}
{Duch{\^e}ne}, G., {Lacour}, S., {Moraux}, E., {Goodwin}, S., \& {Bouvier}, J. 2018, \mnras, 478, 1825

\bibitem[{{Dunstall} {et~al.}(2015){Dunstall}, {Dufton}, {Sana}, {Evans}, {Howarth}, {Sim{\'o}n-D{\'\i}az}, {de Mink}, {Langer}, {Ma{\'\i}z Apell{\'a}niz}, \& {Taylor}}]{Dunstall+2015}
{Dunstall}, P.~R., {Dufton}, P.~L., {Sana}, H., {et~al.} 2015, \aap, 580, A93

\bibitem[{{Ellerbroek} {et~al.}(2013){Ellerbroek}, {Bik}, {Kaper}, {Maaskant}, {Paalvast}, {Tramper}, {Sana}, {Waters}, \& {Balog}}]{Ellerbroek+2013}
{Ellerbroek}, L.~E., {Bik}, A., {Kaper}, L., {et~al.} 2013, \aap, 558, A102

\bibitem[{{Elliott} \& {Bayo}(2016)}]{ElliottBayo+2016}
{Elliott}, P. \& {Bayo}, A. 2016, \mnras, 459, 4499

\bibitem[{{Elliott} {et~al.}(2015){Elliott}, {Hu{\'e}lamo}, {Bouy}, {Bayo}, {Melo}, {Torres}, {Sterzik}, {Quast}, {Chauvin}, \& {Barrado}}]{Elliott+2015}
{Elliott}, P., {Hu{\'e}lamo}, N., {Bouy}, H., {et~al.} 2015, \aap, 580, A88

\bibitem[{{Flaccomio} {et~al.}(2023){Flaccomio}, {Micela}, {Peres}, {Sciortino}, {Salvaggio}, {Prisinzano}, {Guarcello}, {Venuti}, {Bonito}, \& {Pillitteri}}]{Flaccomio+2023}
{Flaccomio}, E., {Micela}, G., {Peres}, G., {et~al.} 2023, \aap, 670, A37

\bibitem[{{Frost} {et~al.}(2019){Frost}, {Oudmaijer}, {de Wit}, \& {Lumsden}}]{Frost+2019}
{Frost}, A.~J., {Oudmaijer}, R.~D., {de Wit}, W.~J., \& {Lumsden}, S.~L. 2019, \aap, 625, A44

\bibitem[{{Frost} {et~al.}(2021){Frost}, {Oudmaijer}, {de Wit}, \& {Lumsden}}]{Frost+2021a}
{Frost}, A.~J., {Oudmaijer}, R.~D., {de Wit}, W.~J., \& {Lumsden}, S.~L. 2021, \aap, 648, A62

\bibitem[{{Fujii} \& {Portegies Zwart}(2011)}]{Fujii+2011}
{Fujii}, M.~S. \& {Portegies Zwart}, S. 2011, Science, 334, 1380

\bibitem[{{Giannini} {et~al.}(2005){Giannini}, {Massi}, {Podio}, {Lorenzetti}, {Nisini}, {Caratti o Garatti}, {Liseau}, {Lo Curto}, \& {Vitali}}]{Giannini+2005}
{Giannini}, T., {Massi}, F., {Podio}, L., {et~al.} 2005, \aap, 433, 941

\bibitem[{{Ginsburg} {et~al.}(2018){Ginsburg}, {Bally}, {Goddi}, {Plambeck}, \& {Wright}}]{Ginsburg+2018}
{Ginsburg}, A., {Bally}, J., {Goddi}, C., {Plambeck}, R., \& {Wright}, M. 2018, \apj, 860, 119

\bibitem[{{Goodwin} {et~al.}(2004){Goodwin}, {Whitworth}, \& {Ward-Thompson}}]{Goodwin+2004}
{Goodwin}, S.~P., {Whitworth}, A.~P., \& {Ward-Thompson}, D. 2004, \aap, 414, 633

\bibitem[{{Grave} \& {Kumar}(2009)}]{Grave+2009}
{Grave}, J.~M.~C. \& {Kumar}, M.~S.~N. 2009, \aap, 498, 147

\bibitem[{{Gravity Collaboration} {et~al.}(2018){Gravity Collaboration}, {Karl}, {Pfuhl}, {Eisenhauer}, {Genzel}, {Grellmann}, {Habibi}, {Abuter}, {Accardo}, {Amorim}, {Anugu}, {{\'A}vila}, {Benisty}, {Berger}, {Blind}, {Bonnet}, {Bourget}, {Brandner}, {Brast}, {Buron}, {Caratti O Garatti}, {Chapron}, {Cl{\'e}net}, {Collin}, {Coud{\'e} Du Foresto}, {de Wit}, {de Zeeuw}, {Deen}, {Delplancke-Str{\"o}bele}, {Dembet}, {Derie}, {Dexter}, {Duvert}, {Ebert}, {Eckart}, {Esselborn}, {F{\'e}dou}, {Finger}, {Garcia}, {Garcia Dabo}, {Garcia Lopez}, {Gao}, {Gendron}, {Gillessen}, {Gont{\'e}}, {Gordo}, {Gr{\"o}zinger}, {Guajardo}, {Guieu}, {Haguenauer}, {Hans}, {Haubois}, {Haug}, {Hau{\ss}mann}, {Henning}, {Hippler}, {Horrobin}, {Huber}, {Hubert}, {Hubin}, {Jakob}, {Jochum}, {Jocou}, {Kaufer}, {Kellner}, {Kendrew}, {Kern}, {Kervella}, {Kiekebusch}, {Klein}, {K{\"o}hler}, {Kolb}, {Kulas}, {Lacour}, {Lapeyr{\`e}re}, {Lazareff}, {Le Bouquin}, {L{\'e}na}, {Lenzen}, {L{\'e}v{\^e}que}, {Lin}, {Lippa}, {Magnard}, {Mehrgan},
  {M{\'e}rand}, {Moulin}, {M{\"u}ller}, {M{\"u}ller}, {Neumann}, {Oberti}, {Ott}, {Pallanca}, {Pand uro}, {Pasquini}, {Paumard}, {Percheron}, {Perraut}, {Perrin}, {Pfl{\"u}ger}, {Duc}, {Plewa}, {Popovic}, {Rabien}, {Ram{\'\i}rez}, {Ramos}, {Rau}, {Riquelme}, {Rodr{\'\i}guez-Coira}, {Rohloff}, {Rosales}, {Rousset}, {Sanchez-Bermudez}, {Scheithauer}, {Sch{\"o}ller}, {Schuhler}, {Spyromilio}, {Straub}, {Straubmeier}, {Sturm}, {Suarez}, {Tristram}, {Ventura}, {Vincent}, {Waisberg}, {Wank}, {Widmann}, {Wieprecht}, {Wiest}, {Wiezorrek}, {Wittkowski}, {Woillez}, {Wolff}, {Yazici}, {Ziegler}, \& {Zins}}]{GravityCollab+2018}
{Gravity Collaboration}, {Karl}, M., {Pfuhl}, O., {et~al.} 2018, \aap, 620, A116

\bibitem[{{Green} {et~al.}(2012){Green}, {Caswell}, {Fuller}, {Avison}, {Breen}, {Ellingsen}, {Gray}, {Pestalozzi}, {Quinn}, {Thompson}, \& {Voronkov}}]{Green+2012}
{Green}, J.~A., {Caswell}, J.~L., {Fuller}, G.~A., {et~al.} 2012, \mnras, 420, 3108

\bibitem[{{Green} \& {McClure-Griffiths}(2011)}]{Green+2011}
{Green}, J.~A. \& {McClure-Griffiths}, N.~M. 2011, \mnras, 417, 2500

\bibitem[{{Grellmann} {et~al.}(2011){Grellmann}, {Ratzka}, {Kraus}, {Linz}, {Preibisch}, \& {Weigelt}}]{Grellmann+2011}
{Grellmann}, R., {Ratzka}, T., {Kraus}, S., {et~al.} 2011, \aap, 532, A109

\bibitem[{{Hwang} {et~al.}(2022){Hwang}, {Ting}, \& {Zakamska}}]{Hwang+2022}
{Hwang}, H.-C., {Ting}, Y.-S., \& {Zakamska}, N.~L. 2022, \mnras, 512, 3383

\bibitem[{{Ilee} {et~al.}(2018){Ilee}, {Cyganowski}, {Brogan}, {Hunter}, {Forgan}, {Haworth}, {Clarke}, \& {Harries}}]{Ilee+2018}
{Ilee}, J.~D., {Cyganowski}, C.~J., {Brogan}, C.~L., {et~al.} 2018, \apjl, 869, L24

\bibitem[{{Ilee} {et~al.}(2013){Ilee}, {Wheelwright}, {Oudmaijer}, {de Wit}, {Maud}, {Hoare}, {Lumsden}, {Moore}, {Urquhart}, \& {Mottram}}]{Ilee+2013}
{Ilee}, J.~D., {Wheelwright}, H.~E., {Oudmaijer}, R.~D., {et~al.} 2013, \mnras, 429, 2960

\bibitem[{{Johnston} {et~al.}(2020){Johnston}, {Hoare}, {Beuther}, {Kuiper}, {Kee}, {Linz}, {Boley}, {Maud}, {Ahmadi}, \& {Robitaille}}]{Johnston+2020}
{Johnston}, K.~G., {Hoare}, M.~G., {Beuther}, H., {et~al.} 2020, \aap, 634, L11

\bibitem[{{Johnston} {et~al.}(2015){Johnston}, {Robitaille}, {Beuther}, {Linz}, {Boley}, {Kuiper}, {Keto}, {Hoare}, \& {van Boekel}}]{Johnston+2015}
{Johnston}, K.~G., {Robitaille}, T.~P., {Beuther}, H., {et~al.} 2015, \apjl, 813, L19

\bibitem[{{Kamezaki} {et~al.}(2014){Kamezaki}, {Imura}, {Omodaka}, {Handa}, {Tsuboi}, {Nagayama}, {Hirota}, {Sunada}, {Kobayashi}, {Chibueze}, {Kawai}, \& {Nakano}}]{Kamezaki+2014}
{Kamezaki}, T., {Imura}, K., {Omodaka}, T., {et~al.} 2014, \apjs, 211, 18

\bibitem[{{Kawamura} {et~al.}(1998){Kawamura}, {Onishi}, {Yonekura}, {Dobashi}, {Mizuno}, {Ogawa}, \& {Fukui}}]{Kawamura+1998}
{Kawamura}, A., {Onishi}, T., {Yonekura}, Y., {et~al.} 1998, \apjs, 117, 387

\bibitem[{{Koumpia} {et~al.}(2019){Koumpia}, {Ababakr}, {de Wit}, {Oudmaijer}, {Caratti o Garatti}, {Boley}, {Linz}, {Kraus}, {Vink}, \& {Le Bouquin}}]{Koumpia+2019}
{Koumpia}, E., {Ababakr}, K.~M., {de Wit}, W.~J., {et~al.} 2019, \aap, 623, L5

\bibitem[{{Koumpia} {et~al.}(2021){Koumpia}, {de Wit}, {Oudmaijer}, {Frost}, {Lumsden}, {Caratti o Garatti}, {Goodwin}, {Stecklum}, {Mendigut{\'\i}a}, {Ilee}, \& {Vioque}}]{Koumpia+2021}
{Koumpia}, E., {de Wit}, W.~J., {Oudmaijer}, R.~D., {et~al.} 2021, \aap, 654, A109

\bibitem[{{Kounkel} {et~al.}(2019){Kounkel}, {Covey}, {Moe}, {Kratter}, {Su{\'a}rez}, {Stassun}, {Rom{\'a}n-Z{\'u}{\~n}iga}, {Hernandez}, {Kim}, {Pe{\~n}a Ram{\'\i}rez}, {Roman-Lopes}, {Stringfellow}, {Jaehnig}, {Borissova}, {Tofflemire}, {Krolikowski}, {Rizzuto}, {Kraus}, {Badenes}, {Longa-Pe{\~n}a}, {G{\'o}mez Maqueo Chew}, {Barba}, {Nidever}, {Brown}, {De Lee}, {Pan}, {Bizyaev}, {Oravetz}, \& {Oravetz}}]{Kounkel+2019}
{Kounkel}, M., {Covey}, K., {Moe}, M., {et~al.} 2019, \aj, 157, 196

\bibitem[{{Kozai}(1962)}]{Kozai1962}
{Kozai}, Y. 1962, \aj, 67, 591

\bibitem[{{Kratter} \& {Matzner}(2006)}]{Kratter+2006}
{Kratter}, K.~M. \& {Matzner}, C.~D. 2006, \mnras, 373, 1563

\bibitem[{{Kratter} {et~al.}(2008){Kratter}, {Matzner}, \& {Krumholz}}]{Kratter+2008}
{Kratter}, K.~M., {Matzner}, C.~D., \& {Krumholz}, M.~R. 2008, \apj, 681, 375

\bibitem[{{Kratter} {et~al.}(2010){Kratter}, {Matzner}, {Krumholz}, \& {Klein}}]{Kratter2010}
{Kratter}, K.~M., {Matzner}, C.~D., {Krumholz}, M.~R., \& {Klein}, R.~I. 2010, \apj, 708, 1585

\bibitem[{{Kraus} {et~al.}(2011){Kraus}, {Ireland}, {Martinache}, \& {Hillenbrand}}]{Kraus+2011}
{Kraus}, A.~L., {Ireland}, M.~J., {Martinache}, F., \& {Hillenbrand}, L.~A. 2011, \apj, 731, 8

\bibitem[{{Kraus} {et~al.}(2010){Kraus}, {Hofmann}, {Menten}, {Schertl}, {Weigelt}, {Wyrowski}, {Meilland}, {Perraut}, {Petrov}, {Robbe-Dubois}, {Schilke}, \& {Testi}}]{Kraus+2010}
{Kraus}, S., {Hofmann}, K.-H., {Menten}, K.~M., {et~al.} 2010, \nat, 466, 339

\bibitem[{{Kraus} {et~al.}(2017){Kraus}, {Kluska}, {Kreplin}, {Bate}, {Harries}, {Hofmann}, {Hone}, {Monnier}, {Weigelt}, {Anugu}, {de Wit}, \& {Wittkowski}}]{Kraus+2017}
{Kraus}, S., {Kluska}, J., {Kreplin}, A., {et~al.} 2017, \apjl, 835, L5

\bibitem[{{Krishnan} {et~al.}(2017){Krishnan}, {Ellingsen}, {Reid}, {Bignall}, {McCallum}, {Phillips}, {Reynolds}, \& {Stevens}}]{Krishnan+2017}
{Krishnan}, V., {Ellingsen}, S.~P., {Reid}, M.~J., {et~al.} 2017, \mnras, 465, 1095

\bibitem[{{Krumholz} \& {Bonnell}(2007)}]{Krumholz+2007}
{Krumholz}, M.~R. \& {Bonnell}, I.~A. 2007, arXiv e-prints, arXiv:0712.0828

\bibitem[{{Krumholz} {et~al.}(2009){Krumholz}, {Klein}, {McKee}, {Offner}, \& {Cunningham}}]{Krumholz+2009}
{Krumholz}, M.~R., {Klein}, R.~I., {McKee}, C.~F., {Offner}, S. S.~R., \& {Cunningham}, A.~J. 2009, Science, 323, 754

\bibitem[{{Kuiper} {et~al.}(2010){Kuiper}, {Klahr}, {Beuther}, \& {Henning}}]{Kuiper+2010}
{Kuiper}, R., {Klahr}, H., {Beuther}, H., \& {Henning}, T. 2010, \apj, 722, 1556

\bibitem[{{Lee} {et~al.}(2019){Lee}, {Offner}, {Kratter}, {Smullen}, \& {Li}}]{Lee+2019}
{Lee}, A.~T., {Offner}, S. S.~R., {Kratter}, K.~M., {Smullen}, R.~A., \& {Li}, P.~S. 2019, \apj, 887, 232

\bibitem[{{Lee} \& {Hennebelle}(2018)}]{LeeHennebelle2018}
{Lee}, Y.-N. \& {Hennebelle}, P. 2018, \aap, 611, A89

\bibitem[{{Lenzen} {et~al.}(2003){Lenzen}, {Hartung}, {Brandner}, {Finger}, {Hubin}, {Lacombe}, {Lagrange}, {Lehnert}, {Moorwood}, \& {Mouillet}}]{Lenzen+2003}
{Lenzen}, R., {Hartung}, M., {Brandner}, W., {et~al.} 2003, in Society of Photo-Optical Instrumentation Engineers (SPIE) Conference Series, Vol. 4841, Instrument Design and Performance for Optical/Infrared Ground-based Telescopes, ed. M.~{Iye} \& A.~F.~M. {Moorwood}, 944--952

\bibitem[{{Lidov}(1962)}]{Lidov1962}
{Lidov}, M.~L. 1962, \planss, 9, 719

\bibitem[{{Liseau} {et~al.}(1992){Liseau}, {Lorenzetti}, {Nisini}, {Spinoglio}, \& {Moneti}}]{Liseau+1992}
{Liseau}, R., {Lorenzetti}, D., {Nisini}, B., {Spinoglio}, L., \& {Moneti}, A. 1992, \aap, 265, 577

\bibitem[{{Liu} {et~al.}(2019){Liu}, {Tan}, {De Buizer}, {Zhang}, {Beltr{\'a}n}, {Staff}, {Tanaka}, {Whitney}, \& {Rosero}}]{Liu+2019}
{Liu}, M., {Tan}, J.~C., {De Buizer}, J.~M., {et~al.} 2019, \apj, 874, 16

\bibitem[{{Lumsden} {et~al.}(2002){Lumsden}, {Hoare}, {Oudmaijer}, \& {Richards}}]{lums2002}
{Lumsden}, S.~L., {Hoare}, M.~G., {Oudmaijer}, R.~D., \& {Richards}, D. 2002, \mnras, 336, 621

\bibitem[{{Lumsden} {et~al.}(2013){Lumsden}, {Hoare}, {Urquhart}, {Oudmaijer}, {Davies}, {Mottram}, {Cooper}, \& {Moore}}]{Lumsden+2013}
{Lumsden}, S.~L., {Hoare}, M.~G., {Urquhart}, J.~S., {et~al.} 2013, \apjs, 208, 11

\bibitem[{{Ma{\'\i}z Apell{\'a}niz} {et~al.}(2017){Ma{\'\i}z Apell{\'a}niz}, {Sana}, {Barb{\'a}}, {Le Bouquin}, \& {Gamen}}]{MaizApellaniz+2017}
{Ma{\'\i}z Apell{\'a}niz}, J., {Sana}, H., {Barb{\'a}}, R.~H., {Le Bouquin}, J.~B., \& {Gamen}, R.~C. 2017, \mnras, 464, 3561

\bibitem[{{Maud} {et~al.}(2019){Maud}, {Cesaroni}, {Kumar}, {Rivilla}, {Ginsburg}, {Klaassen}, {Harsono}, {S{\'a}nchez-Monge}, {Ahmadi}, {Allen}, {Beltr{\'a}n}, {Beuther}, {Galv{\'a}n-Madrid}, {Goddi}, {Hoare}, {Hogerheijde}, {Johnston}, {Kuiper}, {Moscadelli}, {Peters}, {Testi}, {van der Tak}, \& {de Wit}}]{Maud+2019}
{Maud}, L.~T., {Cesaroni}, R., {Kumar}, M.~S.~N., {et~al.} 2019, \aap, 627, L6

\bibitem[{{Meyer} {et~al.}(2018){Meyer}, {Kuiper}, {Kley}, {Johnston}, \& {Vorobyov}}]{Meyer+2018}
{Meyer}, D.~M.~A., {Kuiper}, R., {Kley}, W., {Johnston}, K.~G., \& {Vorobyov}, E. 2018, \mnras, 473, 3615

\bibitem[{{Michaely} \& {Perets}(2014)}]{Michaely+2014}
{Michaely}, E. \& {Perets}, H.~B. 2014, \apj, 794, 122

\bibitem[{{Mignon-Risse} {et~al.}(2021){Mignon-Risse}, {Gonz{\'a}lez}, {Commer{\c{c}}on}, \& {Rosdahl}}]{Mignon-Risse+2021a}
{Mignon-Risse}, R., {Gonz{\'a}lez}, M., {Commer{\c{c}}on}, B., \& {Rosdahl}, J. 2021, \aap, 652, A69

\bibitem[{{Mignon-Risse} {et~al.}(2023){Mignon-Risse}, {Oliva}, {Gonz{\'a}lez}, {Kuiper}, \& {Commer{\c{c}}on}}]{Mignon-Risse+2023}
{Mignon-Risse}, R., {Oliva}, A., {Gonz{\'a}lez}, M., {Kuiper}, R., \& {Commer{\c{c}}on}, B. 2023, \aap, 672, A88

\bibitem[{{Moe} \& {Di Stefano}(2017)}]{Moe+2017}
{Moe}, M. \& {Di Stefano}, R. 2017, \apjs, 230, 15

\bibitem[{{Moe} \& {Kratter}(2018)}]{Moe+2018}
{Moe}, M. \& {Kratter}, K.~M. 2018, \apj, 854, 44

\bibitem[{{Mottram} {et~al.}(2011){Mottram}, {Hoare}, {Davies}, {Lumsden}, {Oudmaijer}, {Urquhart}, {Moore}, {Cooper}, \& {Stead}}]{Mottram+2011}
{Mottram}, J.~C., {Hoare}, M.~G., {Davies}, B., {et~al.} 2011, \apjl, 730, L33

\bibitem[{{Mottram} {et~al.}(2007){Mottram}, {Hoare}, {Lumsden}, {Oudmaijer}, {Urquhart}, {Sheret}, {Clarke}, \& {Allsopp}}]{Mottram+2007}
{Mottram}, J.~C., {Hoare}, M.~G., {Lumsden}, S.~L., {et~al.} 2007, \aap, 476, 1019

\bibitem[{{Mu{\~n}oz} {et~al.}(2019){Mu{\~n}oz}, {Miranda}, \& {Lai}}]{Munoz+2019}
{Mu{\~n}oz}, D.~J., {Miranda}, R., \& {Lai}, D. 2019, \apj, 871, 84

\bibitem[{{Murillo} {et~al.}(2016){Murillo}, {van Dishoeck}, {Tobin}, \& {Fedele}}]{Murillo+2016}
{Murillo}, N.~M., {van Dishoeck}, E.~F., {Tobin}, J.~J., \& {Fedele}, D. 2016, \aap, 592, A56

\bibitem[{{Myers} {et~al.}(2013){Myers}, {McKee}, {Cunningham}, {Klein}, \& {Krumholz}}]{Myers+2013}
{Myers}, A.~T., {McKee}, C.~F., {Cunningham}, A.~J., {Klein}, R.~I., \& {Krumholz}, M.~R. 2013, \apj, 766, 97

\bibitem[{{Navarete} {et~al.}(2015){Navarete}, {Damineli}, {Barbosa}, \& {Blum}}]{Navarete+2015}
{Navarete}, F., {Damineli}, A., {Barbosa}, C.~L., \& {Blum}, R.~D. 2015, \mnras, 450, 4364

\bibitem[{{Netterfield} {et~al.}(2009){Netterfield}, {Ade}, {Bock}, {Chapin}, {Devlin}, {Griffin}, {Gundersen}, {Halpern}, {Hargrave}, {Hughes}, {Klein}, {Marsden}, {Martin}, {Mauskopf}, {Olmi}, {Pascale}, {Patanchon}, {Rex}, {Roy}, {Scott}, {Semisch}, {Thomas}, {Truch}, {Tucker}, {Tucker}, {Viero}, \& {Wiebe}}]{Netterfield+2009}
{Netterfield}, C.~B., {Ade}, P. A.~R., {Bock}, J.~J., {et~al.} 2009, \apj, 707, 1824

\bibitem[{{Offner} {et~al.}(2016){Offner}, {Dunham}, {Lee}, {Arce}, \& {Fielding}}]{Offner+2016}
{Offner}, S. S.~R., {Dunham}, M.~M., {Lee}, K.~I., {Arce}, H.~G., \& {Fielding}, D.~B. 2016, \apjl, 827, L11

\bibitem[{{Offner} {et~al.}(2010){Offner}, {Kratter}, {Matzner}, {Krumholz}, \& {Klein}}]{Offner+2010}
{Offner}, S. S.~R., {Kratter}, K.~M., {Matzner}, C.~D., {Krumholz}, M.~R., \& {Klein}, R.~I. 2010, \apj, 725, 1485

\bibitem[{{Offner} {et~al.}(2022){Offner}, {Moe}, {Kratter}, {Sadavoy}, {Jensen}, \& {Tobin}}]{Offner+2022}
{Offner}, S. S.~R., {Moe}, M., {Kratter}, K.~M., {et~al.} 2022, arXiv e-prints, arXiv:2203.10066

\bibitem[{{Oliva} \& {Kuiper}(2020)}]{Oliva+2020}
{Oliva}, G.~A. \& {Kuiper}, R. 2020, \aap, 644, A41

\bibitem[{{Parker} \& {Meyer}(2014)}]{Parker+2014}
{Parker}, R.~J. \& {Meyer}, M.~R. 2014, \mnras, 442, 3722

\bibitem[{{Parker} \& {Schoettler}(2022)}]{Parker+2022}
{Parker}, R.~J. \& {Schoettler}, C. 2022, \mnras, 510, 1136

\bibitem[{{Pauwels} {et~al.}(2023){Pauwels}, {Reggiani}, {Sana}, {Rainot}, \& {Kratter}}]{Pauwels+2023}
{Pauwels}, T., {Reggiani}, M., {Sana}, H., {Rainot}, A., \& {Kratter}, K. 2023, \aap, 678, A172

\bibitem[{{Pomohaci} {et~al.}(2019){Pomohaci}, {Oudmaijer}, \& {Goodwin}}]{Pomohaci+2019}
{Pomohaci}, R., {Oudmaijer}, R.~D., \& {Goodwin}, S.~P. 2019, \mnras, 484, 226

\bibitem[{{Raghavan} {et~al.}(2010){Raghavan}, {McAlister}, {Henry}, {Latham}, {Marcy}, {Mason}, {Gies}, {White}, \& {ten Brummelaar}}]{Raghavan+2010}
{Raghavan}, D., {McAlister}, H.~A., {Henry}, T.~J., {et~al.} 2010, \apjs, 190, 1

\bibitem[{{Rainot} {et~al.}(2022){Rainot}, {Reggiani}, {Sana}, {Bodensteiner}, \& {Absil}}]{Rainot+2022}
{Rainot}, A., {Reggiani}, M., {Sana}, H., {Bodensteiner}, J., \& {Absil}, O. 2022, \aap, 658, A198

\bibitem[{{Rainot} {et~al.}(2020){Rainot}, {Reggiani}, {Sana}, {Bodensteiner}, {Gomez-Gonzalez}, {Absil}, {Christiaens}, {Delorme}, {Almeida}, {Caballero-Nieves}, {De Ridder}, {Kratter}, {Lacour}, {Le Bouquin}, {Pueyo}, \& {Zinnecker}}]{Rainot+2020}
{Rainot}, A., {Reggiani}, M., {Sana}, H., {et~al.} 2020, \aap, 640, A15

\bibitem[{{Ram{\'\i}rez-Tannus} {et~al.}(2021){Ram{\'\i}rez-Tannus}, {Backs}, {de Koter}, {Sana}, {Beuther}, {Bik}, {Brandner}, {Kaper}, {Linz}, {Henning}, \& {Poorta}}]{Ramirez+2021}
{Ram{\'\i}rez-Tannus}, M.~C., {Backs}, F., {de Koter}, A., {et~al.} 2021, arXiv e-prints, arXiv:2101.01604

\bibitem[{{Reggiani} {et~al.}(2022){Reggiani}, {Rainot}, {Sana}, {Almeida}, {Caballero-Nieves}, {Kratter}, {Lacour}, {Le Bouquin}, \& {Zinnecker}}]{Reggiani+2022}
{Reggiani}, M., {Rainot}, A., {Sana}, H., {et~al.} 2022, \aap, 660, A122

\bibitem[{{Reid} {et~al.}(2009){Reid}, {Menten}, {Zheng}, {Brunthaler}, {Moscadelli}, {Xu}, {Zhang}, {Sato}, {Honma}, {Hirota}, {Hachisuka}, {Choi}, {Moellenbrock}, \& {Bartkiewicz}}]{Reid+2009}
{Reid}, M.~J., {Menten}, K.~M., {Zheng}, X.~W., {et~al.} 2009, \apj, 700, 137

\bibitem[{{Robin} {et~al.}(2003){Robin}, {Reyl{\'e}}, {Derri{\`e}re}, \& {Picaud}}]{Robin+2003}
{Robin}, A.~C., {Reyl{\'e}}, C., {Derri{\`e}re}, S., \& {Picaud}, S. 2003, \aap, 409, 523

\bibitem[{{Rousset} {et~al.}(2003){Rousset}, {Lacombe}, {Puget}, {Hubin}, {Gendron}, {Fusco}, {Arsenault}, {Charton}, {Feautrier}, {Gigan}, {Kern}, {Lagrange}, {Madec}, {Mouillet}, {Rabaud}, {Rabou}, {Stadler}, \& {Zins}}]{Rousset+2003}
{Rousset}, G., {Lacombe}, F., {Puget}, P., {et~al.} 2003, in Society of Photo-Optical Instrumentation Engineers (SPIE) Conference Series, Vol. 4839, Adaptive Optical System Technologies II, ed. P.~L. {Wizinowich} \& D.~{Bonaccini}, 140--149

\bibitem[{{Sana} {et~al.}(2013){Sana}, {de Koter}, {de Mink}, {Dunstall}, {Evans}, {H{\'e}nault-Brunet}, {Ma{\'\i}z Apell{\'a}niz}, {Ram{\'\i}rez-Agudelo}, {Taylor}, {Walborn}, {Clark}, {Crowther}, {Herrero}, {Gieles}, {Langer}, {Lennon}, \& {Vink}}]{Sana+2013}
{Sana}, H., {de Koter}, A., {de Mink}, S.~E., {et~al.} 2013, \aap, 550, A107

\bibitem[{{Sana} {et~al.}(2012){Sana}, {de Mink}, {de Koter}, {Langer}, {Evans}, {Gieles}, {Gosset}, {Izzard}, {Le Bouquin}, \& {Schneider}}]{Sana+2012}
{Sana}, H., {de Mink}, S.~E., {de Koter}, A., {et~al.} 2012, Science, 337, 444

\bibitem[{{Sana} \& {Evans}(2011)}]{Sana+2011}
{Sana}, H. \& {Evans}, C.~J. 2011, in Active OB Stars: Structure, Evolution, Mass Loss, and Critical Limits, ed. C.~{Neiner}, G.~{Wade}, G.~{Meynet}, \& G.~{Peters}, Vol. 272, 474--485

\bibitem[{{Sana} {et~al.}(2014){Sana}, {Le Bouquin}, {Lacour}, {Berger}, {Duvert}, {Gauchet}, {Norris}, {Olofsson}, {Pickel}, {Zins}, {Absil}, {de Koter}, {Kratter}, {Schnurr}, \& {Zinnecker}}]{Sana+2014}
{Sana}, H., {Le Bouquin}, J.~B., {Lacour}, S., {et~al.} 2014, \apjs, 215, 15

\bibitem[{{Sana} {et~al.}(2017){Sana}, {Ram{\'\i}rez-Tannus}, {de Koter}, {Kaper}, {Tramper}, \& {Bik}}]{Sana+2017}
{Sana}, H., {Ram{\'\i}rez-Tannus}, M.~C., {de Koter}, A., {et~al.} 2017, \aap, 599, L9

\bibitem[{{Schreyer} {et~al.}(2003){Schreyer}, {Stecklum}, {Linz}, \& {Henning}}]{Schreyer+2003}
{Schreyer}, K., {Stecklum}, B., {Linz}, H., \& {Henning}, T. 2003, \apj, 599, 335

\bibitem[{{Shenton} {et~al.}(2023){Shenton}, J., \& D.}]{Shenton+2023}
{Shenton}, R.~G., J., H.~R., \& D., O.~R. 2023, Searching for binarity with the largest sample of known MYSOs, submitted for publication to \emph{MNRAS} June 2023

\bibitem[{{Thompson} {et~al.}(1998){Thompson}, {Corbin}, {Young}, \& {Schneider}}]{Thompson+1998}
{Thompson}, R.~I., {Corbin}, M.~R., {Young}, E., \& {Schneider}, G. 1998, \apjl, 492, L177

\bibitem[{{Tobin} {et~al.}(2016){Tobin}, {Looney}, {Li}, {Chandler}, {Dunham}, {Segura-Cox}, {Sadavoy}, {Melis}, {Harris}, {Kratter}, \& {Perez}}]{Tobin+2016}
{Tobin}, J.~J., {Looney}, L.~W., {Li}, Z.-Y., {et~al.} 2016, \apj, 818, 73

\bibitem[{{Tobin} {et~al.}(2018){Tobin}, {Looney}, {Li}, {Sadavoy}, {Dunham}, {Segura-Cox}, {Kratter}, {Chandler}, {Melis}, {Harris}, \& {Perez}}]{Tobin+2018}
{Tobin}, J.~J., {Looney}, L.~W., {Li}, Z.-Y., {et~al.} 2018, \apj, 867, 43

\bibitem[{{Tobin} {et~al.}(2022){Tobin}, {Offner}, {Kratter}, {Megeath}, {Sheehan}, {Looney}, {Diaz-Rodriguez}, {Osorio}, {Anglada}, {Sadavoy}, {Furlan}, {Segura-Cox}, {Karnath}, {van't Hoff}, {van Dishoeck}, {Li}, {Sharma}, {Stutz}, \& {Tychoniec}}]{Tobin+2022}
{Tobin}, J.~J., {Offner}, S. S.~R., {Kratter}, K.~M., {et~al.} 2022, \apj, 925, 39

\bibitem[{{Tohline}(2002)}]{Tohline2002}
{Tohline}, J.~E. 2002, \araa, 40, 349

\bibitem[{{Tokovinin}(2020)}]{Tokovinin_ecc+2020}
{Tokovinin}, A. 2020, \mnras, 496, 987

\bibitem[{{Tokovinin} \& {Moe}(2020)}]{Tokovinin+2020}
{Tokovinin}, A. \& {Moe}, M. 2020, \mnras, 491, 5158

\bibitem[{{Toomre}(1964)}]{Toomre1964}
{Toomre}, A. 1964, \apj, 139, 1217

\bibitem[{{Toonen} {et~al.}(2020){Toonen}, {Portegies Zwart}, {Hamers}, \& {Bandopadhyay}}]{Toonen+2020}
{Toonen}, S., {Portegies Zwart}, S., {Hamers}, A.~S., \& {Bandopadhyay}, D. 2020, \aap, 640, A16

\bibitem[{{Urquhart} {et~al.}(2007){Urquhart}, {Busfield}, {Hoare}, {Lumsden}, {Clarke}, {Moore}, {Mottram}, \& {Oudmaijer}}]{Urquhart+2007}
{Urquhart}, J.~S., {Busfield}, A.~L., {Hoare}, M.~G., {et~al.} 2007, \aap, 461, 11

\bibitem[{{Urquhart} {et~al.}(2011){Urquhart}, {Morgan}, {Figura}, {Moore}, {Lumsden}, {Hoare}, {Oudmaijer}, {Mottram}, {Davies}, \& {Dunham}}]{Urquhart+2011}
{Urquhart}, J.~S., {Morgan}, L.~K., {Figura}, C.~C., {et~al.} 2011, \mnras, 418, 1689

\bibitem[{{Venuti} {et~al.}(2019){Venuti}, {Damiani}, \& {Prisinzano}}]{Venuti+2019}
{Venuti}, L., {Damiani}, F., \& {Prisinzano}, L. 2019, \aap, 621, A14

\bibitem[{{Walsh} {et~al.}(2001){Walsh}, {Bertoldi}, {Burton}, \& {Nikola}}]{Walsh+2001}
{Walsh}, A.~J., {Bertoldi}, F., {Burton}, M.~G., \& {Nikola}, T. 2001, \mnras, 326, 36

\bibitem[{{Walsh} {et~al.}(1999){Walsh}, {Burton}, {Hyland}, \& {Robinson}}]{Walsh+1999}
{Walsh}, A.~J., {Burton}, M.~G., {Hyland}, A.~R., \& {Robinson}, G. 1999, \mnras, 309, 905

\bibitem[{{Ward-Duong} {et~al.}(2015){Ward-Duong}, {Patience}, {De Rosa}, {Bulger}, {Rajan}, {Goodwin}, {Parker}, {McCarthy}, \& {Kulesa}}]{WardDuong+2015}
{Ward-Duong}, K., {Patience}, J., {De Rosa}, R.~J., {et~al.} 2015, \mnras, 449, 2618

\bibitem[{{Wheelwright} {et~al.}(2012){Wheelwright}, {de Wit}, {Oudmaijer}, {Hoare}, {Lumsden}, {Fujiyoshi}, \& {Close}}]{Wheelwright+2012}
{Wheelwright}, H.~E., {de Wit}, W.~J., {Oudmaijer}, R.~D., {et~al.} 2012, \aap, 540, A89

\bibitem[{{Wheelwright} {et~al.}(2010){Wheelwright}, {Oudmaijer}, \& {Goodwin}}]{Wheelwright+2010}
{Wheelwright}, H.~E., {Oudmaijer}, R.~D., \& {Goodwin}, S.~P. 2010, \mnras, 401, 1199

\bibitem[{{Zapata} {et~al.}(2019){Zapata}, {Garay}, {Palau}, {Rodr{\'\i}guez}, {Fern{\'a}ndez-L{\'o}pez}, {Estalella}, \& {Guzm{\'a}n}}]{Zapata+2019}
{Zapata}, L.~A., {Garay}, G., {Palau}, A., {et~al.} 2019, \apj, 872, 176

\bibitem[{{Zhang} {et~al.}(2019){Zhang}, {Tan}, {Tanaka}, {De Buizer}, {Liu}, {Beltr{\'a}n}, {Kratter}, {Mardones}, \& {Garay}}]{Zhang+2019}
{Zhang}, Y., {Tan}, J.~C., {Tanaka}, K. E.~I., {et~al.} 2019, Nature Astronomy, 3, 517

\bibitem[{{Zhu} {et~al.}(2012){Zhu}, {Hartmann}, {Nelson}, \& {Gammie}}]{Zhu+2012}
{Zhu}, Z., {Hartmann}, L., {Nelson}, R.~P., \& {Gammie}, C.~F. 2012, \apj, 746, 110

\bibitem[{{Z{\'u}{\~n}iga-Fern{\'a}ndez} {et~al.}(2021){Z{\'u}{\~n}iga-Fern{\'a}ndez}, {Bayo}, {Elliott}, {Zamora}, {Corval{\'a}n}, {Haubois}, {Corral-Santana}, {Olofsson}, {Hu{\'e}lamo}, {Sterzik}, {Torres}, {Quast}, \& {Melo}}]{Zuniga+2021}
{Z{\'u}{\~n}iga-Fern{\'a}ndez}, S., {Bayo}, A., {Elliott}, P., {et~al.} 2021, \aap, 645, A30

\end{thebibliography}

\begin{appendix}

\section{Overview of the observed multiples}
\label{appendix:images}
We provide here an overview of the observed multiple systems. The left-hand panels correspond to the full FoV of 28\arcsec$\times$ 28\arcsec and the right-hand panels display a zoom-in (15$\times$15 pixels) on the individual detected sources. 


\FloatBarrier
\begin{figure*}[!h]
        \centering
        \subfigure[G203.3166]{\includegraphics[scale=0.75]{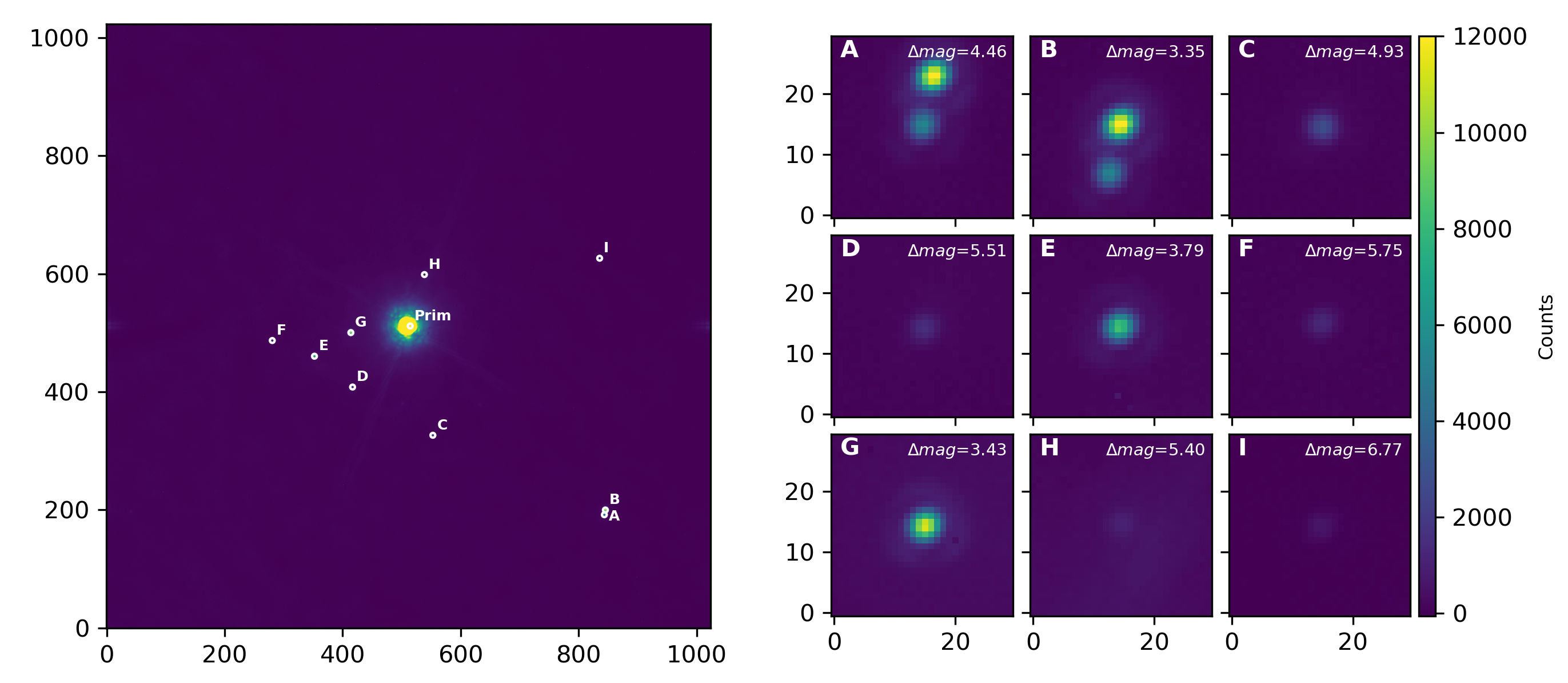}}
        \hfill
    \subfigure[G232.6207]{\includegraphics[scale=0.75]{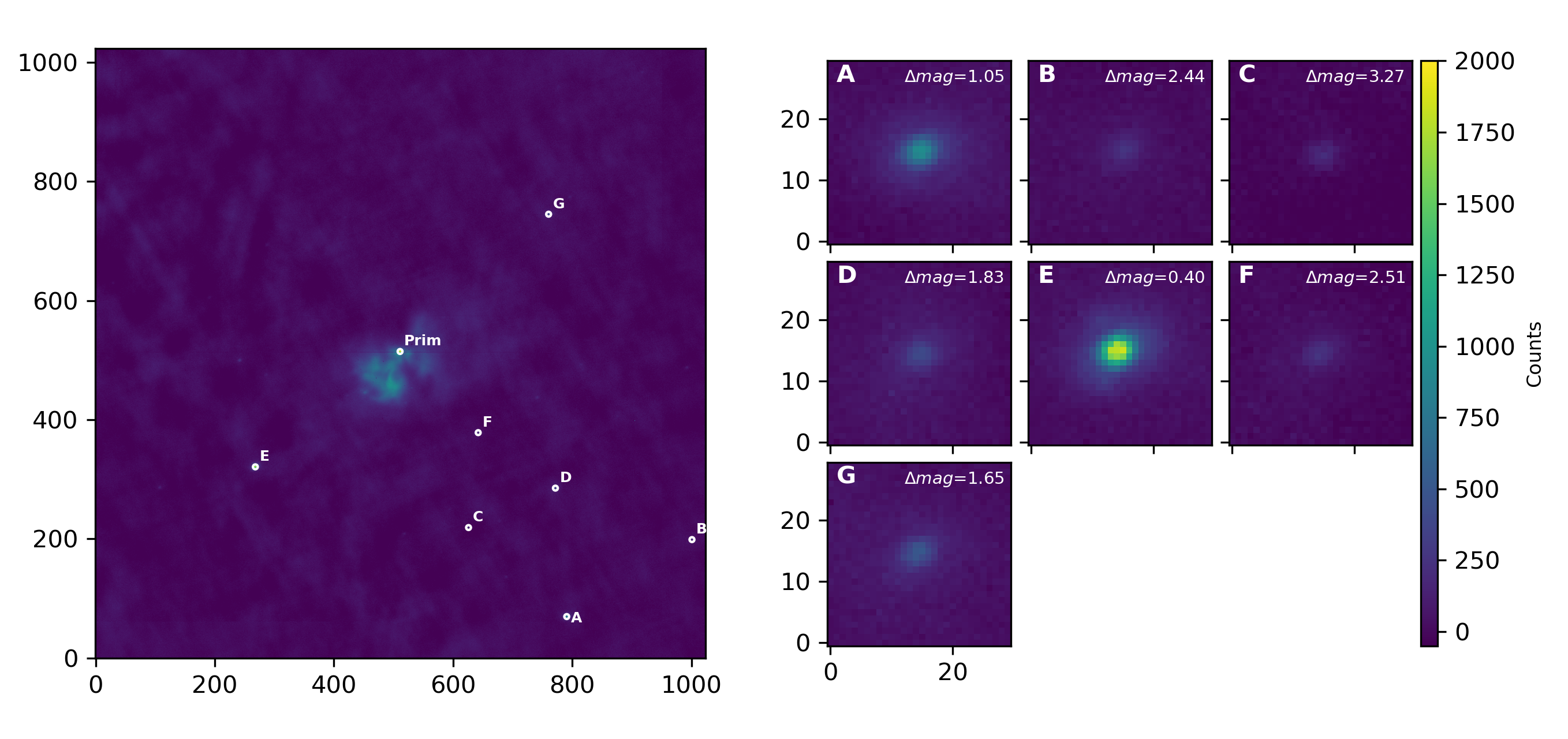}}
    \hfill
    \subfigure[G263.7759]{\includegraphics[scale=0.75]{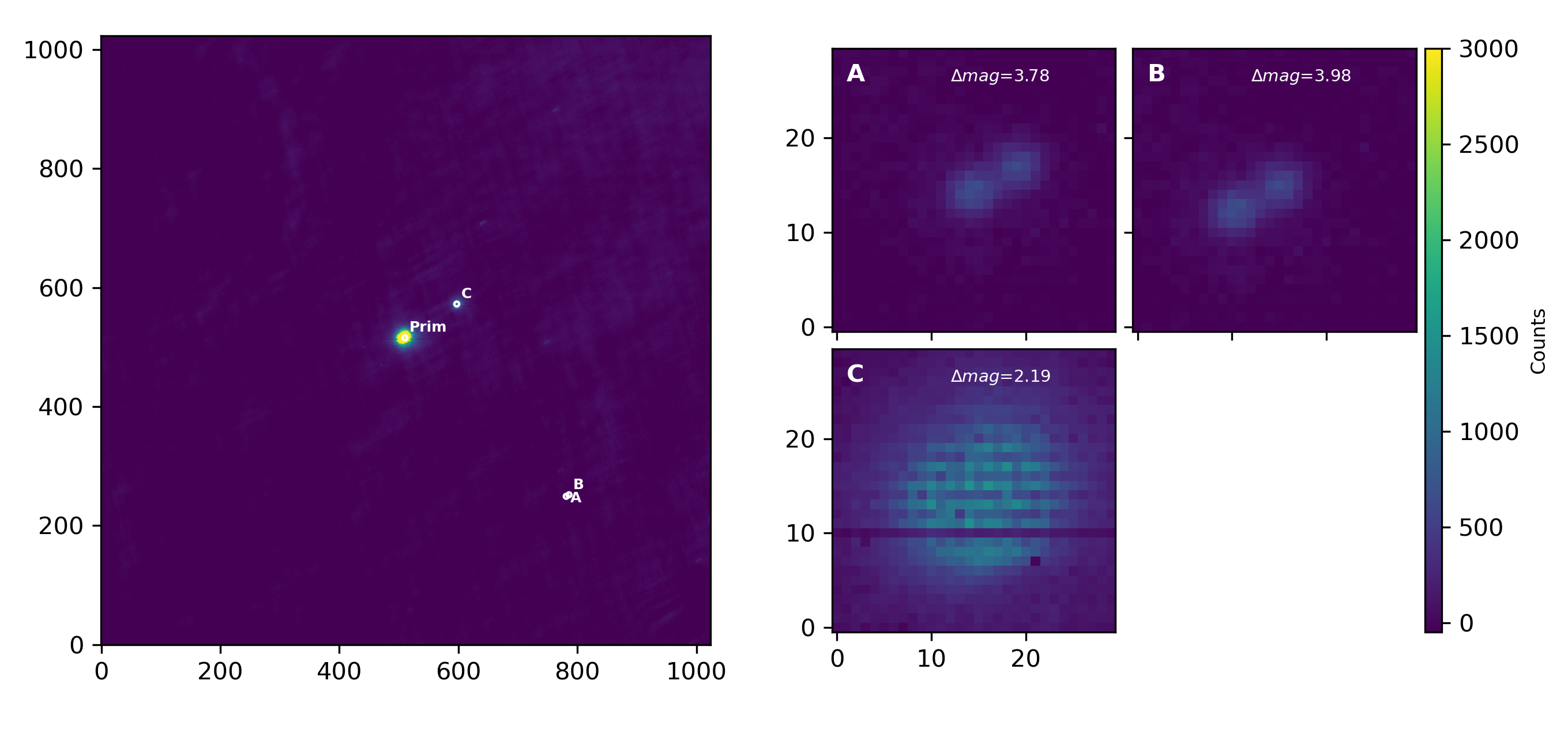}}
        \caption{ 28\arcsec$\times$ 28\arcsec reduced NACO $L'$-band images. The Right Ascension (RA) is on the x-axis and the Declination is displayed on the y-axis. For reference, north is up and east is left. }
    \label{fig:mYSO_comp_1}
        \end{figure*}

\begin{figure*}[!h]
        \centering
        \subfigure[G265.1438]{\includegraphics[scale=0.80]{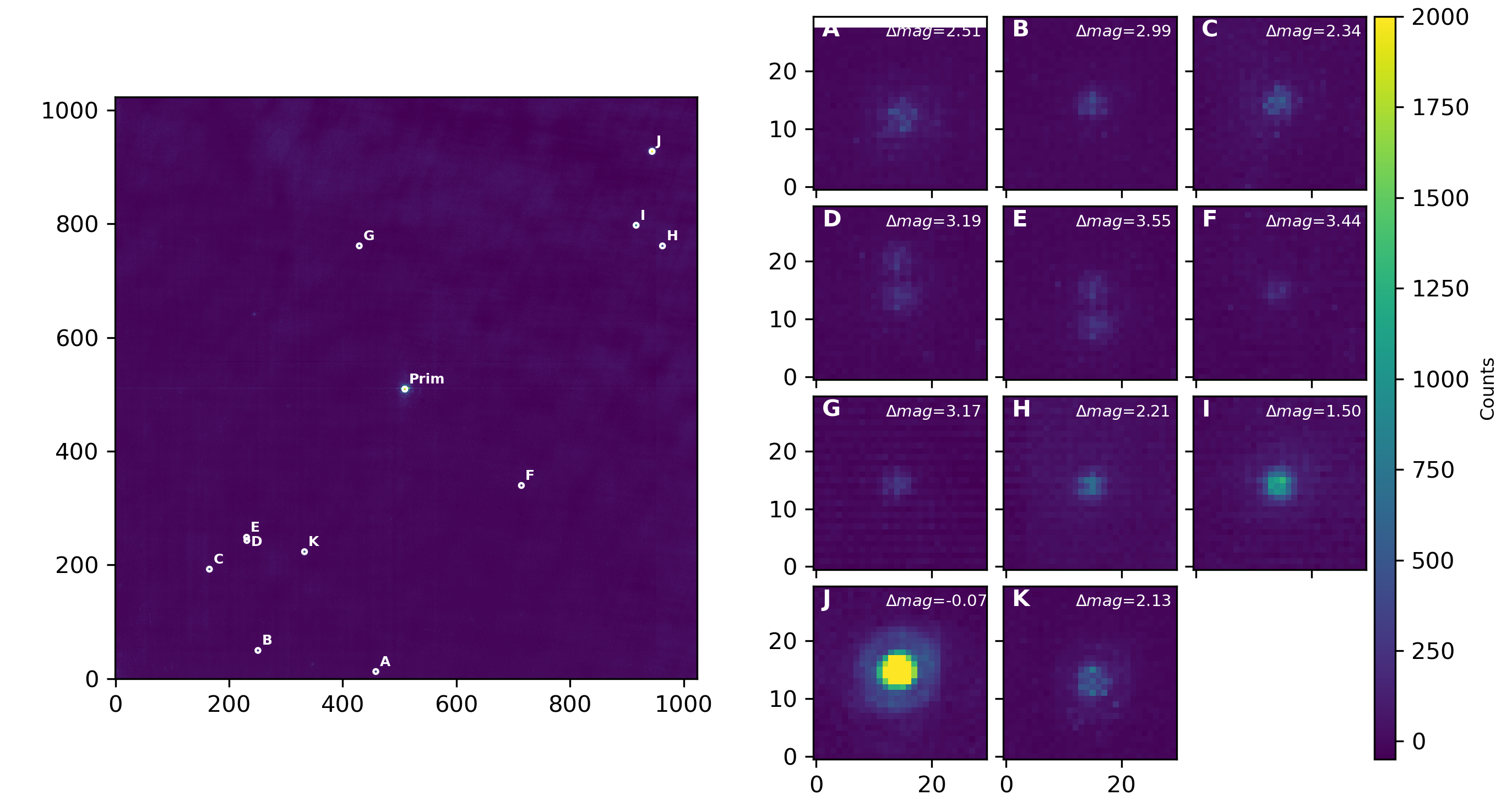}}
        \hfill
    \subfigure[G268.3957]{\includegraphics[scale=0.80]{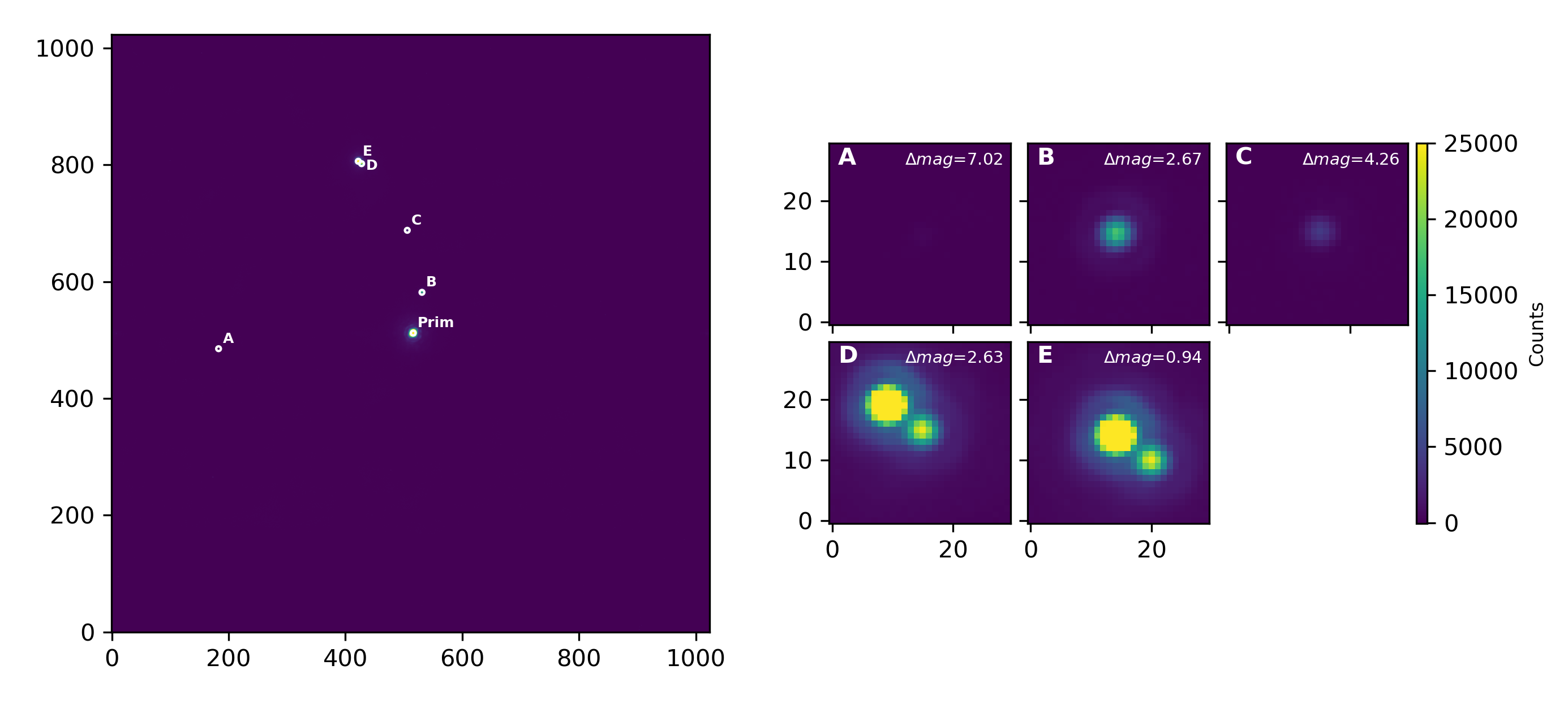}}
    \hfill
    
        \caption{ 28\arcsec$\times$ 28\arcsec reduced NACO $L'$-band images. The Right Ascension (RA) is on the x-axis and the Declination is displayed on the y-axis. For reference, north is up and east is left. }
    \label{fig:mYSO_comp_2}
        \end{figure*}

\begin{figure*}[!h]
        \centering
        \subfigure[G269.1586]{\includegraphics[scale=0.75]{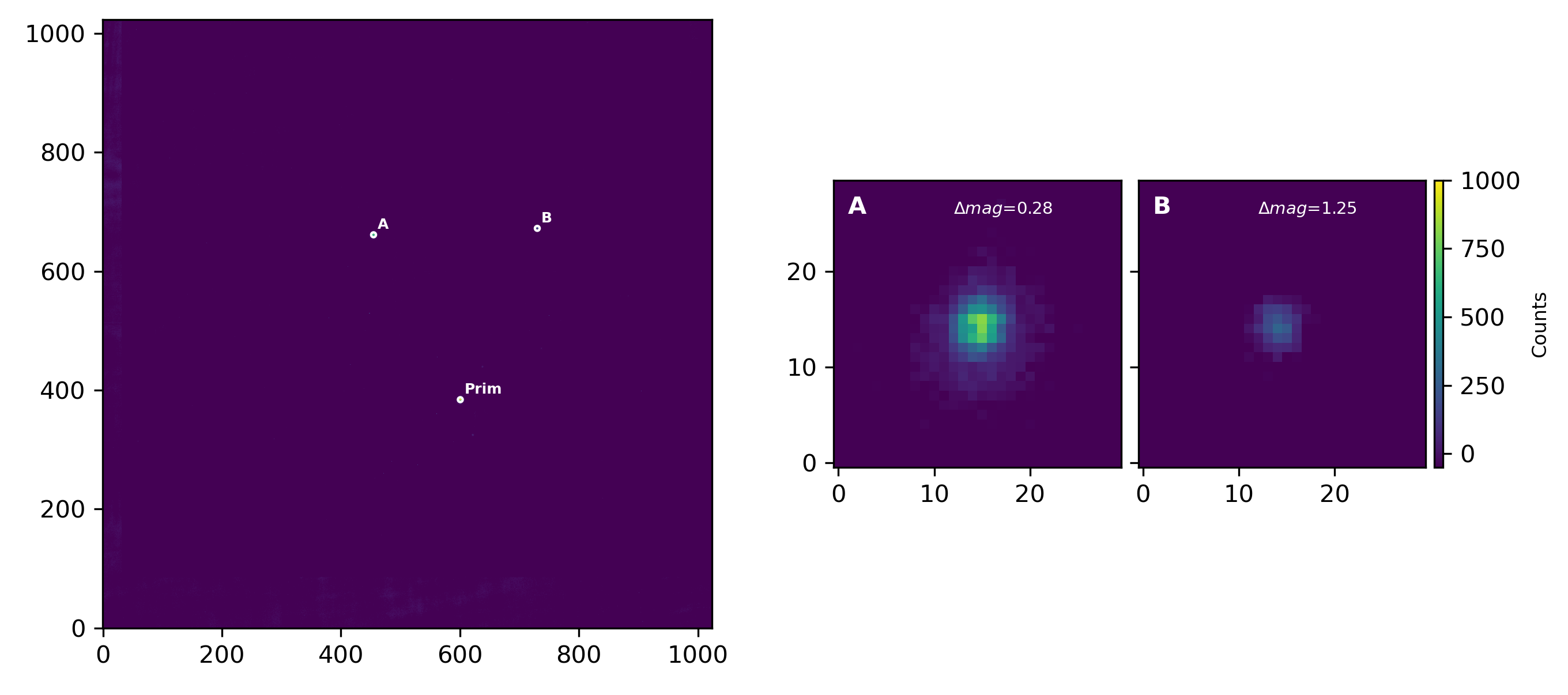}}
        \hfill
    \subfigure[G305.2017]{\includegraphics[scale=0.75]{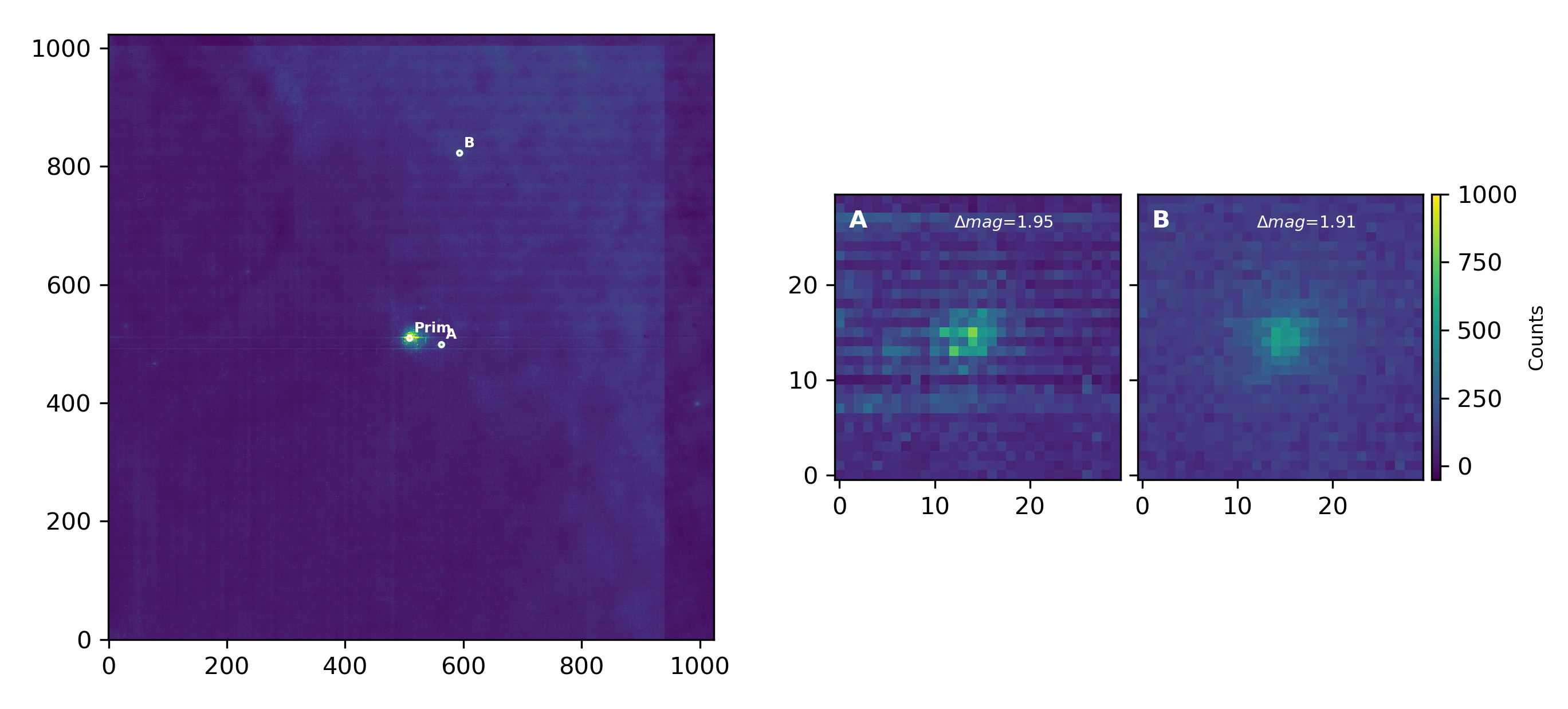}}
    \hfill
    \subfigure[G310.0135]{\includegraphics[scale=0.75]{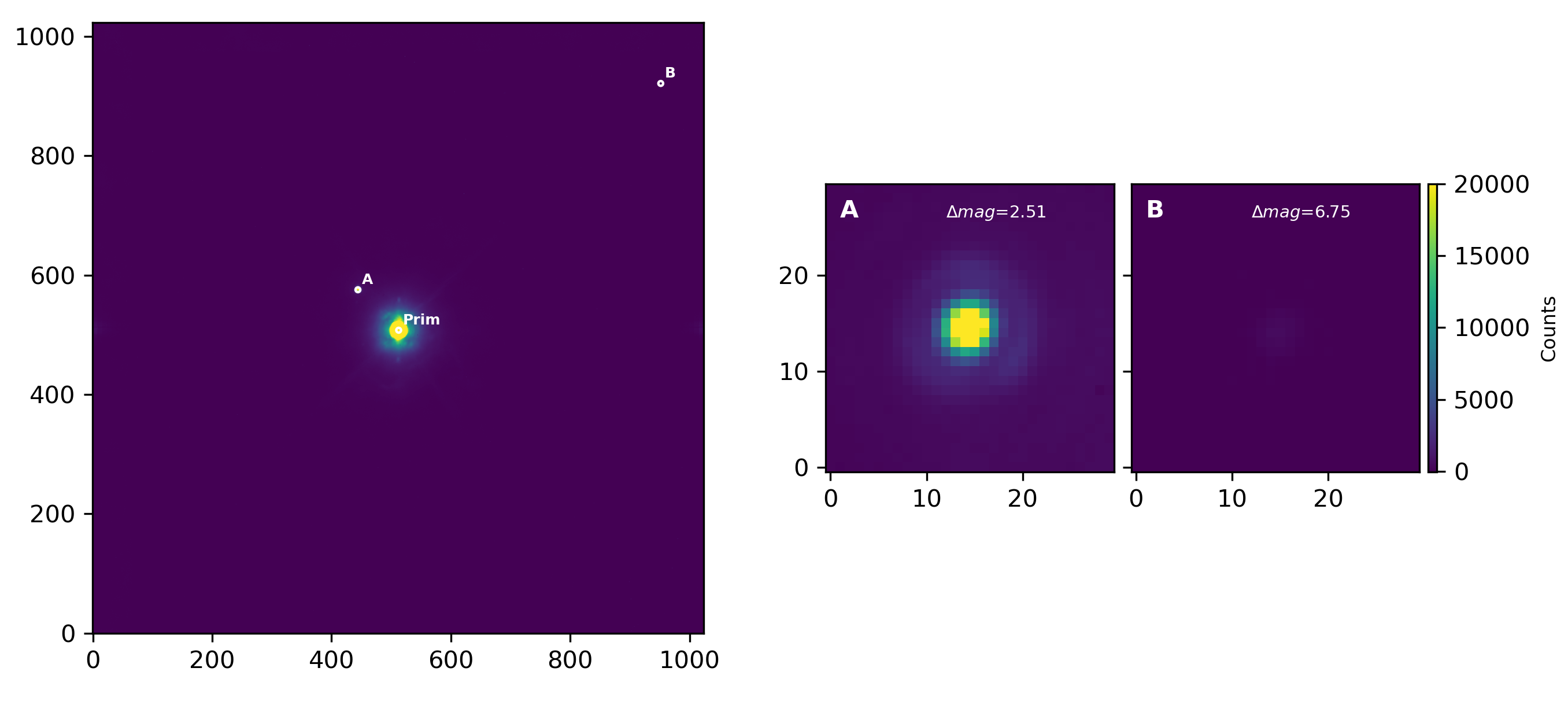}}
        \caption{ 28\arcsec$\times$ 28\arcsec reduced NACO $L'$-band images. The Right Ascension (RA) is on the x-axis and the Declination is displayed on the y-axis. For reference, north is up and east is left. }
    \label{fig:mYSO_comp_3}
        \end{figure*}

\section{Summary of single objects}
We display the $28\arcsec\times28\arcsec$ reduced NACO $L'-$band images of sources for which we do not detect companions. 

\begin{figure*}[!h]
        \centering
        \subfigure[G194.9349]{\includegraphics[scale=0.37]{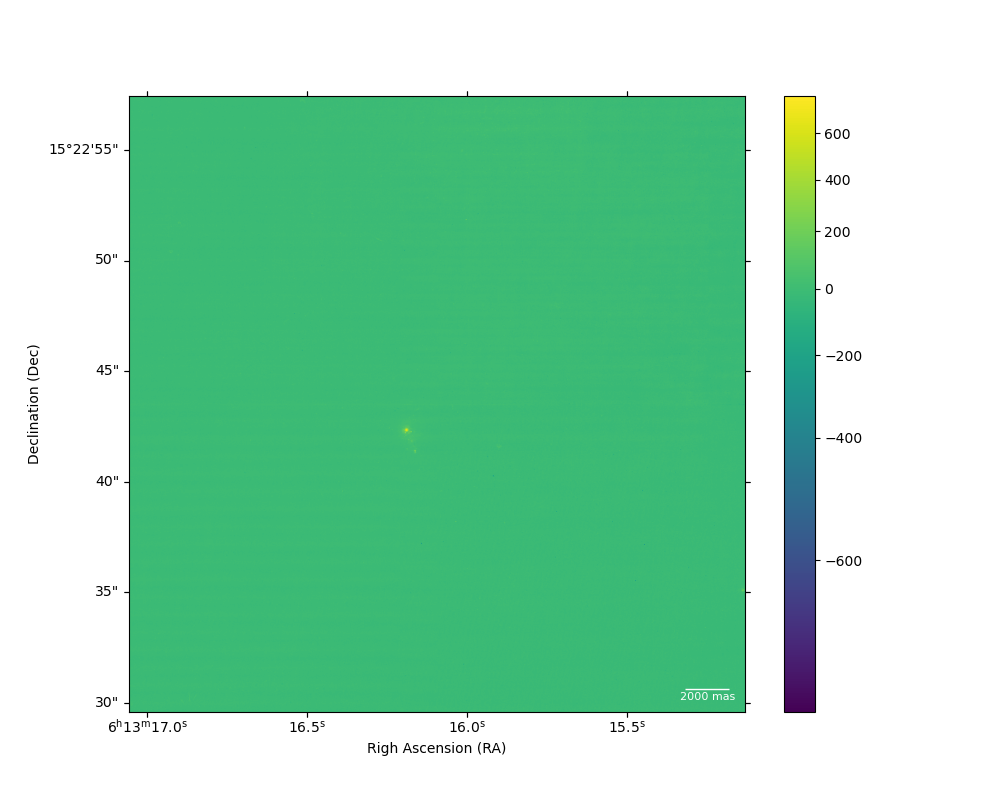}}
        \hspace{-0.6cm}
        \subfigure[G254.0548]{\includegraphics[scale=0.37]{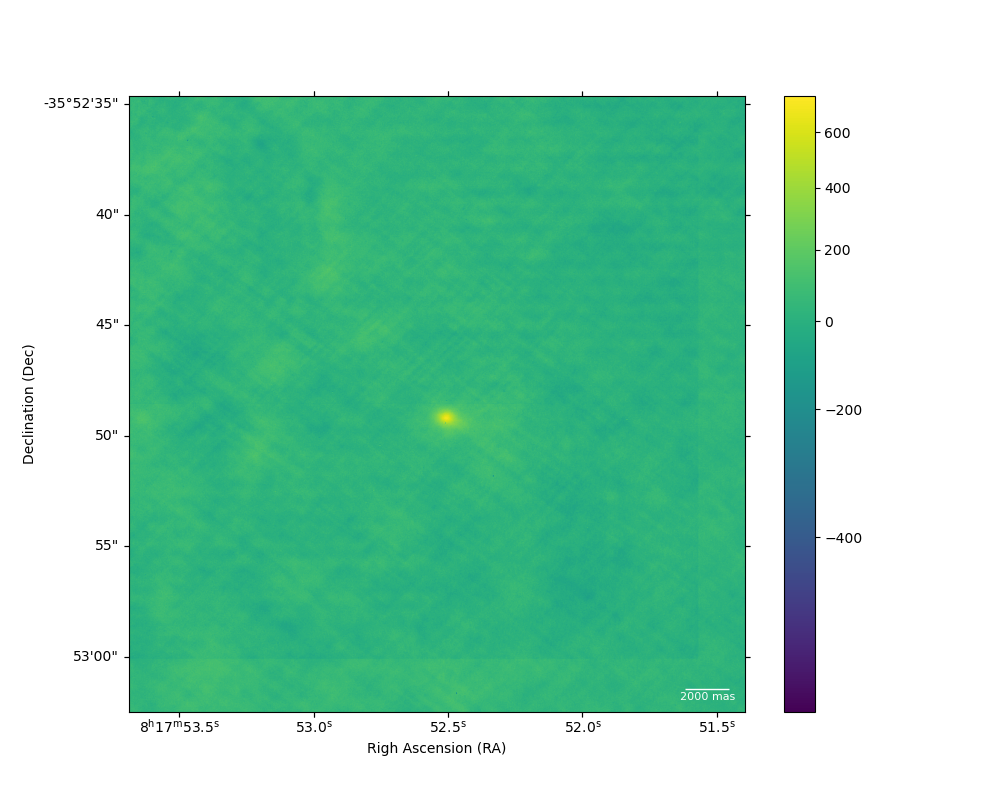}}
        \subfigure[G263.7434]{\includegraphics[scale=0.37]{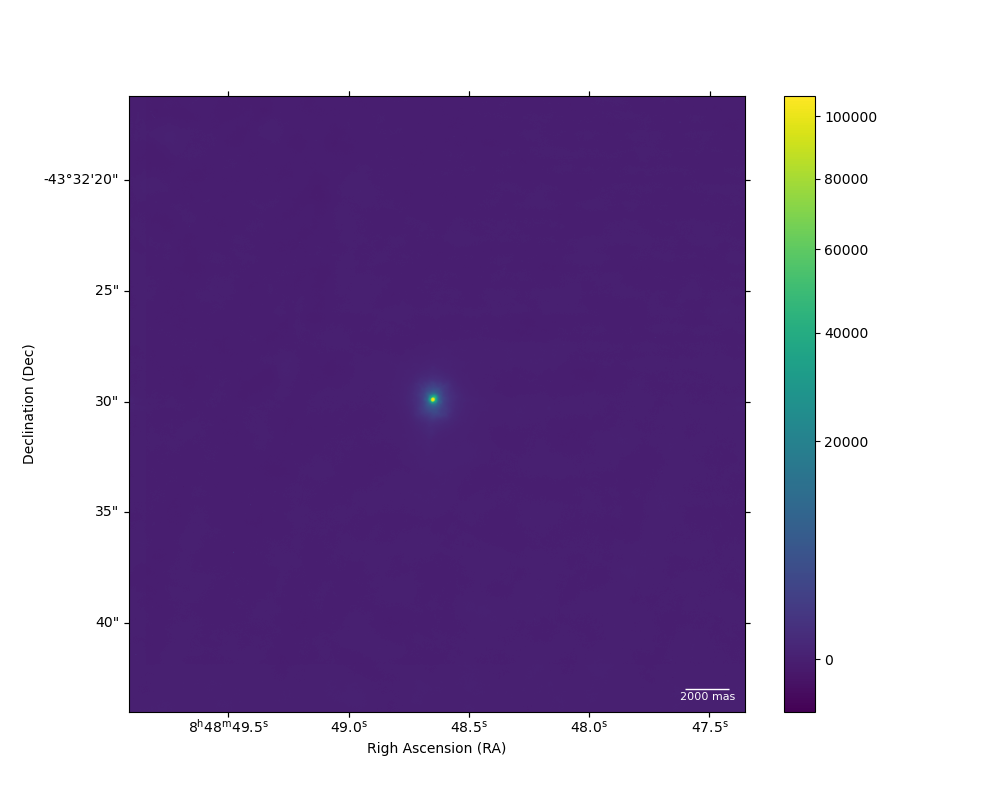}}
        \hspace{-0.6cm}
    \subfigure[G314.3197]{\includegraphics[scale=0.37]{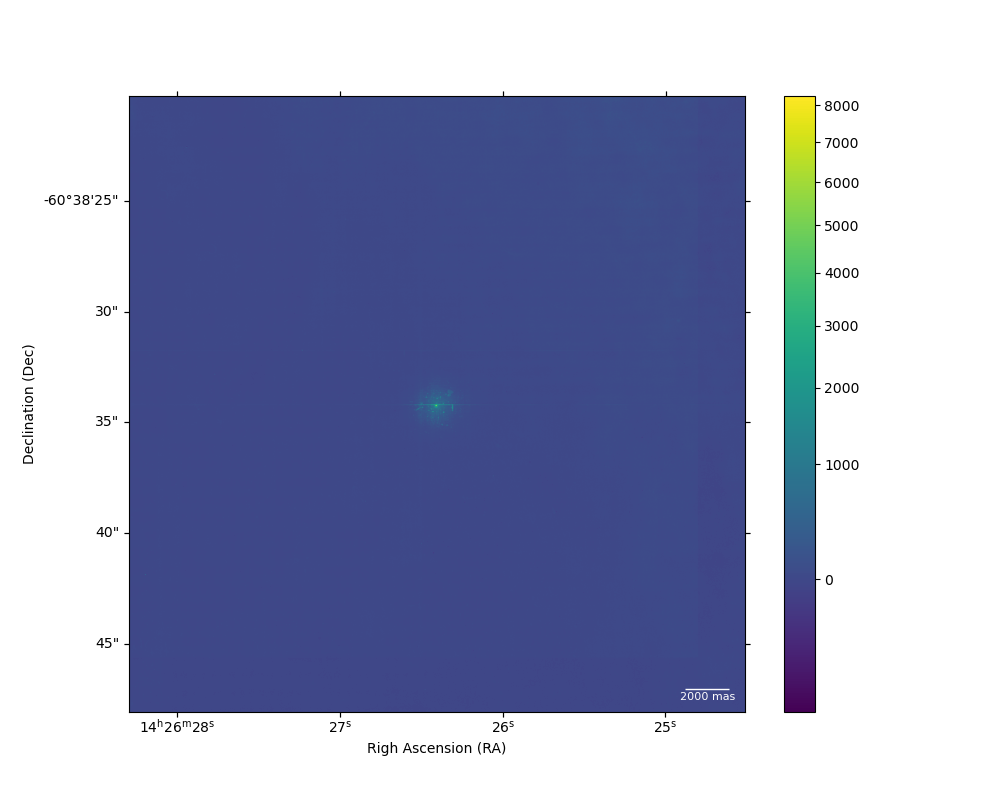}}
    \subfigure[G318.0489]{\includegraphics[scale=0.40]{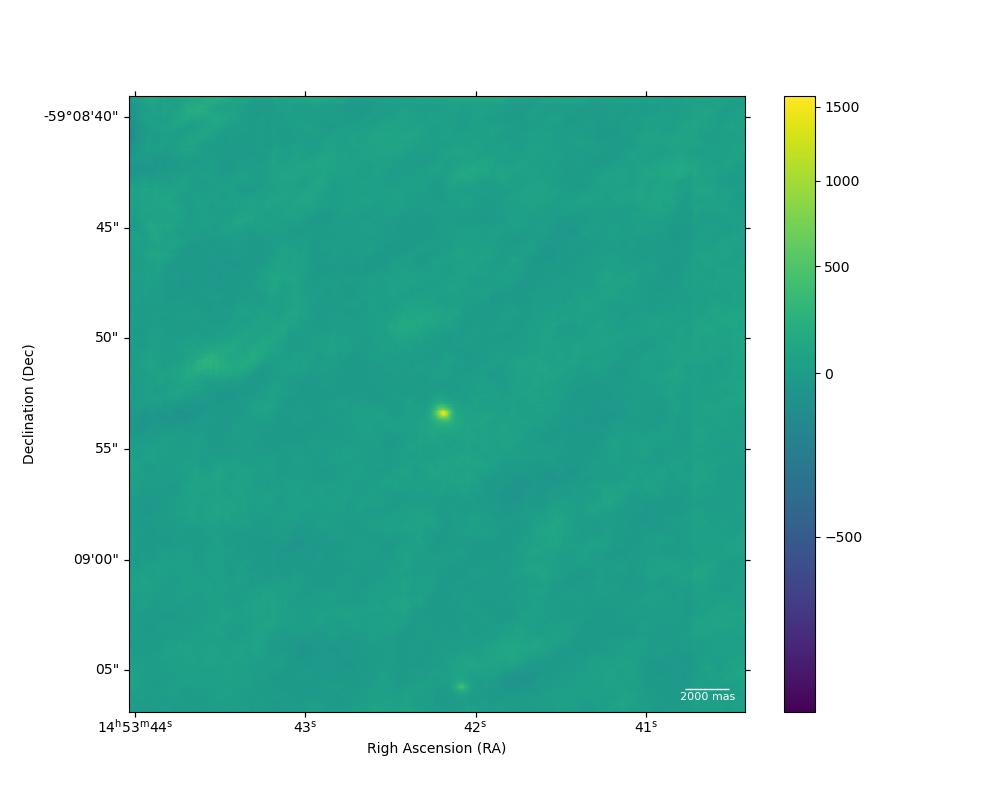}}
        \caption{ 28\arcsec$\times$ 28\arcsec reduced NACO $L'$-band images. The Right Ascension (RA) is on the x-axis and the Declination is displayed on the y-axis. For reference, north is up and east is left. }
    \label{fig:mYSO_single}
    \end{figure*}

\section{Detection limits}
\label{appendix:detection_limits}

Here we show the limiting magnitude determination by injecting artificial sources in all the images. Purple points correspond to artificially injected sources, not detected by our source detection algorithm at a $5\sigma$ level, while the light blue ones show the non-detections. For reference, we added to the plot all the detected sources (orange symbol of a star) in the real image and provided for both groups the lowest detected magnitude. The y-axis values are the differences in magnitudes between the artificial sources and the primary source, and the x-axis is the distance in arcseconds between these two. Using that method we ensure that all the sources within the sensitivity of the instrument and matching the requirements of our source detection algorithm are detected. Any missed source would either be fainter than the limiting magnitude, masked by the background variations or untrustworthy detections. 

\begin{figure*}[!h]
        \centering
        \subfigure[G194.9349]{\includegraphics[scale=0.55]{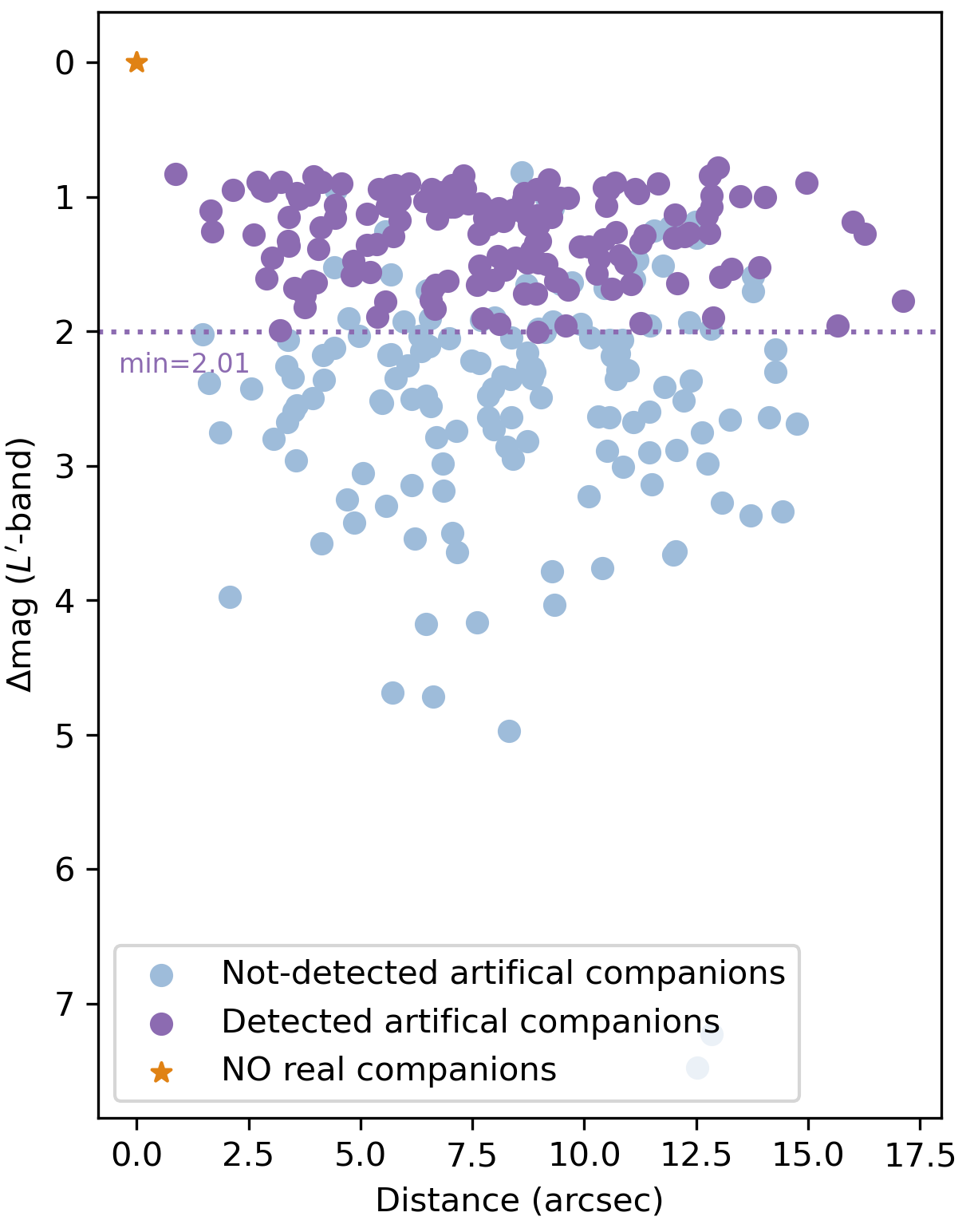}}
        \subfigure[G203.3166]{\includegraphics[scale=0.55]{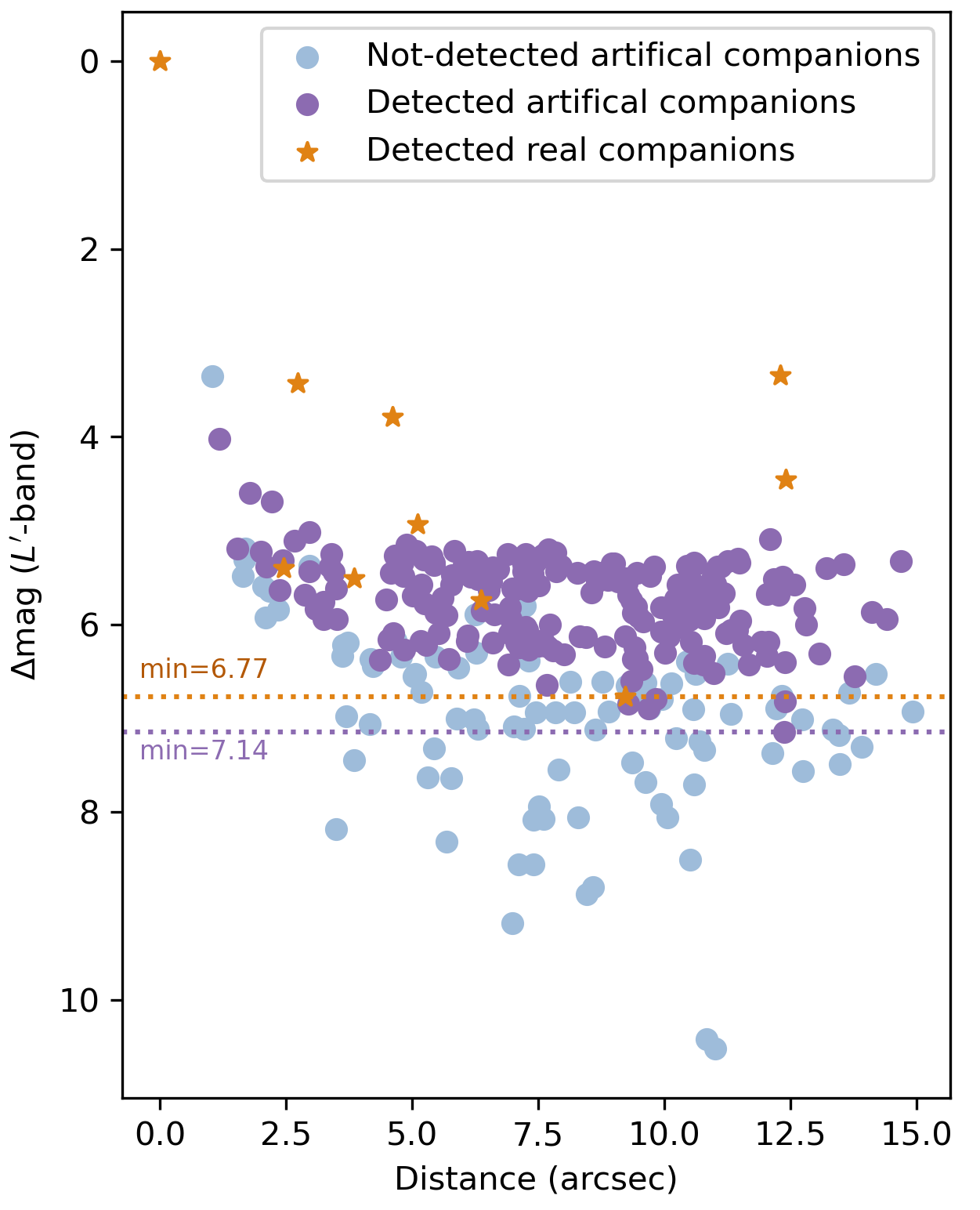}}
    \hspace{0.5cm}
        \subfigure[G232.6207]{\includegraphics[scale=0.55]{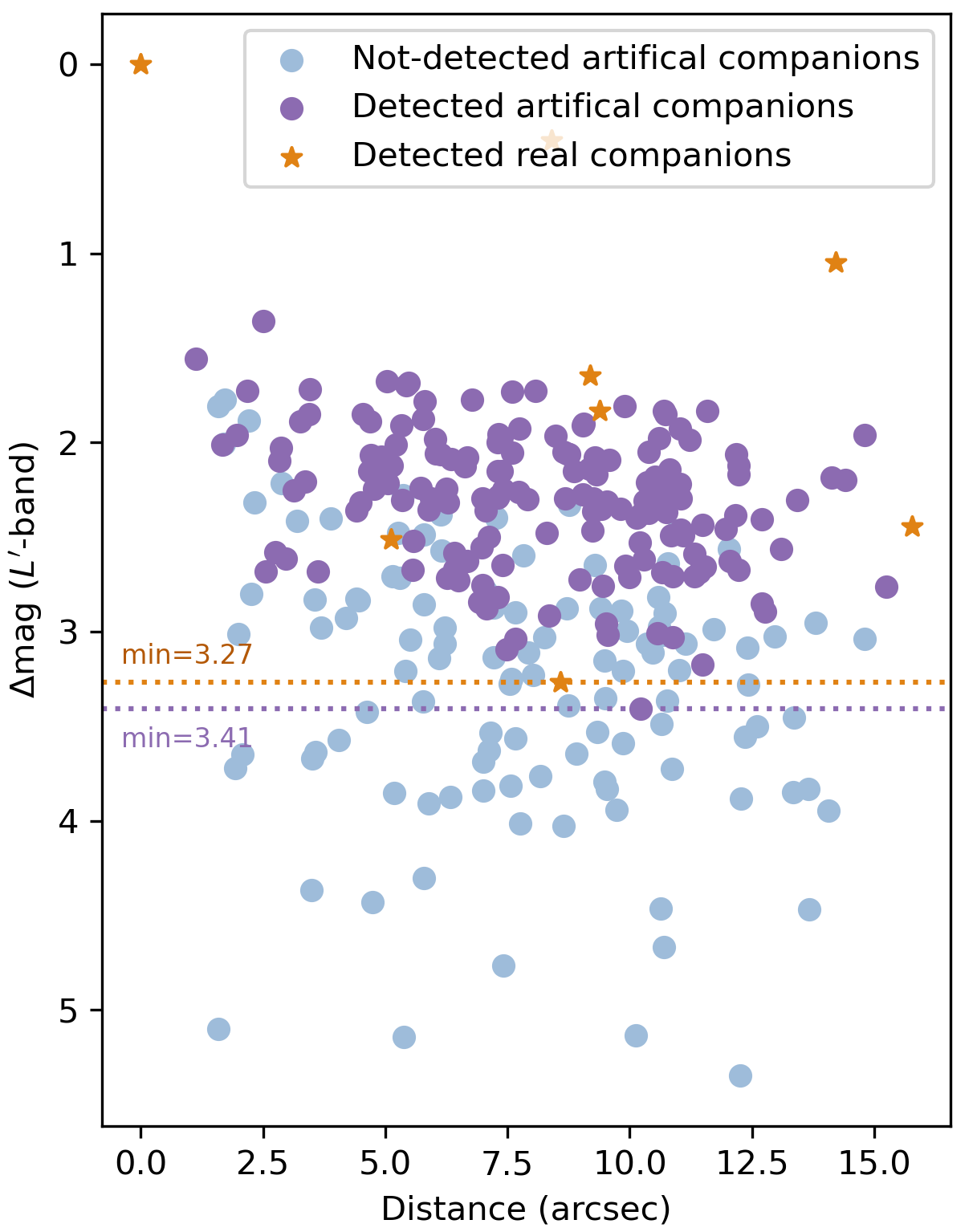}}
    \subfigure[G254.0548]{\includegraphics[scale=0.55]{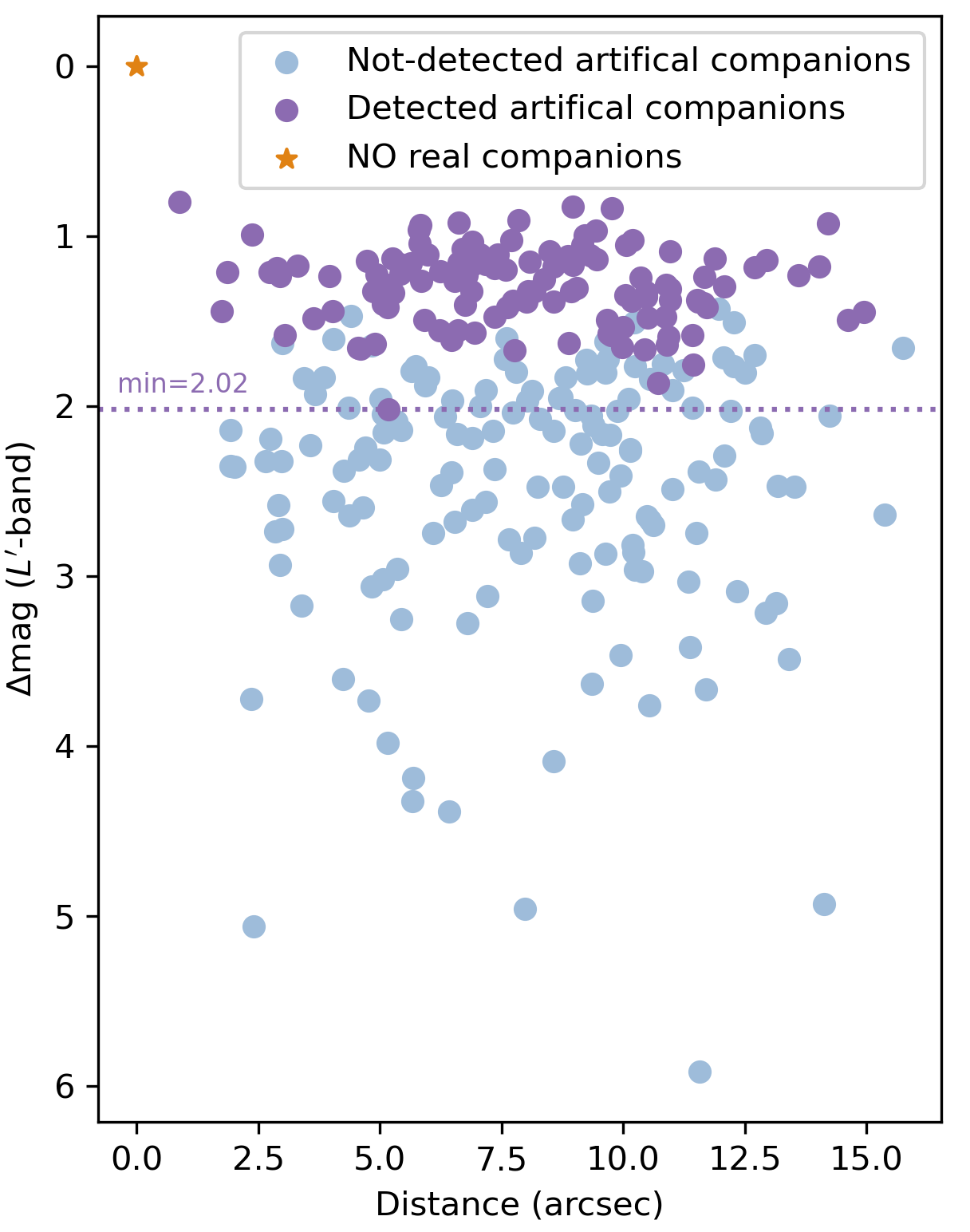}}
        \subfigure[G263.7434]{\includegraphics[scale=0.55]{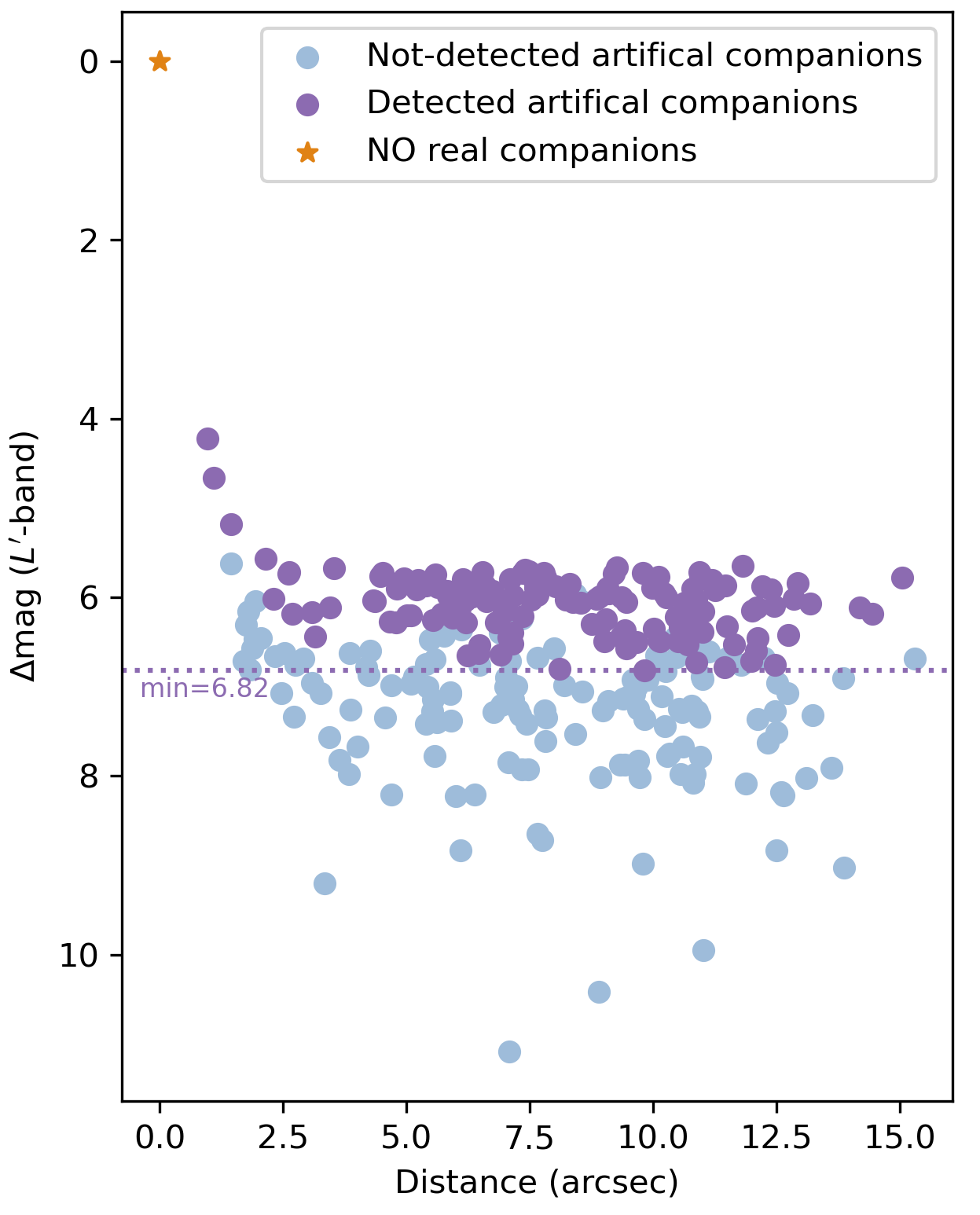}}
    \subfigure[G263.7759]{\includegraphics[scale=0.55]{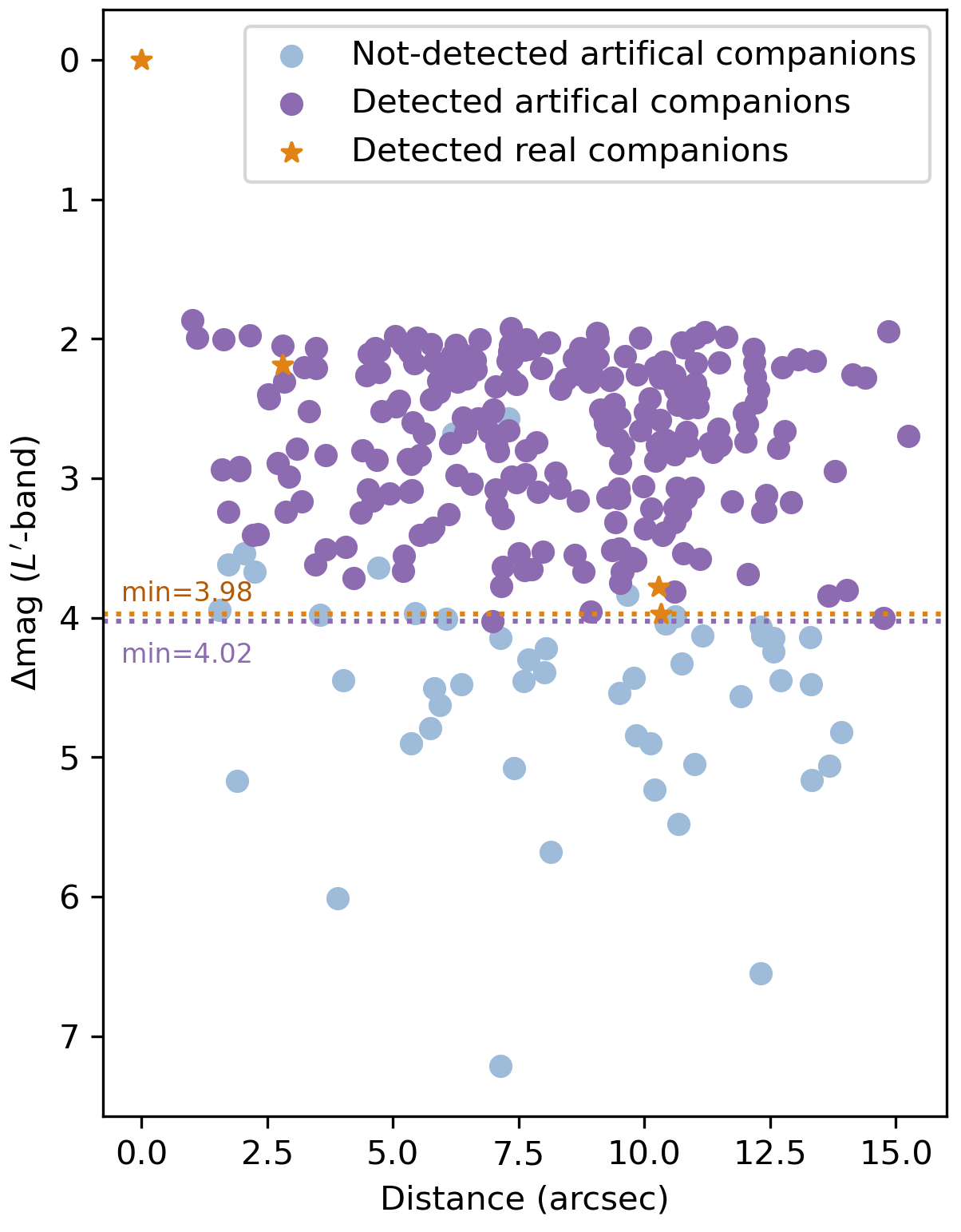}}
    \subfigure[G265.1438]{\includegraphics[scale=0.55]{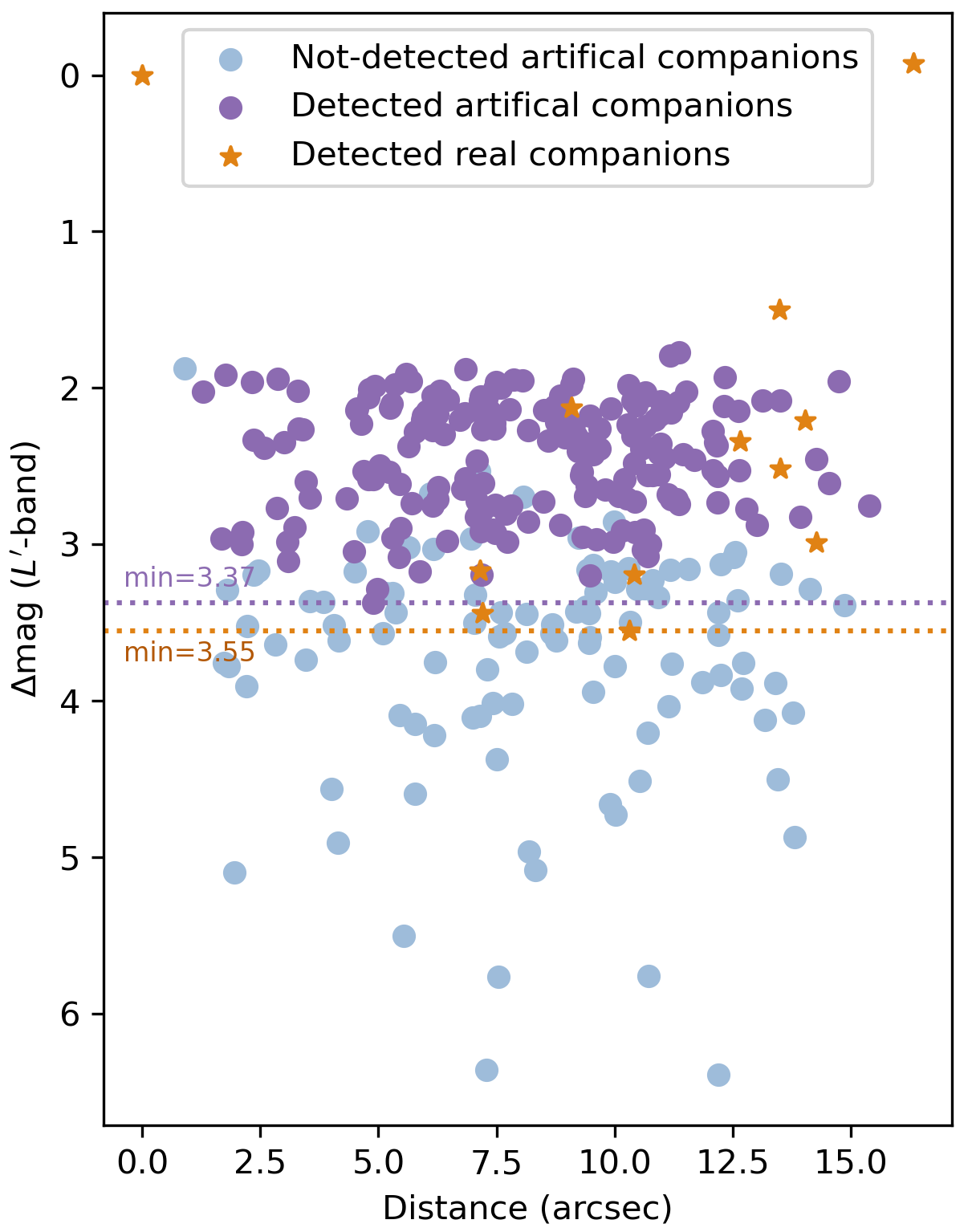}}
        \subfigure[G268.3957]{\includegraphics[scale=0.55]{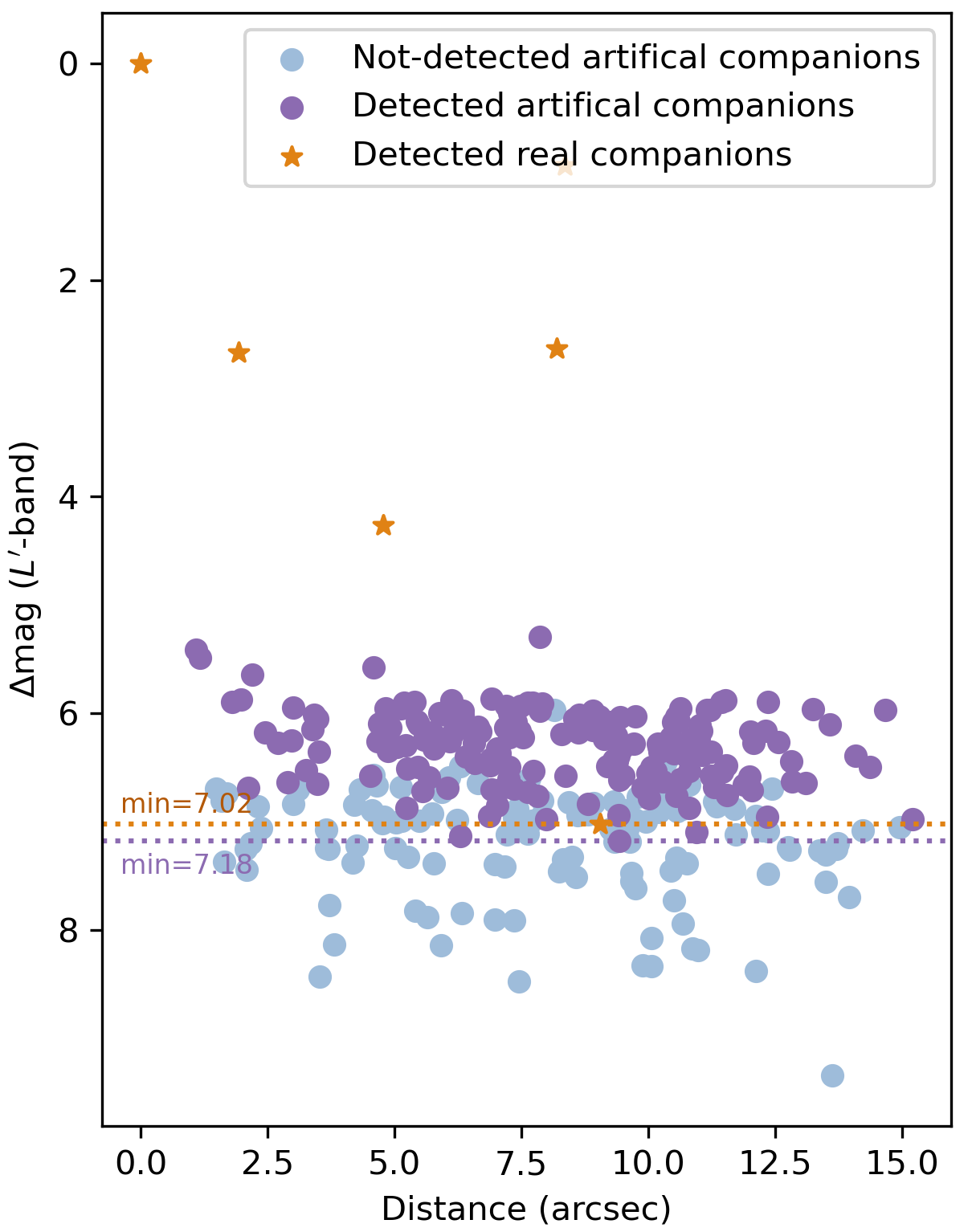}}
        \subfigure[G269.1586]{\includegraphics[scale=0.55]{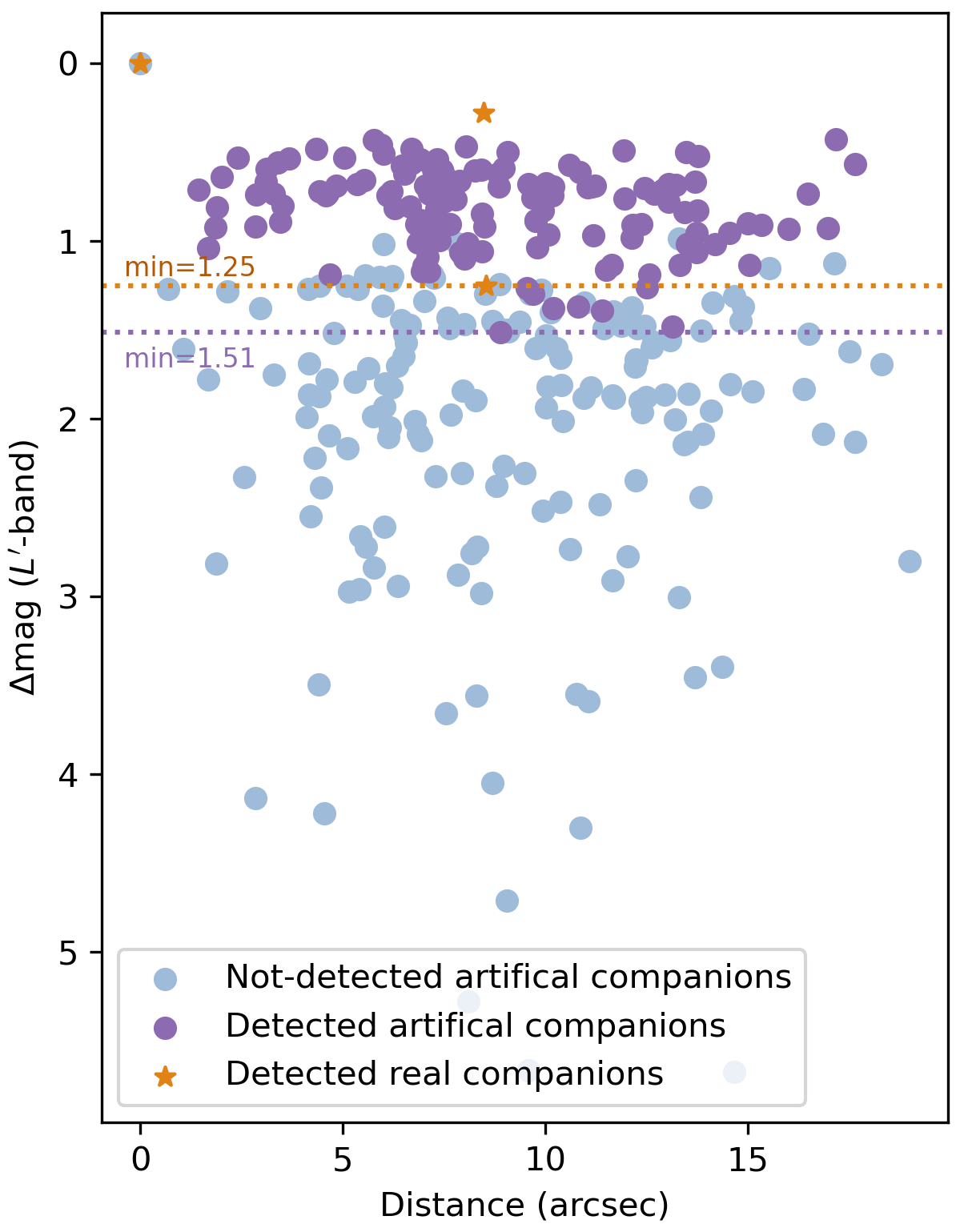}}
        \caption{Detection limit artificial source tests for all the targets. We display only one iteration. The orange star symbols stand for the real detected companions (if any) and the plain dot symbols refer to artificially injected sources. We display the magnitude contrast of the faintest (i) source detected in the real image (orange dot and line) and (ii) artificial source (purple dot and line).}
        \end{figure*}

\begin{figure*}[!h]
        \centering
        \subfigure[G305.2017]{\includegraphics[scale=0.54]{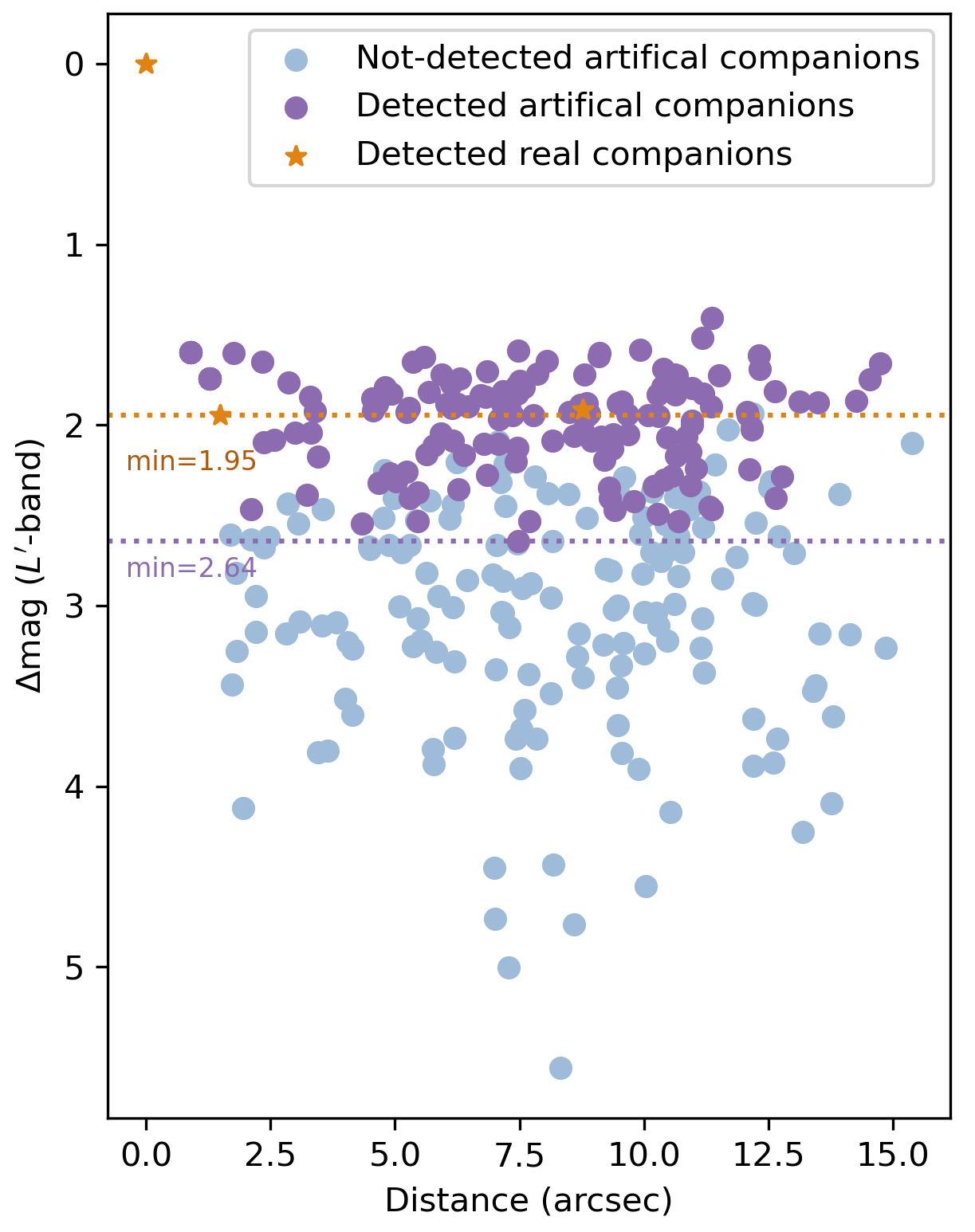}}
    \hspace{0.5cm}
        \subfigure[G310.0135]{\includegraphics[scale=0.54]{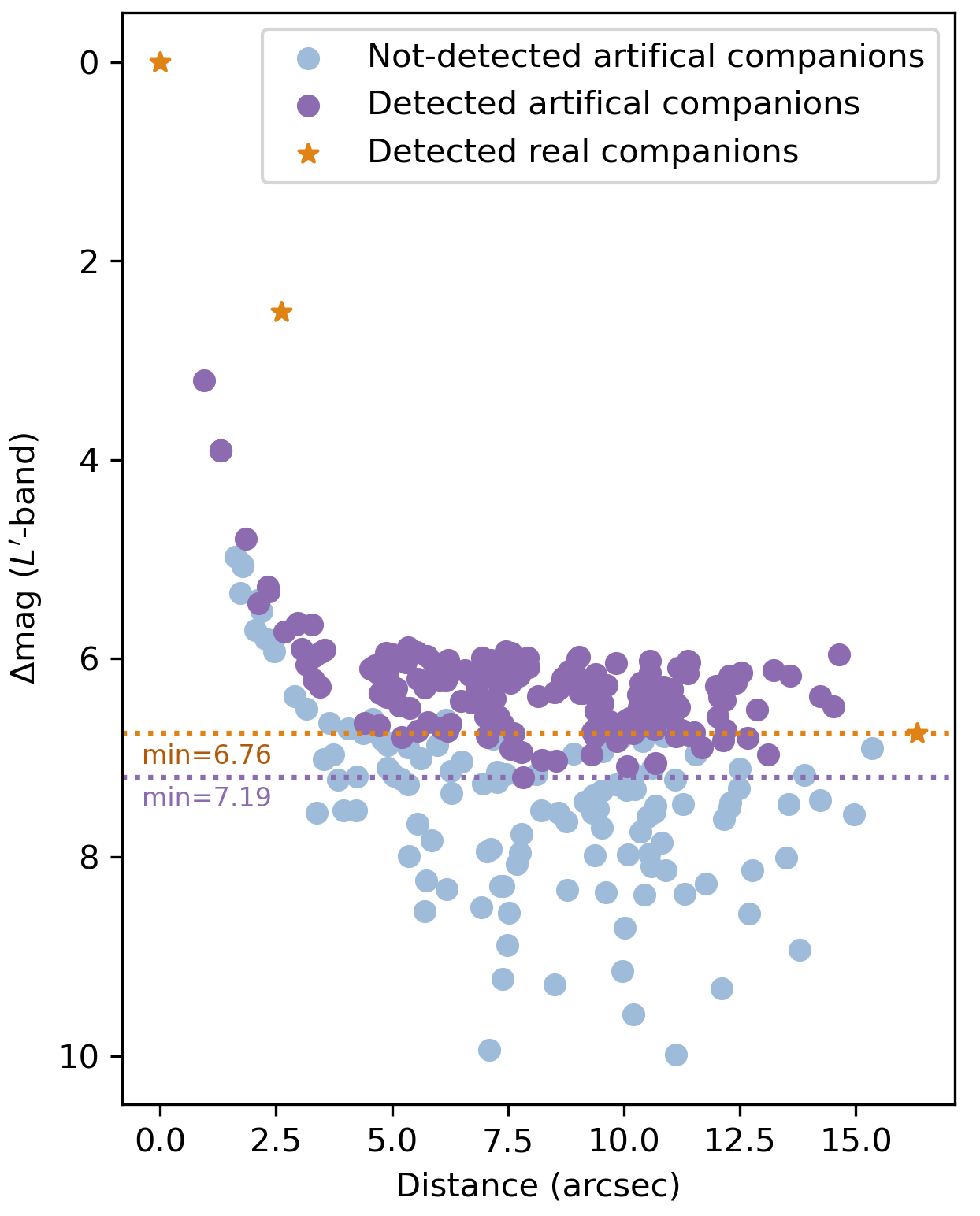}}
    \hspace{0.5cm}
    \subfigure[G314.3197]{\includegraphics[scale=0.54]{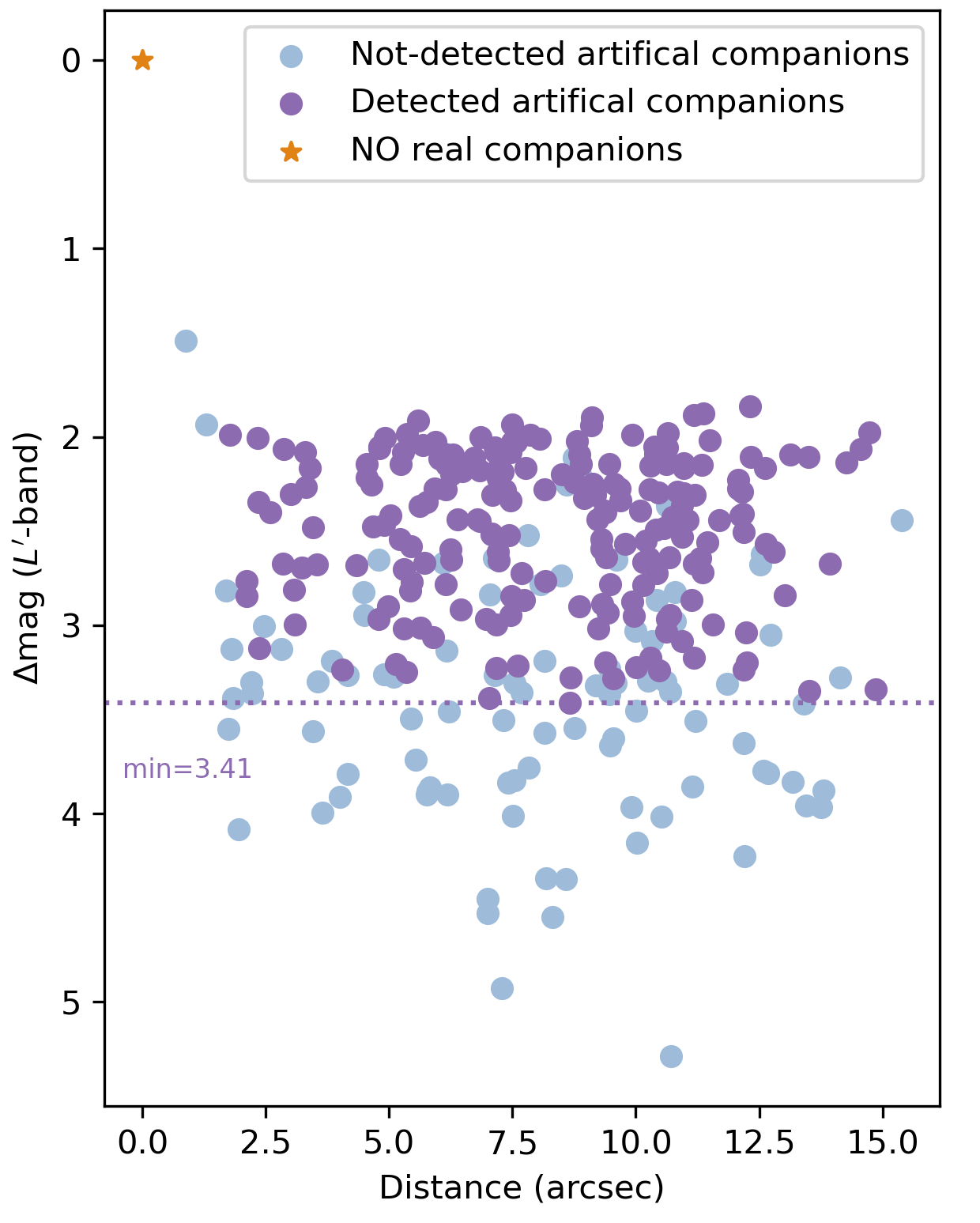}}
    \hspace{5cm}
    \subfigure[G318.0489]{\includegraphics[scale=0.55]{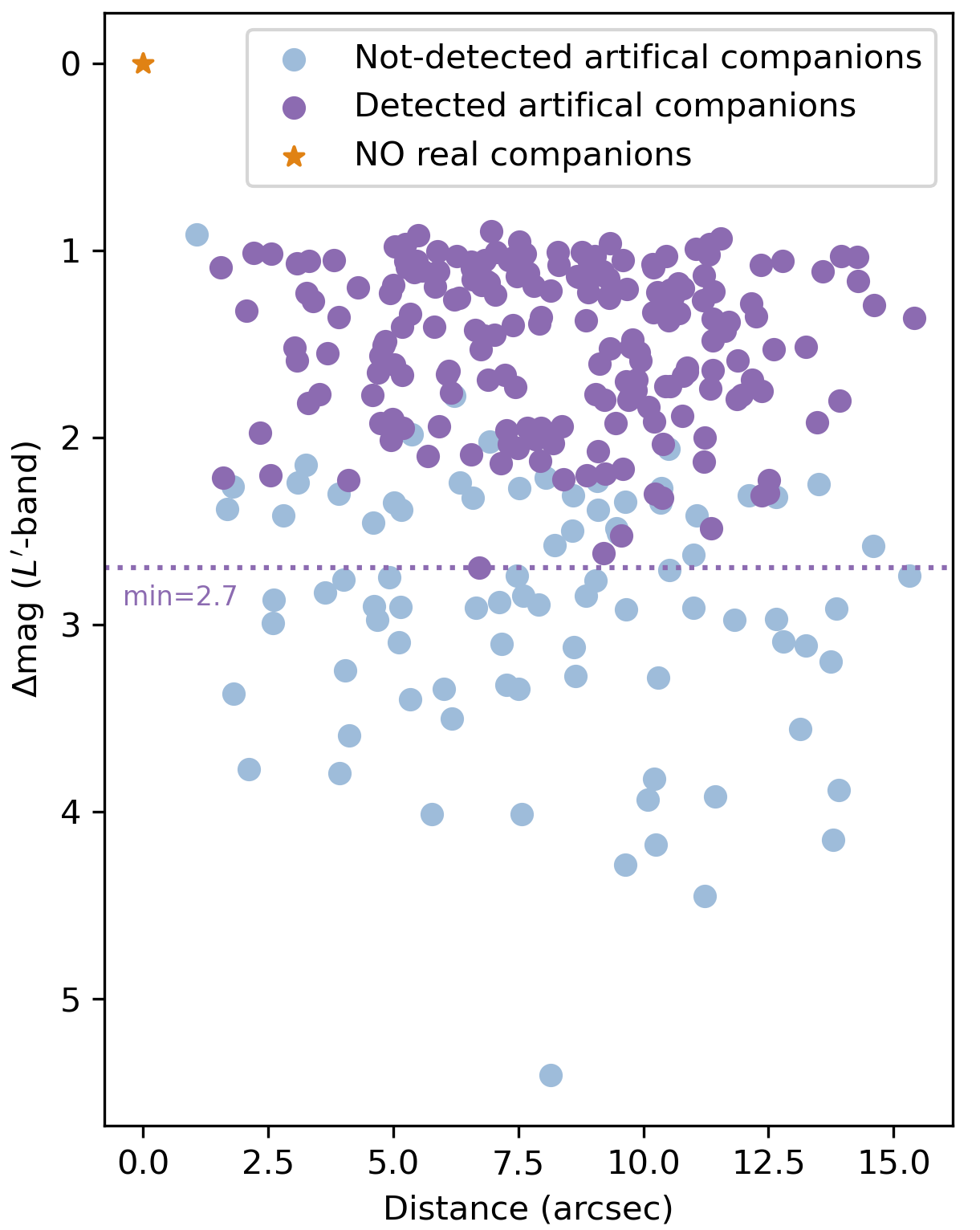}}
        \caption{ Detection limit artificial source tests for all the targets. We display only one iteration. The orange star symbols stand for the real detected companions (if any) and the plain dot symbols refer to artificially injected sources. We display the magnitude contrast of the faintest (i) source detected in the real image (orange dot and line) and (ii) artificial source (purple dot and line). }
        \end{figure*}

\section{Notes on the previously studied MYSOs}
\label{appendix:Notes}

\textbf{G194.9349: }\\
This source has been the object of NIR spectroscopy \citet{Cooper+2013} and far-IR observations to detect the presence of outflows and characterise the infall dynamics through the SiO(8-7) emission line on one side, and the $HCO^{+}/H^{13}CO^{+}$ emission on the other hand \citep{Cunningham+2018}. They report a non-detection of SiO but the presence of both $HCO^{+}$ and $H^{13}CO^{+}$. The former non-detection could either be associated with the absence of an active outflow or a potential fossil remnant-driven outflow. The structures associated with G194.9349 cannot be confirmed at this stage and may require higher spatial resolution observations where infall motions onto the protostar and outflows can be resolved. 

\textbf{G203.3166: }\\
G203.3166 known under various names (NGC224 IRS1, AFGL989, Allen's source) has been studied at different wavelengths and angular resolutions. \citet{Schreyer+2003} described G203.3166 as a young B-type star, devoid of disk but surrounded by low-mass companions located in a low-density cavity and surrounded by denser gas. As part of the commissioning of the NICMOS ($K-$band) camera on board Hubble Space Telescope (HST), \citet{Thompson+1998} report the discovery of six point objects at projected separations of 2.6\arcsec to 4.9\arcsec from the central source. 
In the $L-$band NACO images it seems that we observe four out of the six objects reported in \citet{Thompson+1998}: objects G (\#6), E(\#5), D(\#4) and C(\#2). \citet{Schreyer+2003} further complete the picture with $K-$ band (SOFI at ESO's NTT) and mid-infrared (4.6 and 11.9~\um, TIMMI2 st ESO's 3.6m telescope) observations. Within the field of view of our interest, they confirm the presence of the 6 objects detected by \citet{Thompson+1998} and report the detection of two new objects (\#7 and \#8), that respectively coincide with object F and the binary system (A and B) in our images. The $K-$band polarisation map displayed in Figure 2, clearly shows that object \#8 has its own polarisation pattern confirming the presence of circumstellar dust. Object \#8 being brighter at longer wavelengths further supports the presence of dusty circumstellar material. 
The advent of AO-assisted images allowed us to resolve the binary system as shown in our images, which was not the case in the NICMOS/Hubble images. \citet{deWit+2009} looked at the 24.5~\um  emission of G203.3166 featuring COMICS observations. In their $6\arcsec\times6\arcsec$ image, they do not detect companions. \\
\citet{Grellmann+2011} explored the closer IR environment of G203.3166 seeking for closer companions. The interferometric observations (MIDI/VLTI) do not provide any hint of resolved companions around G203.3166, in the 30-230~au separation range. However, a model of a geometrically flat but optically thick circumstellar disk successfully reproduces the short-wavelength region of an SED and visibilities (SED taken with SWS/ISO satellite). Finally, a multi-scale analysis by \citet{Frost+2021a} fitting these visibilities, the 24.5$\mu$m image profile and the SED simultaneously found that the source is likely to harbour a small dust disk (0.3\Msun with a maximum radius of 500\Rsun) which contains a gap-like substructure, surrounded by a dusty envelope with outflow cavities.


\textbf{G232.6207: }\\
NIR and MIR observations ($K-$,$L-$ and $N$-band) conducted by \citet{Walsh+1999} and \citet{Walsh+2001} have shown that the source is composed of a compact centre with a very reddened object and diffuse emission. \citet{DeBuizer+2017} further confirm the presence of a NIR/MIR bright compact core and extended emission, using 11.7~\si{\micro\meter} Gemini South images, \emph{Spitzer} 8~\si{\micro\meter} and 31 and 37~\si{\micro\meter} SOFIA data. Both \citet{Walsh+2001} and \citet{DeBuizer+2017} highlight the influence of a putative outflow that is driven by a radio continuum source, coinciding with the NIR/MIR source. More recently, \citet{Zhang+2019} reported that high angular resolution observations with ALMA of the 1.3mm continuum and the H30$\alpha$ recombination line emission confirmed the central object to be a massive protobinary with apparent separation of 180~au. The massive central binary is expected to have a total mass of 18\Msun and a maximum period of 570~years. They also report features exhibiting stream-like structures, at the $10^4$~au scales. Their measurements show evidence for ongoing accretion at different spatial scales: from infalling streams at $1000-10000$~au, to forming a massive binary on $100-1000$~au scales via a circumbinary disk and finally to circumstellar accretion disks feeding each star separately on 10~au scales. Apart from the central object, they detect three other compact sources within 17000~au, two of them being low-mass protostars (0.022 and 0.014\Msun) and lying respectively at 6\arcsec and 9\arcsec away from the central massive binary. The third one 9\arcsec away (0.25\Msun) from the binary was not retained as a protostellar because of its non-point-like core shape.

\textbf{G254.0548: }\\
G263.7434 has no radio counterpart as found by \citet{Urquhart+2007}. They report the protostar to be unresolved with spherical morphology. It would seem that this object has not been further studied, there is no archive with images at NIR or MIR wavelengths. 

\textbf{G263.7434: }\\
G263.7434 has been observed at 20~\um as part as the VISIR and COMICS campaign lead by \citet{Wheelwright+2012}. However, they report the protostar to be unresolved and could not conduct the modelling of a putative envelope as they did for the other sources in their sample. 

\textbf{G263.7759: }\\
In the resolved 20~\um\,  images, \citet{Wheelwright+2012} noticed a bipolar morphology that contrasts with the circular symmetric morphology revealed by most of the resolved MYSOs in their sample. They suggest that the morphology traces the cavities of a bipolar outflow seen close to edge on. Indeed, the extension from the NW to the SE direction they found is in clear agreement with a well-collimated $H_{2}$ jet detected by \citep{Giannini+2005}. The jet seems to be driven by several different sources that they detected thanks to ISAAC/VLT. The structure of the jets and the various associated knots is also well seen in the OSIRIS/SOAR narrow-band $H_{2}$ images \citet{Navarete+2015}. Being a resolved object, \citet{Wheelwright+2012} could model the SED and intensity profile of G263.7759 with an infalling envelope of mass 15.5\Msun and a mass infall rate of $1.5\times10^{-4}$\Msun$yr^{-1}$. 

\textbf{G265.1438: }\\
G265.1438 is classified as an accreting Class 0/I YSO of about 1.1~Myr \citep{Ellerbroek+2013}. The latter authors lead a multi-wavelength study of the RCW 36 region which includes NIR/MIR imaging and photometry (NTT/SOFI and Spitzer/IRAC), as well as NIR integral field spectroscopy (VLT/SINFONI) and optical spectroscopy (VLT/X-SHOOTER). They report that G265.1438 shows a highly reddened spectrum with many emission lines and outflow features. They also detect the presence of a dust shell or disk that might responsible for the severe reddening. \citet{Wheelwright+2012} had already reported the possible presence of outflow and their model is compatible with an envelope of mass 15.5\Msun and similar mass infall rates as G263.7759, viewed at relatively low inclination ($i=32^{\circ}$). 

\textbf{G268.3957: }\\
G268.3957 was observed as part of a VISIR/VLT campaign and was spatially resolved in the MIR. \citet{Wheelwright+2012} explore the nature of its circumstellar structure using radiative transfer models. G268.3957 exhibits a rather circular symmetric morphology in the MIR, a consequence of the low inclination at which it is observed ($i=30^{\circ}$). The spatial extension of its 20~\um emission and its SED is better reproduced with a model featuring an infalling ($3.5\times10^{-4}$\Msun$yr^{-1}$) rotating envelope ($9.4\times10^{3}$~\Msun) with outflow cavities ($10^{\circ}$ opening angle). No companions were reported within the $30\arcsec\times30\arcsec$ field of view of the VISIR (20~\um) observations. 

\textbf{G269.1586: }\\
\citet{Wheelwright+2012} observed G269.1586 as part of their VISIR/VLT campaign and report the object to be spatially resolved in the MIR. They also report a distinct localised emission at the NE in the 20~\um\, image, that they associate with the \Hii region detected by \citet{Urquhart+2007}, in the radio regime. They describe a cometary-like morphology for both objects. \citet{Wheelwright+2012} suggest that the exotic morphology, for which they cannot model the intensity profile, may indicate that the object is experiencing a different phase of massive star formation than the other MYSOs observed in their sample, a fact that is supported by the nearby presence of the \Hii region. As for G268.3957, \citet{Wheelwright+2012} could model the SED with an infalling envelope model. This is the most massive envelope they derive, with a mass of 32.2~\Msun and a mass infall rate of $7.5\times10^{-4}$\Msun$yr^{-1}$. 

\textbf{G305.2017: }\\
Using two sets of spatially resolved data in the MIR (MIDI/VLTI and VISIR/VLT) together with an SED, \citet{Frost+2019} constrained the geometry of the MYSO. Using a 2.5D radiative transfer models fit to N-band interferometric data, Q-band imaging data and an SED simultaneously, they show that the best model implies the source is surrounded by a large extended dusty envelope featuring bipolar cavities and a dusty disk. They further concluded that an inner hole is likely present in the disk and the low envelope density indicates that the protostar is at an advanced stage in its formation, similar to a transition disk phase of low-mass star formation. 

\textbf{G310.0135: }\\
G310.0135 is one of the first MYSOs around which a disk of scales small enough to be an accretion disk was found which has been studied using diverse multi-wavelengths techniques \citep{Kraus+2010,Ilee+2013,Boley+2016,Caratti+2015a,Caratti+2016}. The $\sim17.3$~au-diameter compact disk revealed by NIR $K-$band interferometry \citep{Kraus+2010} is in Keplerian rotation \citep{Ilee+2013}. Perpendicular to the disk plane, \citep{Kraus+2010} observed a molecular outflow and two bow shocks, showing that a bipolar outflow originates from the inner regions of the MYSO. The MIR observations lead by \citep{Wheelwright+2012} that trace the warm dust show an extended emission along the direction of the outflow. Being resolved at $\sim20$~\um the latter authors could model the SED and intensity profile. The model predicts an infalling envelope of 9.8~\Msun with a mass infall rate of $7.5\times10^{-4}$\Msun$yr^{-1}$, and an inclination of $i$=32$^{\circ}$, in agreement with the intermediate inclination of 45 found by \citet{Kraus+2010}. 

\textbf{G314.3197: }\\
\citet{Wheelwright+2012} observed G314.3197 as part of their VISIR/COMICS campaign. The 20~\um  image shows a spherically symmetric object whose circumstellar environment is unresolved at the resolution of 0.6\arcsec, most likely as a result of the great distance to the protostar. 

\textbf{G318.0489: }\\
G318.0489 was observed during the VISIR/COMICS observing campaign lead by \citet{Wheelwright+2012}. They report a slight cometary morphology emanating from the 20~\um  image. Facing the inaccuracy around the distance to this source, \citet{Wheelwright+2012} chose to not attempt to reproduce its intensity profile and SED as they cannot be unambiguously modelled. \citet{Navarete+2015} shows the continuum-subtracted $H_{2}$ maps of G318.0489. The area surrounding the MYSO does not show any particular excess in the $H_{2}$ emission nor any excess in the continuum filter. However, the protostar is associated with a radio-quiet source.

\end{appendix}

\end{document}